\documentclass[a4paper,fleqn,usenatbib]{mnras}

\usepackage{newtxtext,newtxmath}

\usepackage[T1]{fontenc}
\usepackage{ae,aecompl}
\usepackage[caption=false]{subfig}
\usepackage{footnote}

\usepackage{graphicx}	
\usepackage{amsmath}	
\usepackage{amssymb}	
\usepackage{textcmds}  
\usepackage[normalem]{ulem} 
\usepackage{lscape}

\title[Radio-loud AGN properties]{A VLA$-$GMRT Look at 11 Powerful FR~$\mathtt{II}$ Quasars}

\author[Vaddi et al.]{
S. Vaddi,$^{1}$\thanks{E-mail: sravani.vaddi@gmail.com}
P. Kharb, $^{1}$
R. A. Daly, $^{2}$
C. P. O'Dea, $^{3,4}$
S. A. Baum, $^{3,4}$
D. K. Deo, $^{5,7}$
\newauthor
T.C. Barbusca, $^{2}$
C. Murali $^{6,7}$
\\
$^{1}$National Centre for Radio Astrophysics - Tata Institute of Fundamental Research, Ganeshkhind, Pune 411007, India\\
$^{2}$Penn State University, Berks Campus, P. O. Box 7009, Reading, PA 19610-6009\\
$^{3}$Physics and Astronomy, University of Manitoba, Winnipeg, MB R3T 2N2 Canada\\
$^{4}$Rochester Institute of Technology, Rochester, New York, 14623, USA\\
$^{5}${University of Missouri-Kansas City, Kansas City, Missouri, 64110, USA}\\
$^{6}${University of Texas at Dallas, Texas, 75080, USA}\\
$^{7}$Indian Institute of Astrophysics, II Block, Koramangala, Bangalore 560034, India\thanks{Work towards this project was performed at this institute.}
}

\date{Accepted XXX. Received YYY; in original form ZZZ}

\pubyear{2018}

\begin{document}
\label{firstpage}
\pagerange{\pageref{firstpage}--\pageref{lastpage}}
\maketitle

\begin{abstract}
		We present results from 1.4 and 5~GHz observations at matched resolution with the Karl G. Jansky Very Large Array (VLA) of 11 powerful 3C FR~$\mathtt{II}$ quasars. We examine the 11 quasars along with a sample of 13 narrow-line FR~$\mathtt{II}$ radio galaxies and find that radio-loud unification largely holds but environmental effects cannot be ignored. {The radio core prominence, largest linear size, and axial ratio parameter values indicate that quasars are at relatively smaller angles compared to the radio galaxies and thus probe orientation.  Lack of correlation between statistical orientation indicators such as misalignment angle and radio core prominence, and larger lobe distortions in quasars  compared to radio galaxies suggest that intrinsic/environment effects are also at play. Some of 150~MHz observations with the TGSS-GMRT reveal peculiar lobe morphologies in these FR~$\mathtt{II}$ sources, suggesting complex past lives and possibly restarted AGN activity. Using the total 150~MHz flux density we estimate the time-averaged jet kinetic power in these sources and this ranges from $(1 - 38)\times 10^{45}$~erg~s$^{-1}$, with 3C~470 having the highest jet kinetic power. }

\end{abstract}




\section{Introduction}

Active galactic nuclei (AGNs) are the centres of a special class of galaxies that consist of actively accreting central supermassive black hole (SBH). High-speed collimated outflows are present in a small fraction of AGNs. AGNs show a dichotomy in their radio power - these are referred to as radio-loud (RL) and radio-quiet (RQ) AGNs \citep{Strittmatter1980, Miller1993}.  Historically, AGNs with higher radio-to-optical flux $R\ge10$ where  $R \equiv S_{5~GHz} / S_{B-band}$ are termed as RL AGNs and RQ otherwise \citep{Kellermann1989}. This distinction however, is suggested to be valid only for broad-lined, unobscured AGNs since AGN luminosity estimates may be affected by dust, and therefore  \cite{Padovani2017} suggested a nomenclature based on a fundamental physical difference, namely the presence or absence of relativistic jets.

Based on the radio morphology and power, extended RL AGNs are classified into FR~$\mathtt{I}$ and FR~$\mathtt{II}$ \citep{Fanaroff1974}. FR~$\mathtt{II}$ sources have radio structures consisting of a core, collimated jets, lobes, and hotspots located at the edges of these lobes. FR~$\mathtt{I}$s show extended plume-like structures and tails with no distinct collimated jets or terminal hotspots \cite[{more detailed differences in}][]{Bridle_Perley1984}.  Based on the optical spectra RL AGNs are classified into Type 1 (have broad and narrow emission lines) and Type 2 (have only narrow emission lines).  Type 2 AGNs comprise narrow-line radio galaxies (NLRG); Type 1 AGNs {comprise} broad-line radio galaxies (BLRG) at low luminosity and radio-loud quasars at high luminosity.  The latter are further divided into flat spectrum radio-loud quasars (FSRQ) ($\alpha>0.5$) and steep spectrum radio-loud quasars (SSRQ) ($\alpha<0.5$) (see \cite{Urry1995, Tadhunter2008, Netzer2015, Tadhunter2016} for a review).  SSRQs tend to have lobe-dominated radio structure while FSRQs often have core-dominated structure. Although we see different types of AGN, it is hypothesised that these are intrinsically similar objects but appear different depending on their orientation with respect to the line of sight.  This has come to be known as orientation-based AGN Unification Scheme \citep{Antonucci1993, Urry1995}.  

There is compelling evidence in the literature that is consistent with the predictions of the unification scheme.  Some of the evidences include detection of broad emission lines in the polarized intensity spectra of Type 2 AGNs indicating a hidden broad line region obscured from direct view \citep{Antonucci1984, Antonucci_Miller1985, Tran1995, Young1996, Ogle1997, Cohen1999}, detection of ionisation cones of the narrow line region in optical images of Type 2 AGNs \citep{Pogge1988, Wilson1993, Jackson1998}, detection of excess near-IR emission and polarized light perpendicular to the jet axis \citep{Antonucci1990}, apparently superluminal velocities in VLBI observations of compact radio sources \citep{Cohen1977, Kellermann2007}, detection of extended halos around compact sources representing radio lobes viewed face-on \citep{Browne1982, Antonucci_Ulvestad1985}, {and} observed anisotropy in the radio structure {\citep{Laing1983, Barthel1989, Smith1990, Mullin2008}. According to unification, FSRQ form the beamed counterparts of FR~{\tt II} radio galaxies, and SSRQ (and BLRG) are oriented at angles that are in between NLRG and FSRQ.}

This paper is the first in a series of papers in our study of FR~{\tt II} radio-loud AGNs.  Here we present the results from our VLA study of 11 SSRQ that show classical double lobes at GHz frequencies. We have analysed the properties of these quasars along with a sample of 13 FR~$\mathtt{II}$ radio galaxies from \citet{Kharb2008}. Interestingly, several of these quasars and radio galaxies reveal {extended}, winged or asymmetric radio morphologies in the TIFR GMRT Sky Survey (TGSS) at 150~MHz.  {This is in contrast to} the symmetric structures detected with the VLA at GHz frequencies.

Throughout this paper, we assume a cosmology with H$_0$ = 73~km~s$^{-1}$~Mpc$^{-1}$, $\Omega_{mat}$ = 0.27, $\Omega_{vac}$ =  0.73. Spectral index, $\alpha$, is defined such that flux density at frequency $\nu$ is $S_\nu\propto\nu^\alpha$.

\section{The Sample}
\label{sec:data}
The 11 quasars in this study were selected to augment the sample of 55 radio sources with spin determinations described by \citet{Daly2011}.  The parent population is powerful FR~{\tt II} sources that are identified as quasars, where the original published optical classification was used to identify them as quasars, and this identification was confirmed by the follow-up observations of \cite{Mclure2006}; these 11 sources were identified based on the following criterion: (i) the quasar has a black hole mass determination (from \cite{Mclure2006}); (ii) at least one side of the source does not exhibit a radio jet, so that a spectral ageing analysis can be carried out on the non-jetted side; (iii) the non-jetted lobe(s) have angular sizes large enough to allow the spectra of the lobe to be measured at several locations. 

\subsection{Observations and Data Reduction}
These 11 quasars were observed with the VLA at 1.4 GHz (L-band) and 5~GHz (C-band) with A and CnB-array configurations, respectively (Project ID: 10C-178) for a total of 22 hours. 5~GHz observations were not scheduled for 3C~336\footnote{One of the observations (for 3C432) had technical issues and was therefore re-observed. However, due to confusion in the sequencing of observations, we missed obtaining data for 3C336. This was realised {late} and could not be rectified}.  Hence, we used archival VLA data\footnote{Project ID: AB0454; C-band B-array configuration} for this source.  Also, due to the noisy 1.4 GHz data on 3C~351, we chose archival data\footnote{Project ID: AD0429; L-band A-array configuration} for this source as well.  For 3C~208 at 5~GHz, since there were issues with amplitude calibration using 3C~48,  the source flux was rescaled using the \textsc{aips} task {\tt RESCALE} and the core flux density from \cite{Yuan2012}.

The data in the SDMset format were pre-calibrated using the \textsc{casa} calibration pipeline\footnote{\textsc{casa} release 4.5.3 with pipeline release 4.6.0}. The final imaging and self-calibration were done in the Astronomical Image Processing System (\textsc{aips}), by iteratively running the tasks {\tt IMAGR} and {\tt CALIB}.  The 1.4$-$5 GHz spectral index images were made using task {\tt COMB} after first creating L and C-band images with identical circular beams (at the poorer L-band resolution), which were then positionally aligned using the {\tt AIPS} task {\tt OGEOM}. We used 3$\sigma$ r.m.s. noise of the respective L and C-band images as the clipping level input to {\tt COMB}. Final flux density and spectral index values were estimated using the {\tt AIPS} verbs {\tt TVWIN} and {\tt IMSTAT}, as well as the Gaussian-fitting task {\tt JMFIT}. All extents were measured in {\tt AIPS} using {\tt IMDIST}. The basic properties and derived parameters are listed in Tables ~\ref{tab:quasar_basic_prop} through ~\ref{tab:quasar_hotspots}. 

\subsection{Data from the Literature}
In order to put the results from the 11 quasars in perspective, we have analysed their results along with the sample of {13} FR~$\mathtt{II}$ {narrow-line} radio galaxies that were studied in \citet{Kharb2008}; the radio galaxy study had been carried out at similar frequencies and resolutions {as} the 11 quasars presented in this paper. Some properties of the radio galaxies have been re-estimated (like $R_c$, 1.4 GHz flux density from NVSS) to compare them with the current quasar sample, and are tabulated in Table~\ref{tab:radiogalaxy_prop}. We note that the combined quasar and radio galaxy sample studied here is statistically small and eclectic. Some statistical tests, therefore, suffer from small number statistics. 

\subsection{Physical Parameters Estimation Method}
\label{sec:param}
The spectral index is estimated from the spectral index images.  The spectral index of the core is obtained by placing a box around the core using {\tt TVWIN} and {estimating} the mean using {\tt IMSTAT}. The hotspot spectral index is measured at the image position corresponding to the peak intensity in {1.4 and} 5 GHz image, where the peak intensity is determined using {\tt TVMAXFIT}.

The arm-length ratio is defined as the ratio of the longer core-hotspot distance to the shorter core-hotspot distance. The distance is measured from the peak flux position at the core to the peak flux position at the hotspot. The peak flux density was estimated using the procedure {\tt TVMAXFIT}. The 5 GHz radio maps were used in the determination of the arm-length ratio.

The axial ratio is defined as the ratio of the length to the width of the lobe \citep{Leahy1989}. The core-hotspot distance is taken as the length of the lobe.  Although lobe emission is not observed to be extending all the way to the core in the 1.4 GHz emission, lower frequency images (610 MHz GMRT observations, Vaddi et al. 2018  in preparation) show extended emission and thus it is safe to assume the lobe length to be the core-hotspot distance. The FWHM of the lobe is measured using the Gaussian-fitting task {\tt SLFIT} to a one-dimensional slice across the lobe.  The slice is chosen by-eye {to be} perpendicular to and at half the distance between the core and hotspot. 
{To account for the finite resolution of the observing beam, we subtract the beam in quadrature from the measured Gaussian fit: $\Theta_{decon}^2 = \Theta_G^2 - \Theta_b^2$ where $\Theta_G$ is the measured FWHM from the Gaussian fit, and $\Theta_b$ is the average FWHM of the observing beam ($\sqrt(FWHM_{major~axis} * FWHM_{minor~axis})$), and $\Theta_{decon}$ is the FWHM of the Gaussian fit deconvolved with the beam.  The lobe width is then taken as 2/$\sqrt 3$ of the deconvolved Gaussian fit \citep{Leahy1989}. Correction for the radio galaxies is negligible since its lobe widths are large compared to the beam size \citep{Daly2010}.}

{The largest linear size (LLS) is defined as the distance between two hotspots (definition consistent with \cite{Kharb2008}).  The distance is estimated by first obtaining the position of the hotspot peak intensity with {\tt TVMAXFIT} and then measuring the distance between the hotspot peak intensity using {\tt IMDIST}.  The 5~GHz maps were used in the estimates.}

{The misalignment angle is the difference between the position angles of the core-hotspot segments on either side of the core. When there is more than one hotspot on one side of the core, the other side is used to define the core-hotspot line segment and misalignment angles to each of the two hotspots on the opposite side are obtained. Figure~\ref{fig:example_fig} in the appendix identifies different components that were estimated using AIPS task {\tt TVDIST}.}


\subsection{Errors}
Uncertainties in the flux measurements were estimated by combining all errors in quadrature. {The errors considered are the calibration errors, rms, and random errors ($\sigma = \sqrt{\sigma^2_{cal} + \sigma^2_{rms} + \sigma^2_{random}}$). We assumed that the calibration errors are 10 per cent of the measured flux densities.  RMS is the error reported in the tasks ({\tt JMFIT}, {\tt IMSTAT}) used for flux measurement.  Random error of 5 per cent accounts for the different values we get when taking repeated measurements; {an average of five} measurements were obtained for a few sources and the per cent standard deviation from the average values was noted. This gives the uncertainties of the order of few mJy~beam$^{-1}$. 

The error in the spectral index is estimated by adding in quadrature the noise map produced in the {\tt COMB} task and a 5 per cent random error for $\alpha_{HS}$, 15 per cent random error for the $\alpha_{core}$, and a 10 per cent random error for the $\alpha_{lobe}$. 

The uncertainty on the measurement of the lobe width is obtained by adding in quadrature the 5 per cent random error on the measured width and the error in the Gaussian fit to the slice. {The uncertainty of the deconvolved lobe width is obtained using the expression { $\delta\Theta_{decon}^2 = (\Theta_{decon} /\Theta_{G})^2 \delta\Theta_G^2 + (\Theta_{decon}/\Theta_{b})^2 \delta\Theta_b^2$} where $\delta\Theta_G$ is the uncertainty of the lobe width and $\delta\Theta_b$ is the uncertainty of the observing beam.  Lobes whose widths are similar to the beam have larger uncertainty.

The uncertainties on angular size, arm-length ratios are estimated by repeated measurements (five times) on few sources and taking the standard deviation of these measurements.  The ratios are calculated using the usual error propagation. These uncertainties are between 5$-$20 per cent.

The LLS is estimated from peak hotspot-to-hotspot distance which need not be the true LLS and true LLS is usually larger than the hotspot-hotspot distance.  To quote the uncertainty in the LLS due to this, we estimate an average of several lobe edge-to-edge distance measurement obtained using {\tt TVDIST} and the deviation of this estimate with the hotspot-hotspot distance is taken as the uncertainty. The uncertainty of the misalignment angle is estimated based on the uncertainties of the hotspot and core positions.
}

\begin{landscape}
	\begin{table}
		\caption{Basic properties of the radio-loud quasar sample}
		\label{tab:quasar_basic_prop}
		\begin{tabular}{lcccccccccccc}
			\hline
			Name	&	RA	&	DEC	&	z	&	log~$M_{BH}$	&	{S$^{1.4}_{total}$}	&	{S$^{5}_{total}$} 	&	{S$^{150}_{total}$}	&	{LLS}	&	{$\log R_c$}	&	{Arm-length}	&	{Misalignment} 	&	{$Q_{jet}$}\\
				&	J2000	&	J2000	&	&	M$_{\odot}$	&	{(Jy)}	&	{(Jy)}	&	{(Jy)}	&	{(kpc)}	&	{}	&	{Ratio}	&	{Angle (deg)}	&	{($10^{45}$ erg s$^{-1}$)}\\
				(1)&(2)&(3)&(4)&(5)&(6)&(7)&(8)&(9)&(10)&(11)&(12)&(13)\\
			\hline
			3C14 	&	00:36:06.5	&	+18:37:59	&	1.469	&	9.4 	& 	1.97$\pm$0.03	&	0.68$\pm$0.01	&	13.51$\pm$2.03	&	201.9	&	-2.65$\pm$0.05	&	1.59	&	9.3, 24.8	&	9.71$\pm$0.33\\
			3C47	&	01:36:24.4	&	+20:57:27	&	0.425	&	9.2	&	3.59$\pm$0.01	&	1.01$\pm$0.01	&	36.33$\pm$4.46	&	374.1	&	-1.86$\pm$0.05	&	1.23	&	6.4	&	3.63$\pm$0.05\\
			3C109	&	04:13:40.4	&	+11:12:14	&	0.306	&	8.3	&	3.41$\pm$0.02	&	1.26$\pm$0.01	&	20.72$\pm$2.73	&	407.6	&	-1.18$\pm$0.05	&	1.04	&	13.7	&	1.30$\pm$0.04\\
			3C204	&	08:37:44.9	&	+65:13:35	&	1.112	&	9.5	&	1.20$\pm$0.02	&	0.35$\pm$0.01	&	13.88$\pm$1.86	&	282.8	&	-1.68$\pm$0.05	&	1.40	&	3.3, 3.4	&	5.29$\pm$0.24\\
			3C205	&	08:39:06.4	&	+57:54:17	&	1.534	&	9.6	&	2.29$\pm$0.08	&	0.55$\pm$0.02	&	21.53$\pm$3.20	&	130.5	&	-2.33$\pm$0.05	&	1.17	&	0, 7.5	&	15.3$\pm$0.30\\
			3C208	&	08:53:08.8	&	+13:52:55	&	1.11	&	9.4	&	2.39$\pm$0.1	&	0.48$\pm$0.02	&	23.08$\pm$3.56	&	88.60	&	-1.97$\pm$0.05	&	1.28	&	0	&	10.60$\pm$0.23\\
			3C249.1	&	11:04:13.7	&	+76:58:58	&	0.312	&	9.3	&	2.12$\pm$0.01	&	0.68$\pm$0.01	&	18.26$\pm$2.60	&	199.2$^{\dagger}$	&	-1.53$\pm$0.05	&	2.04	&	7.6	&	1.15$\pm$0.05\\
			3C263	&	11:39:57.0	&	+65:47:49	&	0.646	&	9.1	&	2.96$\pm$0.05	&	1.03$\pm$0.02	&	20.37$\pm$3.07	&	320.5	&	-1.53$\pm$0.05	&	1.72	&	1.6	&	4.42$\pm$0.12\\
			3C336	&	16:24:39.1	&	+23:45:12	&	0.927	&	9.2	&	2.64$\pm$0.06	&	0.38$\pm$0.01	&	14.47$\pm$2.23	&	172.0	&	-2.61$\pm$0.05	&	2.00	&	13.6	&	5.83$\pm$0.21\\
			3C351	&	17:04:41.4	&	+60:44:31	&	0.372	&	9.5	&	2.65$\pm$0.03	&	1.08$\pm$0.01	&	16.94$\pm$2.36	&	316.8	&	-2.25$\pm$0.05	&	2.23	&	5.8, 17	&	1.50$\pm$0.06\\
			3C432	&	21:22:46.2	&	+17:04:38	&	1.785	&	10.1	&	1.62$\pm$0.07	&	0.34$\pm$0.01	&	14.64$\pm$2.30	&	109.8	&	-2.87$\pm$0.05	&	1.33	&	11.9	&	11.70$\pm$0.38\\
		\hline
		\multicolumn{13}{p{1.2\textwidth}}{Col.(1-3) give the source name and position. Col(4)- Redshift from NED.  Col.(5)-Logarithm of black hole mass from \citet{Mclure2006}. Col(6-8)- Total flux density at 1.4 GHz, 5 GHz, and 150 MHz respectively in Jy Beam$^{-1}$.  Col(9)- Largest linear size which is the distance between the two hotspots, measured using {\tt IMDIST} {on 5~GHz images}. Col(10)- Logarithm of the radio core prominence.  Col(11)- Arm-length ratio which is the distance between the core to the hotspot, measured using {\tt TVDIST} {on 5~GHz images}.  Col(12)- Misalignment angle in degree.  {Where multiple hotspots are present, angle to each of the hotspots is mentioned separated by a comma. An angle of zero indicates that the hotspots and core all lie on a single line. 5~GHz images are used to get the angles.  The uncertainty on the misalignment angles is about a tenth of a degree.} Col(13)- Jet kinetic power.\newline {$^{\dagger}$ Since this source has only one hotspot, the LLS for this source is measured using {\tt TVDIST} and choosing the end points by eye.}}
		\end{tabular}
	\end{table}
\end{landscape}

	\begin{table}
		\caption{Properties of hotspots}
		\label{tab:quasar_hotspots}
		\begin{tabular}{lccccc}
			\hline
			{Name}	& {Side} &	{S$^{5}_{peak}$}	&	{$\alpha^{5}_{1.4}$}	&	{$\Theta$}\\
			{}	&	{}	&	{(mJy beam$^{-1}$)}	&	{}	&{(arcsec)}\\
			(1)&(2)&(3)&(4)&(5)\\
			\hline
			3C14 	&	N1	&	81.90	&	-0.98$\pm$0.16	&	9.440\\
			&	N2	&	59.00	&	-1.15$\pm$0.18	&	6.960\\
			&	S	&	111.7	&	-0.86$\pm$0.14	&	14.97\\                
			3C47	&	N	&	20.6	&	-0.77$\pm$0.15	&	31.19\\		
			&	S	&	163.0	&	-0.38$\pm$0.16	&	38.2\\				
			3C109	&	N	&	31.9	&	-0.45$\pm$0.11	&	46.09\\
			&	S	&	211.0	&	-0.66$\pm$0.12	&	48.12\\						
			3C204	&	E	&	63.90	&	-0.72$\pm$0.12	&	18.13\\
			&	W1	&	85.60	&	-0.62$\pm$0.10	&	13.01\\				
			&	W2	&	24.10	&	-1.04$\pm$0.17	&	17.12\\
			3C205	&	N	&	104.8	&	-1.22$\pm$0.19	&	7.230\\
			&	S1	&	211.3	&	-0.94$\pm$0.15	&	8.480\\
			&	S2	&	207.0	&	-1.10$\pm$0.18	&	8.650\\
			3C208	&	E	&	1356.	&	-0.62$\pm$0.10	&	6.220\\
			&	W	&	205.3	&	-0.46$\pm$0.07	&	4.860\\
			3C249.1	&	E	&	83.20	&	-0.75$\pm$0.12	&	7.620\\
			&	W	&	85.80	&	-0.81$\pm$0.13	&	15.51\\						
			3C263	&	E	&	27.50	&	-0.52$\pm$0.08	&	27.88\\
			&	W	&	486.7	&	-0.66$\pm$0.11	&	16.24\\				
			3C336	&	N	&	110.7	&	-1.36$\pm$0.21	&	15.01\\
			&	S	&	65.90	&	-1.37$\pm$0.22	&	7.530\\		
			3C351	&	N1	&	330.0	&	-0.73$\pm$0.12	&	27.29\\
			&	N2	&	194.0	&	-0.76$\pm$0.13	&	25.02\\
			&	S	&	3.366	&	-0.91$\pm$0.16	&	60.79\\		
			3C432	&	N	&	130.0	&	-1.31$\pm$0.21	&	5.700\\
			&	S	&	109.0	&	-0.91$\pm$0.15	&	7.590\\
		\hline
			\multicolumn{6}{p{.4\textwidth}}{Col.(2): N, S, E, W denote north, south, east, and west hotspots respectively;N2, S2, W2 are the second brightest hotspots in the north, west, and south sides.  Col.(3):Peak flux density of the hotspot at 5 GHz estimated using {\tt TVMAXFIT}.  {The errors are of the order of a few per cent ($<$5\%).} Col(4): Spectral index of the hotspot between 1.4 and 5 GHz corresponding to the peak flux density position at the hotspot.  This is obtained using {\tt IMVAL} on the spectral images.  Col(5): Distance from core to the hotspot in arcsec obtained using {\tt TVDIST} on 5 GHz intensity images. { The uncertainties are $\ll$1\%.}}
		\end{tabular}
	\end{table}
	\begin{table}
		\caption{Properties of the core}
		
		\begin{tabular}{lrrc}
		\hline
			{Name} &	{S$^{1.4}_{total}$}	&	{S$^{5}_{total}$}	&	{$\alpha^{5}_{1.4}$}\\
			{}	&	{(mJy)}	&	{(mJy)}	&	{}\\
			(1)&(2)&(3)&(4)\\
		\hline
			3C14	&	24.3$\pm$2.7	&	10.9$\pm$1.6	&   $-$0.68$\pm$0.10\\
			3C47	&	58.2$\pm$6.5	&	69.2$\pm$7.5	&	+0.08$\pm$0.09\\
			3C109	&	248.7$\pm$27.8	&	274.0$\pm$30.1	&	$-$0.04$\pm$0.05\\
			3C204	&	58.1$\pm$6.5	&	50.3$\pm$5.8	&	$-$0.29$\pm$0.08\\
			3C205	&	40.4$\pm$4.5	&	26.6$\pm$3.3	&	$-$0.34$\pm$0.05\\
			3C208	&	59.4$\pm$6.6	&	51.0$\pm$5.9	&	$-$0.03$\pm$0.02\\
			3C249.1	&	76.0$\pm$8.5	&	78.9$\pm$8.8	&	+0.14$\pm$0.05\\
			3C263	&	148.1$\pm$16.6	&	137.6$\pm$15.6	&	+0.05$\pm$0.02\\
			3C336	&	38.9$\pm$4.4	&	11.1$\pm$2.2	&	$-$0.62$\pm$0.09\\
			3C351	&	35.5$\pm$4.0	&	20.3$\pm$2.7	&	$-$0.45$\pm$0.12\\
			3C432	&	10.2$\pm$1.2	&	6.1$\pm$0.8	&	$-$0.47$\pm$0.08\\
		\hline
		\multicolumn{4}{p{.4\textwidth}}{Col(2-3): Core flux density at 1.4 and 5 GHz respectively.  Col(4):Core spectral index between 1.4 and 5 GHz.  This is obtained using {\tt TVWIN} + {\tt IMSTAT} on the spectral index image.}
		\end{tabular}
	\end{table}
	
	\begin{table*}
		\caption{{Properties of the radio lobes}}
		\label{tab:quasar_prop_lobes}
		\begin{tabular}{lrrrrr}

			\hline
			{Name}	&	{$\alpha^{5}_{1.4}$}	&	{$\Theta_{width}$}	 &{B$_{min}$}	&	{P$_{min}$}	&	{Axial Ratio}\\
			{}	&	{N,S}	&	{N,S} &	{$\mu$G}	&	{10$^{-11}$dyne~cm$^{-2}$}	&	N,S\\
			(1)&(2)&(3)&(4)&(5)&(6)\\
			\hline
			3C14	&	-1.39$\pm$0.14, -1.39$\pm$0.15	&	2.4$\pm$0.3, 1.8$\pm$0.3	&	109.2$\pm$18.3, 95.2$\pm$18.3	&	59.34, 45.11	&	 4.0$\pm$0.5, 8.5$\pm$1.4\\
            
			3C47	&	-1.28$\pm$0.18, -1.28$\pm$0.18	&	31.0$\pm$4.7, 35.2$\pm$4.9	&	9.5$\pm$1.7, 9.5$\pm$1.7	&	0.45, 0.44	&	1.0$\pm$0.2, 1.1$\pm$0.1\\
            
			3C109	&	-1.18$\pm$0.18, -1.57$\pm$0.22	&	7.7$\pm$0.8, 15.6$\pm$1.6	&	16.9$\pm$2.4, 16.3$\pm$2.4	&	1.43, 1.32	&	 6.0$\pm$0.6, 3.1$\pm$0.3\\
            
			3C204$^\ddagger$	&	-1.42$\pm$0.17, -1.5$\pm$0.17	&	5.9$\pm$0.7, 4.9$\pm$0.6	&	35.6$\pm$7.0, 50.7$\pm$7.0	&	6.29, 12.76	&	3.1$\pm$0.4, 2.7$\pm$0.3\\
            
			3C205	&	-1.42$\pm$0.15, -1.13$\pm$0.11	&	2.1$\pm$0.3, 1.8$\pm$0.3	&	116.4$\pm$22.1, 110.5$\pm$22.1	&	67.44, 60.74	&	3.4$\pm$0.5, 4.7$\pm$0.8\\
            
			3C208$^\ddagger$	&	-0.75$\pm$0.09, -0.71$\pm$0.08	&	2.0$\pm$0.3, 1.3$\pm$0.3	&	79.7$\pm$7.9, 72.7$\pm$7.9	&	31.56, 26.28	&	3.1$\pm$0.4, 3.7$\pm$1.0\\
            
			3C249.1$^\ddagger$	&	..., -1.03$\pm$0.11	&	..., 7.6$\pm$0.8	&		..., 20.8$\pm$2.0	&		..., 2.15	&	..., 2.0$\pm$0.2\\
            
			3C263$^\ddagger$	&	-1.06$\pm$0.14, -0.88$\pm$0.09	&	12.2$\pm$1.5, 9.1$\pm$1.1 	&	18.3$\pm$2.2, 15.0$\pm$2.2	&	1.67, 1.11	&	 2.3$\pm$0.3, 1.8$\pm$0.2\\
            
			3C336	&	-2.01$\pm$0.2, -1.89$\pm$0.20	&	6.3$\pm$0.7, 4.9$\pm$0.6	&	93.7$\pm$25.6, 93.2$\pm$25.6	&	43.68, 43.22	&	2.4$\pm$0.3, 1.5$\pm$0.2\\
            
			3C351	&	-1.10$\pm$0.18, -0.92$\pm$0.18	&	14.3$\pm$1.7, 7.4$\pm$0.8	&	15.4$\pm$2.1, 7.2$\pm$2.1	&	1.19, 0.26	&	1.9$\pm$0.2, 8.2$\pm$0.9\\
            
			3C432	&	-1.20$\pm$0.13, -1.35$\pm$0.15	&	2.8$\pm$0.4, 2.6$\pm$0.4	&	103.5$\pm$16.2, 98.4$\pm$16.2	&	53.27, 48.16	&	2.1$\pm$0.3, 2.9$\pm$0.5\\
			
			\hline
			\multicolumn{6}{p{0.8\textwidth}}{Col(1): Source name.  Col(2): Spectral index of the lobe between 1.4 GHz and 5 GHz for the northern and southern lobes.  Col(3): Lobe width in arcsec {obtained using the 1.4 GHz images}.  Col(4): Magnetic field at minimum pressure condition obtained using Equation ~\ref{eq:bmin}.  Col(5):  Minimum internal pressure in the lobe obtained using Equation ~\ref{eq:pmin}. Col(6) Axial Ratios i.e., ratio of the lobe length (core-to-hotspot distance) to lobe width.\newline $^\ddagger$ The estimates are for the eastern and western lobe respectively.}
		
		\end{tabular}		
	\end{table*}
	
		\begin{table}
		\caption{Quasar properties - Jet sidedness}
        \label{tab:quasar_prop_jet}
		\begin{tabular}{lcccc}
			\hline
				{Name}	&	{Jet Side}	&	{Bright HS}	&	{Ratio of HS}	&	{Shorter}\\
				{}	&	{}	&	{Side}	&	{peak flux}	&	{Side}\\
				(1)&(2)&(3)&(4)&(5)\\
			\hline
	    3C14	&	S	&	S	&	1.36	&	N\\
        3C47	&	S	&	S	&	7.91	&	N\\
        3C109	&	S$^{\dagger}$	&	S	&	6.61	&	S\\
        3C204	&	W	&	W	&	1.34	&	W\\
        3C205	&	S	&	S	&	2.02	&	N\\
        3C208	&	E(?)	&	E	&	6.62	&	W\\
        3C249.1$^{\#}$&	-	&	-	&	-	&	N\\
        3C263	&	E	&	E	&	17.70	&	E\\
        3C336	&	S$^{\dagger}$	&	N	&	1.68	&	S\\
        3C351	&	N(?)	&	N	&	98.03	&	N\\
        3C432	&	S(?)	&	N	&	1.19	&	N\\
			\hline			
			\multicolumn{5}{p{.47\textwidth}}{$\dagger$ Not observed in the current data, but observed in other high-resolution VLA archival images \citep{Giovannini1987}.\newline
			\# This is a hybrid structure and so jet cannot be defined.\newline
			Jet features that are ambiguous are denoted with a `?'.}
		\end{tabular}
	\end{table}

\begin{table*}
	\caption{Newly estimated properties of the \citet{Kharb2008} radio galaxy sample.}
	\label{tab:radiogalaxy_prop}
	\begin{tabular}{llcllcccc}
		\hline
		{Name}	&	{\it z}	&	{Arm-length}	&	{log~R$_c$}&
		{LLS}	&	{S$^{1.4}_{total}$}	&	{S$^{150}_{total}$}	&	{P$_{min}$(N,S)}	&	Q$_{jet}$\\
		{}	&	{}	&	{Ratio}	&	{}	&	{(kpc)}	&	{Jy}	&	{Jy}	&	{10$^{-11}$dyne~cm$^{-2}$}	&	{10$^{45}$ erg s$^{-1}$}\\
		(1)&(2)&(3)&(4)&(5)&(6)&(7)&(8)&(9)\\
		\hline
		3C6.1	&	0.804	&	1.15	&	-2.90	&	241	&	3.8		&	17.86	&	15.70, 7.51	&	11.48\\
		3C13	&	1.351	&	1.36	&	-4.19	&	276	&	1.92	&	17.11	&	... &	28.46\\
		3C34	&	0.69	&	1.14	&	-3.47	&	346	&	1.78	&	16.03	&	7.46, 2.81	&	6.74\\
		3C41	&	0.794	&	1.11	&	-3.40	&	204	&	3.51	&	12.16	&	15.10, 9.27	&	8.21\\
		3C44	&	0.66	&	1.63	&	-3.94	&	502	&	1.37	&	10.98	&	4.24, 5.27	&	4.82\\
		3C54	&	0.8274	&	1.07	&	-3.96	&	458	&	2.03	&	11.48	&	3.87, 3.16	&	8.21\\
		3C114	&	0.815	&	1.09	&	-2.53	&	451	&	1.01	&	11.56	&	5.36, 2.22	&	6.35\\
		3C142.1	&	0.4061	&	1.87	&	-3.57	&	299	&	3.10	&	28.48	&	4.88, 1.42	&	3.00\\
		3C169.1	&	0.633	&	1.44	&	-3.40	&	367	&	1.297	&	8.00	&	2.18, 1.21	&	3.12\\
		3C172	&	0.5191	&	1.13	&	-4.18	&	676	&	2.84	&	18.85	&	1.45, 0.65	&	4.54\\
		3C441	&	0.707	&	2.45	&	-4.32	&	299	&	2.53$^\dagger$ &	14.56	&	1.49, 2.74	&	7.29\\
		3C469.1	&	1.336	&	1.07	&	-3.40	&	712	&	1.72	&	12.89	&	13.20, 4.15	&	25.79\\
		3C470	&	1.653	&	1.36	&	-3.97	&	255	&	1.72	&	11.92	& ...	&	37.56\\
		\hline
		
		\multicolumn{9}{p{.8\linewidth}}{$\dagger$ Taken from \citet{Condon1998} \newline Col.(1)- Name of the source. Col(2)- Redshift from NED.  Col(3)- Arm-length ratio.  Col(4)- K-corrected radio core prominence parameter $S_{5\,GHz}(core)/S_{1.4 \,GHz}(lobe)$.  The 1.4 GHz flux density is taken from \citet{White1992} and the 5 GHz flux density is estimated from  8 GHz assuming $\alpha=0.6$.  Col.(5)-  Projected largest linear separation.  Col.(6)- Total flux density of the source at 1.4 GHz taken from \citet{White1992} in units of (Jy). Col(7)- 150 MHz flux density from TGSS-GMRT in units of (Jy).  {Col(8)- Minimum internal pressure in the lobe obtained using Equation ~\ref{eq:pmin}.} Col(9)- Jet kinetic power estimated as described in Section~\ref{sec:jet_power}.}
	\end{tabular}
\end{table*}

\section{Results}

{In our quasar sample, $\sim$ 70 per cent of sources show one-sided radio jets, while the remaining 30 per cent sources do not clearly show a jet. In all but one source (3C~249.1 which shows a hybrid FR I/FR II morphology) two-sided hotspots located at the end of the radio lobe (i.e. FR II type morphology) are seen. For two sources {(3C~109, 3C~336)}, jet-like features are seen in higher resolution images in the literature \citep{Giovannini1987}. In the radio galaxy sample of \citet{Kharb2008}, 30 per cent of the sources show one-sided jet-like features (see Table 6 of \citep{Kharb2008}). The 5 GHz images of the quasars show the presence of multiple hotspots in some sources. The southern lobe of 3C~263 and northern lobe of 3C~351 have off-axis extended emission i.e., the emission is extended on one side of the jet axis.} See the Appendix for the contour images of the quasars at 1.4 GHz and 5 GHz.
 
We studied various properties of the cores, lobes, and hotspots of quasars and radio galaxies. We have quantified the observed trends with appropriate statistical tests. The significance of the relationship between properties is assessed using the non-parametric Spearman and Kendall rank tests. Both tests produced nearly identical results and so we have quoted here the results from the Spearman rank test. The p-value is a measure of the significance of the null hypothesis that two variables are uncorrelated. To determine if two distributions are the same, we have used the two sample Kolmogorov$-$Smirnov (KS) test. The null hypothesis to be tested in the KS test assumes that both the samples are drawn from the same population. A p-value of 0.05 and lower is considered to be significant to reject the null hypothesis. $1-p$ gives the per cent confidence level at which the null hypothesis is rejected.

\subsection{Structure Asymmetry}
Arm-length ratio is an asymmetry parameter and is defined as the ratio of the longer to the shorter arm-length, where arm-length is the core-to-hotspot distance. For quasars that show two hotspots on one side, we measured the arm-length ratio using both the hotspots. Figure~\ref{fig:redshift_armlenratio} shows the arm-length ratio with respect to the redshift for the quasars and FR~$\mathtt{II}$ radio galaxies. The arm-length ratio distribution for quasars shows that there are both symmetric and asymmetric structures. A larger fraction of the sources in the radio galaxy sample are symmetric sources than in the quasar sample. Combining both the samples, we find that $\sim 65$ per cent of the sources have arm-length ratios lower than the combined average arm-length ratios (=1.49) suggesting symmetric structures. The KS test indicates that there is no substantial difference in arm-length ratios between the two types of AGNs and they may be drawn from the same population ($p=0.15$). The arm-length ratios  appear to be uncorrelated with redshift ($p=0.2$) \citep[see also][]{Kharb2008} suggesting that symmetric and asymmetric structures prevail at all redshifts.

The (brighter) jet side is observed in seven out of 10 quasars (3C~249.1 is excluded because of its hybrid structure). The remaining three sources have ambiguity in the jet feature (see Table~\ref{tab:quasar_prop_jet}).  Of these seven jetted quasars, four quasars have their bright hotspot side on the same side as the shorter lobe. If we also include the ``tentative'' brighter jet sides from our images, 50 per cent (i.e. five out of ten quasars) have their brighter hotspot side on the same side as the shorter lobe. Simple relativistic beaming theories, however expect the longer arm to be the brighter arm {\citep{Ryle_Longair1967}}. Our results suggest that the brighter hotspot side is equally likely to be either on the shorter arm or on the longer arm .  This asymmetry in brightness and morphology may be a result of asymmetric environments, rather than orientation-dependent Doppler effects. The environmental effects that constrained the jet from expanding might also make the hotspot appear brighter {\citep{Saikia1995, gopalkrishna2004}}.  
	\begin{figure}
		\includegraphics[width=\columnwidth]{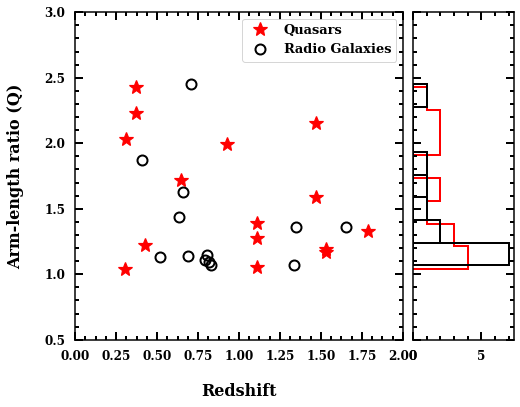}
		\caption{Arm-length ratio (\textit{Q}) {which is the ratio of longer to the shorter core-hotspot distance} vs redshift. Quasars from our sample are indicated by stars and FR~$\mathtt{II}$ radio galaxies by diamond symbols. The radio galaxy sample is from \citet{Kharb2008}.  {Q for sources with multiple hotspots is also plotted}.  The histogram of the \textit{Q} for quasar and radio galaxies is also shown to the right of the y-axis.  {KS test indicates no substantial difference in arm-length ratio between quasars and radio galaxies.}}
		\label{fig:redshift_armlenratio}
	\end{figure}

\subsection{Hotspot Spectral Indices}
	\begin{figure}
	\includegraphics[width=\columnwidth]{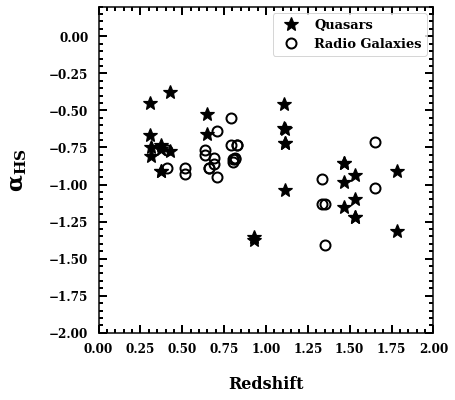}
	\caption{Spectral index of the hotspot between 1.4 and 5 GHz in the observed frame vs. redshift.  {Multiple hotspots are also plotted.}}
	\label{fig:redshift_spix}
	\end{figure}
The correlation of spectral index with redshift or radio luminosity is well established \citep{Tielens1979, Blumenthal_Miley1979, Wellman1997, Dennett1999, Ishwara2000, Kharb2008}. {The tendency for the integrated spectral index to steepen with redshift is conventionally attributed to a ``k-correction''. At higher redshift, the spectrum gets shifted to lower frequencies bringing the steep part of the spectrum closer to the observed frequency \citep{Bolton1966, Blumenthal_Miley1979, Laing_Peacock1980, gopalkrishna1988}.  This interpretation for the $z-\alpha$ is based on the observation that the SEDs of radio galaxies steepen at high frequencies.  However, the k-correction mechanism has been disputed owing to the remarkable linearity and the lack of evidence for a spectral curvature in wide band radio spectrum of high redshift radio galaxies \citep{Chambers1990, Klamer2006, Klamer2007}.   \cite{gopalkrishna2012} however argue that since the steepness of the spectral index anti-correlates with the spectral curvature \citep{Mangalam1995}, the spectral steepening becomes negligible for sources with steep spectra ($\alpha_{1GHz} \le -1.2$, also reflected in \cite{Klamer2006}).  k-correction being insufficient to explain the observed $z-\alpha$ relation, alternative effects have been proposed.  A density-dependent effect in which the jets are working against a denser environment at higher redshift can result in greater synchrotron losses at higher frequencies \citep{Athreya_Kapahi1998, Klamer2006, Klamer2007, Ker2012}.  It has also been suggested that for high-$z$ sources, there can be an enhanced spectral aging due to increased inverse-Compton losses of the relativistic electrons against the cosmic microwave background (CMB), since the CMB energy density increases as $(1+z)^4$ \citep{gopalkrishna1989, Krolik_Chen1991, Morabito2018}.   High power sources have higher magnetic fields in the hotspots than in the less powerful sources. This can result in higher synchrotron losses, thus steepening the spectra \citep{Meisenheimer1989, gopalkrishna1990, Blundell1999A}. Yet another possibility is that the $z-\alpha$ is an indirect manifestation of an intrinsic correlation between the luminosity and spectral index which when coupled to Malmquist bias translates to the $z-\alpha$ correlation \citep{Laing_Peacock1980, Chambers1990, Blundell1999A}. This, however, seems to be not the case in our sample as we discuss ahead.

Unlike previous studies where integrated spectral index was used, we examine the hotspot spectral index.  Since hotspots are located far away from the galaxy, their study can provide physical insight into the environments of the radio source, and thereby test the effect of ambient density on source properties.}
Figure~\ref{fig:redshift_spix} shows the variation of spectral index of the hotspot with redshift in quasars and radio galaxies. We see a spectral steepening in the quasar and radio galaxies with increasing redshift consistent with earlier findings. The Spearman rank test indicates a statistically significant correlation with the correlation coefficient of {-0.42} at a significance level greater than 99.99 per cent. A stronger dependence of spectral index on the luminosity is also observed in our sample and is shown in Figure~\ref{fig:radiopower_spix}. High radio power sources have steeper spectral indices and follow the relation $\log P_{1.4GHz} \propto \alpha_{HS}^{0.9}$ at a significance level of 99.99 per cent. {However, this $P-\alpha$ correlation turns out to be not significant when the partial correlation \footnote{The ppcor package in R programming language has been used in partial correlation statistics.} coefficient ($\rho=0.2$, $p-value=0.2$) was measured considering the effect of redshift.  The observed $P-\alpha$ correlation is likely due to the strong correlations between $z-\alpha$ (correlation coefficient = 0.7) and $z-P$ (correlation coefficient = 0.5) at > 99.9\% significance.}
	\begin{figure}
		\includegraphics[width=\columnwidth]{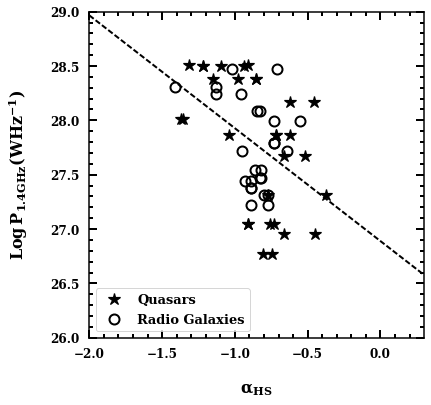}
		\caption{Total 1.4~GHz radio power versus $1.4-5$~GHz hotspot spectral indices. {The dashed line is the best-fit line with a slope of $-0.9$.}}
		\label{fig:radiopower_spix}
	\end{figure}

The average spectral index of the hotspots in quasars is {0.88$\pm$0.04} similar to the radio galaxies {which has $\alpha_{HS}$=0.87$\pm$0.02} . The KS test indicates that the $\alpha_{HS}$ distribution  of quasars and radio galaxies are the same ($p-value=0.40$). In $\sim$80 per cent of the quasars, the hotspot with the flatter spectral index is also the one with higher peak {power} at 1.4 GHz; this fraction is $\sim60$ per cent in the case of the radio galaxies. 
In 70 per cent of the quasars (i.e., 7 out of 10 quasars excluding the hybrid quasar 3C349.1) where one-sided jets are observed, the jet-sided lobe has a significantly brighter hotspot than the non-jetted side in 6 out these 7 quasars. On the other hand, in the radio galaxies, only $\sim 30$ per cent of them show one-sided jet-like features and again the jet side points to the brightest hotspot. These observations (i.e. a similar distribution of hotspot spectral indices, jet-side pointing to the brightest hotspot in quasars and radio galaxies) imply that the brighter hotspot is likely, not due to Doppler boosting. The hotspots of quasars and radio galaxies considered here, {do not have a bulk motion that is relativistic.}
The hotspot power is between 10-20 per cent of the total power at 1.4 GHz. The hotspot power correlates and increases with the total power in quasars and radio galaxies. The statistical significance of the $\alpha - z$ relation for the hotspot ($p-value=5E-5$) is stronger than the $\alpha - z$ relation for core spectral index ($p-value=0.04$). 

\subsection{Radio Core Prominence}

\begin{figure}
	\includegraphics[width=\columnwidth]{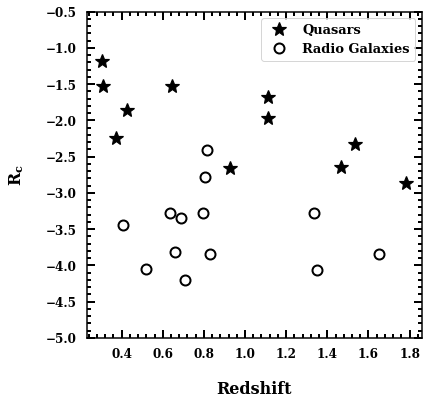}
	\caption{Logarithm of the radio core prominence {(ratio of core flux to the lobe flux density) }{R$_c$}vs. redshift for quasars and radio galaxies. {Quasars have larger R$_c$ than radio galaxies consistent with the orientation-based unification scheme.}}        
	\label{fig:redshift_corepromin}
\end{figure}

The radio core prominence $R_c$, is the ratio of the (beamed) radio core flux density to the (unbeamed) lobe flux density \citep{Orr_Browne1982,Zirbel_Baum1995}. It is a statistical indicator of beaming and thereby orientation \citep{Padovani_Urry1992, Kharb2004}; larger $R_c$ values correspond to smaller viewing angles. The k-corrected radio core prominence is calculated using
\begin{equation}
R_c =\frac{S_{core}}{s_{total}-s_{core}}(1+z)^{\alpha_{core}-\alpha_{ext}}\\
\end{equation} 
where $S_{core}$ is the core flux density at 5~GHz and $S_{total}$ is the total radio emission at 1.4~GHz. Figure~\ref{fig:redshift_corepromin} shows log~$R_c$ with respect to redshift for quasars and radio galaxies. For the radio galaxies, the 5~GHz core flux density is estimated from the 8~GHz measurements assuming a spectral index of $-0.6$ {(average core spectral index in \cite{Kharb2008} sample)}, and the total flux density is taken at 1.4~GHz. It can be seen that quasars have larger $R_c$ values when compared to the radio galaxies. This is consistent with the orientation-based unification of quasars with radio galaxies; quasars are oriented at relatively smaller angles compared to the radio galaxies.

The median log~$R_c$ for the quasars and radio galaxies is $-1.97$ and $-3.44$, respectively, consistent with the RL unification scheme. For quasars, $-2.8 < log~R_c < -1.0$, and for the radio galaxies, $-4.3 < log~R_c < -2.5$. 
\citet{Kharb2004} (see their Figure 3) have estimated log~$R_c$ values for a large number of AGN ranging from the plane-of-sky narrow-line radio galaxies to nearly pole-on blazars. 
{ Using the relations between $R_c$, jet speed and orientation angles presented in Appendix A of \citet{Kharb2004}, a minimum log~$R_c$ ($R_c^{min}$) value of $-4.5$ \citep[see][]{Kharb2004}, bulk jet Lorentz factor of 10 \citep{Kapahi1982}, we deduce that the radio galaxies considered here lie at orientation angles between 34$\pm$1$\degr$ and 74$\pm$2$\degr$, while the quasars lie at orientation angles between 19$\pm$1$\degr$ and 38$\pm$1$\degr$.  85\% of the radio galaxies have orientation angles greater than 45$\degr$.} \cite{Barthel1989} suggested that the average $\theta=31\degr$ for quasars, and $\theta=69\degr$ for radio galaxies. These are, of course, crude representative estimates because $R_c$ is a statistical indicator of orientation and cannot be used to derive orientation angles for individual sources.  It is important to note that while quasars are at relatively small angles compared to the radio galaxies, as indicated by their relative $R_c$ values, the overlapping spread in $R_c$ values indicates that the quasars are not oriented at small enough angles ($\lesssim10\degr$) to qualify as blazars. 

	\begin{figure}
	\includegraphics[width=\columnwidth]{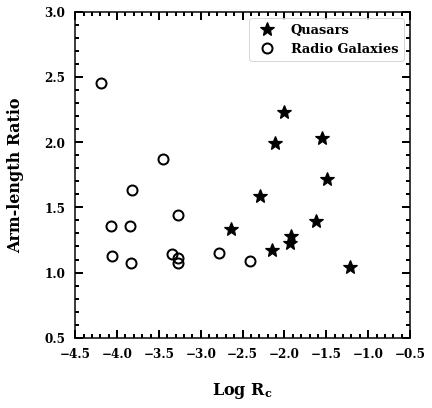}
	\caption{Arm-length ratio vs. logarithm of radio core prominence $R_c$.  {Arm-length ratio from multiple hotspots is plotted.  The KS test indicates a weak correlation suggesting that arm-length ratio is not strongly affected by projection.}}
	\label{fig:rc_armlenratio}
\end{figure}

Figure~\ref{fig:rc_armlenratio} shows log~$R_c$ plotted against the arm-length ratio, \textit{Q}.  We find a weak anti-correlation {in the combined sample} with a correlation coefficient of $-0.38$ at a significance of 95 per cent ($p-value =0.05$). This marginal relation, which also has a large scatter, might suggest that the arm-length ratio is not strongly affected by projection but is an intrinsic asymmetry.  {We suggest that this weak relation is apparent and may be a result of small sample size, and $R_c$ and arm-length ratio may not be correlated.  This is supported by the orientation independent statistics for the arm-length ratio in this sample {(see Section 3.1)}.}

\subsection{Largest Linear Size}
The jet orientation angle with respect to our line of sight influences the observed size of a radio source such that, $LLS_{proj} = LLS_{true}\times sin\theta$. Sources with jets oriented close to the line of sight will have smaller apparent sizes than sources oriented close to the plane of the sky for sources with similar intrinsic sizes. In the present work, the angular sizes of the radio galaxies appear to be larger than the quasars. The average projected LLS of radio galaxies is 1.6 times larger than those of quasars. Following the Unification scheme, if we assume that quasars and radio galaxies are the same objects and have similar distributions in true linear sizes, then $LLS_{proj_{RG}}/LLS_{proj_{RLQ}} = sin\theta_{RG}/sin\theta_{RLQ}$. { Using an average $\theta_{RLQ}=29\degr$ and $\theta_{RG}=54\degr$ for quasars and radio galaxies, respectively, as indicated by their $R_c$ values, gives $sin\theta_{RG}/sin\theta_{RLQ} \sim 1.6$, fully consistent with $LLS_{proj_{RG}}/LLS_{proj_{RLQ}} = 1.6\pm0.3$. 

The LLS ratio can be used to estimate the angle dividing quasars and radio galaxies ($\theta_0$) using the foreshortening calculations given by: 
	\begin{equation}
		LLS_{proj_{RG}}/LLS_{proj_{RLQ}} = \frac{sin(cos^{-1}(0.5cos\theta_0))}{sin(cos^{-1}(0.5(1+cos\theta_0)))}
	\end{equation}
For an LLS ratio of 1.6, this gives $\theta_0=51\degr$.
These are in agreement with the torus opening angle estimated from the projected linear sizes of the radio sources. The average projected LLS of quasars is 240$\pm$28~kpc, while the radio galaxies have an average projected LLS of 390$\pm41$~kpc. Using Eq (7) of \cite{Arshakian2005}, the torus opening angle for the quasars in our sample is 57$\degr\pm$8$\degr$.  This yields an average orientation angle of 74$\pm 4\degr$ for radio galaxies and 39$\pm 5\degr$ for quasars.  Encouragingly, this matches with the upper end of the average orientation angles estimated using R$_c$.}

    
\cite{WanDaly1998b} and \citet{Barthel1989} reported $LLS_{proj_{RG}}/LLS_{proj_{RLQ}}$ ratios of 1.25 and 1.8 respectively {for 3C sources}.  Recently, using galaxies from the Bootes survey observed using LOFAR, \cite{Morabito2017} found the ratio of the average projected linear size of radio galaxies to the quasars to be $3.1\pm1.0$,  and $2.0\pm0.3$ after correcting for redshift evolution.  The LLS estimates in this study are consistent with the LLS estimates in the literature, and minimal difference in the LLS ratio is likely due to the way the samples are selected.  The selection criteria in our study (and \citet{WanDaly1998b}) favours quasars that have well-resolved extended lobes which in turn correspond to quasars at relatively larger viewing angles to the line of sight. 

Overall, our results are consistent with the orientation-based unification scheme for quasars and radio galaxies. However, the range of orientation angles might not be large enough to show clear correlations with respect to orientation-indicators like $R_c$ or misalignment angles.  We find the inverse dependence of the angular size with the redshift (widely known as the D-z relation) consistent with previous studies \citep{Miley1968, Kapahi1985, Blundell1999A, Ker2012, Morabito2017}.

\begin{figure}
	\includegraphics[width=\columnwidth]{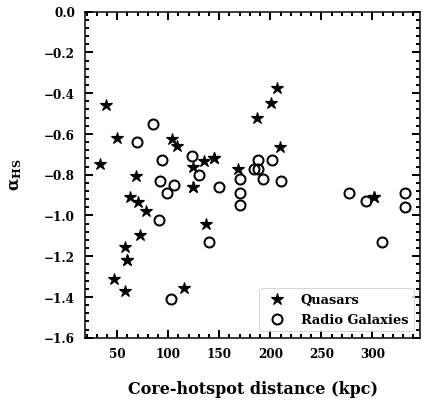}
	\caption{$1.4-5$~GHz hotspot observed spectral index versus projected {core-hotspot distance.} {The $\alpha_{HS}$ for multiple hotspots in each lobe is plotted.}}
	\label{fig:spix_las}
\end{figure}

The dependence of the $1.4-5$~GHz spectral index of the hotspots on the projected {core-hotspot distance} is shown in Figure~\ref{fig:spix_las}. We have plotted the spectral index of the hotspots of both the lobes for each source. We find a correlation between the spectral index and the projected {core-hotspot distance} of the quasars at a {96} per cent significance level. { This apparent correlation, however, disappears once a partial correlation test is performed between $\alpha_{HS}$ and the core-hotspot distance neglecting the effects of redshift.  This is indicative of the evolution of spectral index and size with redshift.}

\subsection{Lobe Axial Ratios}
The axial ratios (lobe length-to-width ratio) for quasars are lower and statistically different from the radio galaxies in our study consistent with results obtained by \cite{Leahy1989}. The intensity maps show more asymmetric and distorted lobes {(such as extended diffuse emission to one side of the jet)} in quasars (e.g. 3C~14, 3C~204, 3C~249.1, 3C~263, 3C~351) than in radio galaxies. The KS test suggests that quasars and radio galaxies have different distributions at 96 per cent significance ($p-value=0.04$). The median axial ratios of quasars and radio galaxies are {3.0} and 4.2 respectively.  The lobe widths of quasars are on an average only {1.3} times fatter (i.e. larger lobe width) than the radio galaxies and their median lobe widths are similar.  Thus, lobe widths of quasars studied here are rather insensitive to projection effects, consistent with \cite{WanDaly1998b} who showed that projection has little effect on lobe width except at very small angles or for source widths measured close to the hotspot. This indicates that the ratio of lobe length to the width i.e. axial ratio is likely governed by the projection of the length.  The average core-hotspot size ratio for similar lobe widths is 1.45. This likely suggests that the different axial ratios can be accounted for by projection effects alone. {\cite{Leahy1989} however find that the differences in axial ratios cannot be explained by projection effects alone. Larger lobe structure distortions seen in quasars (also seen in our quasar sample) can partly explain the low axial ratios.} Such lobe distortions can, for example, arise due to variations in the outflow angles, as discussed in Section~\ref{sec:misalignment}.  {Any evolutionary relationship between quasars and radio galaxies leading to smaller linear sizes in quasars compared to radio galaxies, however, cannot be ruled out ~\citep[e.g.][]{Miley1971,Wardle_Miley1974}.}

The axial ratios show a marginal correlation with the spectral index of the lobes (correlation coefficient = {0.3, $p-value=0.03$}) (see Figure ~\ref{fig:axialratio_lobeflux}) and an anti-correlation with the lobe flux density ({$\rho$=-0.4, $p-value=0.01$}), both at a greater significance level i.e. fatter lobes are brighter and have steeper spectra. This is believed to be due to confinement of the lobes which enhances the radio emission through synchrotron losses while reducing the adiabatic expansion losses \citep[e.g.,][]{Roland1985, Barthel1996}. We further compare the axial ratios against the structure asymmetry using the arm-length ratio in Figure \ref{fig:axialratio_armlenratio}. Individually, the quasars and radio galaxies do not show any correlation, but the combined sample shows an anti-correlation at a significance level of 96 per cent ($p-value=0.04$). This anti-correlation implies that sources that are narrower i.e. with larger axial ratios are also more symmetric in terms of their lobe sizes (i.e., have arm-length ratios closer to unity). The shorter or fatter sources are the more asymmetric. Such an anti-correlation was also observed in the combined radio galaxy sample in \citet{Kharb2008}, who found that the shorter lobe was fatter and had a steeper spectral index. 

	\begin{figure}
		\includegraphics[width=\columnwidth]{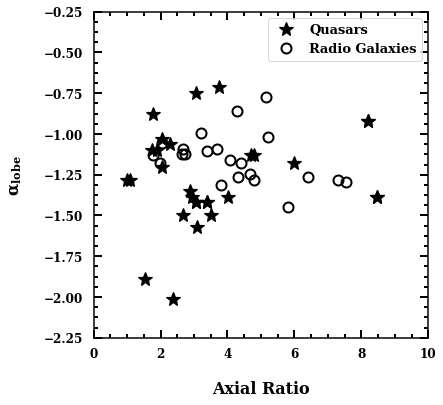}
		\caption{Axial ratio vs lobe spectral index shows significant marginal anti-correlation ($\rho$=-0.3).}
		\label{fig:axialratio_lobeflux}
	\end{figure}

	\begin{figure}
		\includegraphics[width=\columnwidth]{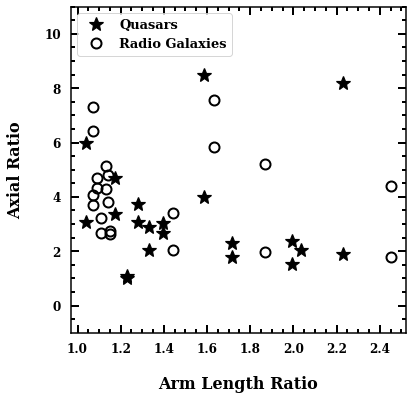}
		\caption{Plot of the axial ratio vs arm-length ratio for the quasar and radio galaxy sample.}
		\label{fig:axialratio_armlenratio}
	\end{figure}

\subsection{Jet Axis Misalignment}
\label{sec:misalignment}
Misalignment angle is the complement of the angle between the lines joining the core to the two hotspots. This is an asymmetry parameter and is sensitive to the orientation of the source to the line of sight. An intrinsic jet axis misalignment can appear enhanced for small angles to the line of sight, due to projection effects \citep{Reynolds1980, Conway_Murphy1993}. However, the misalignment angle is also found to be sensitive to environmental effects \citep{Kharb2008}. The misalignment angles for quasars are typically greater than $5\degr$ and the average misalignment angle for quasars is about 1.8 times greater than radio galaxies  {(See Table ~\ref{tab:quasar_basic_prop}),  however, we} cannot reject the hypothesis that both the samples have similar distribution at 28 per cent ($p-value=0.28$).  {Although differences in misalignment angles can become amplified due to a geometrical projection effect, the difference seen in our sample is marginal  because amplification is likely to be significant only in sources that are oriented close to the line of sight \citep{Moore1981, Saikia1995}.}

Figure~\ref{fig:misalign_rc} shows the misalignment angle versus the radio core prominence, log~$R_c$. Statistical tests show no correlation for the quasars and radio galaxies taken together ($p-value=0.8$) or separately for quasars ($p-value=0.2$). \citet{Kharb2008} had found a weak positive correlation for a large sample of radio galaxies. The radio core prominence$-$misalignment angle correlation has also been observed in quasars by \citet{Kapahi1982,Hough1989}. This implies that either $R_c$ or the misalignment angles or both are not uniquely orientation-dependent. Environmental factors may be instrumental in this lack of a correlation. $R_c$ could be influenced by environmental asymmetries on parsec-scales, while misalignment angles could be influenced by environmental asymmetries on parsec as well as kpc-scales \citep[][]{Kharb2010}.

\begin{figure}
	\includegraphics[width=\columnwidth]{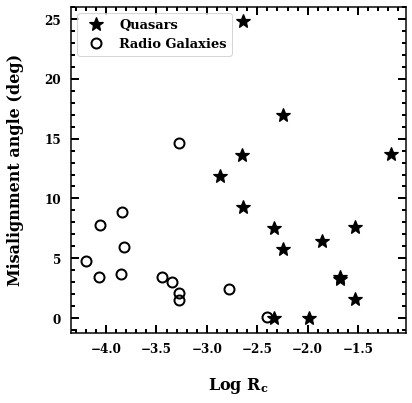}
	\caption{Misalignment angle vs logarithm of the radio core prominence. {Misalignment angle obtained for multiple hotspots is also plotted.}}
	\label{fig:misalign_rc}
\end{figure}

Figure~\ref{fig:misalign_axialratio} shows the misalignment angles against axial ratios (longer and narrower sources have larger axial ratios).  Quasars show smaller axial ratios and larger misalignment angles. This is consistent with the idea that environment plays an important role in tracing the jet and lobe structure. When the jet and the counter-jet plough into {a} medium with different densities, it will cause the jet to {bend} leading to greater misalignment angles {\citep{McCarthy1991}}. {It is also possible that outflow angle is changing with time, as in the \qq{dentist drill} model \citep{Scheuer1982} $-$ the jet axis random walks sending out the jet in different directions.}

\begin{figure}
	\includegraphics[width=\columnwidth]{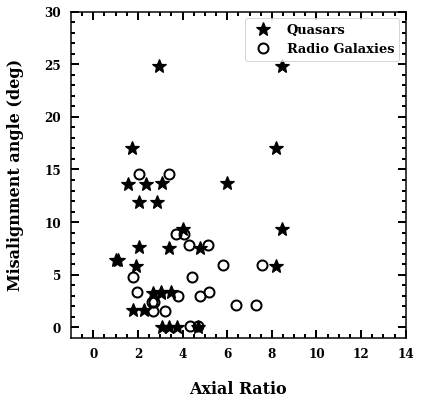}
	\caption{Misalignment angle vs axial ratio for the quasar and radio galaxy sample.}
	\label{fig:misalign_axialratio}
\end{figure}

\subsection{Minimum Internal Pressures}

The internal pressure of the jet when compared with the external gas pressure of the ambient medium can be used to understand whether the ambient medium is sufficient to confine the outer parts of the radio structure.  The non-thermal internal pressure in the radio jets contributed by the relativistic particles and magnetic field can be crudely estimated from the minimum energy arguments for synchrotron radiation.  It assumes equipartition of energy density in the particles and in the magnetic field \citep[e.g.,][]{Burbidge1956, Miley1980}.  This internal pressure will be a lower limit and is given by (following \cite{Odea1987})

\begin{equation}
P_{min} = \frac{B_{min}^2}{8\pi} + \frac{E_{min}}{3\phi V} ~~~ dyne~cm^{-2}
\label{eq:pmin}
\end{equation}
where,
\begin{equation}
B_{min} = [2\pi(1+k)C_{12}L_{rad}(V\phi)^{-1}]^{2/7}		~~~ 10 ~\mu G
\label{eq:bmin}
\end{equation}
is the magnetic field at minimum pressure condition, and
\begin{equation}
E_{min} = \left[\frac{V\phi}{2\pi}\right]^{3/7}[L_{rad}(1+k)C_{12}]^{4/7} ~~~ erg
\end{equation}

is the energy of the particles at minimum pressure. $k$ is the ratio of proton to electron energy and is assumed to be 1, $C_{12}$ is a constant from synchrotron theory that depends on the spectral index, and the upper and lower cut-off frequencies, $L_{rad}$ is the radio luminosity within the upper and lower cut-off frequencies. We use $C_{12}$= 2.8 $\times ~10^7$ by assuming a spectral index of $-0.7$, 10 MHz and 100 GHz as the lower and upper cut-off frequencies, respectively. $\phi$ is the volume filling factor and is assumed to be equal to 1. To estimate the volume V (in cm$^{-3}$), we make an assumption that the radio lobe is cylindrical with half the lobe width as the radius and core-to-hotspot distance as the height of the cylinder. The radio luminosity $L_{rad}$ is calculated using \citep{Odea1987}
\begin{equation}
\begin{split}
L_{rad} = 1.2 \times 10^{27}D_{Mpc}^2S_0\nu_0^{-\alpha}(1+z)^{-(1+\alpha)} \\ \times (\nu_u^{(1+\alpha)} - \nu_l^{(1+\alpha)})(1+\alpha)^{-1}	~~erg~s^{-1}
\end{split}
\end{equation}

The average $P_{min}$ for quasars is $2.60\times 10^{-10}$ dyne cm$^{-2}$.  We follow the same procedure and estimate minimum pressure for the radio galaxies studied here. The average $P_{min}=0.58\times10^{10}$~dynes~cm$^{-2}$ for radio galaxies. This is roughly {a} factor {of} 4 lower than the $P_{min}$ in quasars. {Given that a greater fraction of radio galaxies are below redshift of 1, the lower $P_{min}$ of radio galaxies may likely be a result of the increasing $P_{min}$ with redshift as seen in Figure ~\ref{fig:redshift_pmin} and described below.}

Figure ~\ref{fig:redshift_pmin} shows the minimum lobe pressure as a function of redshift. The lobe pressure correlates strongly with redshift with a correlation coefficient of 0.83 obtained at a significance greater than 99.99 per cent. The slope of the linear fit is 1.32. The plot suggests that high redshift sources have lobes with higher minimum non-thermal internal pressure \citep[see also][]{Odea2009}. This could be an outcome of the increasing radio luminosity and decreasing lobe widths observed in this sample with redshift. This also results in increasing $B_{min}$ with redshift.  \citet{WanDaly1996} suggest that inner regions on radio lobes (close to the core) are in pressure equilibrium with their surroundings and the ambient gas densities are found to have generally no dependence on redshift \citep{Odea2009}. This implies that the actual source magnetic field strengths are lower than the minimum energy magnetic field strengths. Alternatively, high redshift sources are likely interacting with the ambient medium and have not yet reached pressure equilibrium with their surrounding. High redshift sources have denser ambient gas than at low redshift. This requires high pressures in the radio lobes to maintain pressure equilibrium with the ambient dense gas.

\begin{figure}
	\includegraphics[width=\columnwidth]{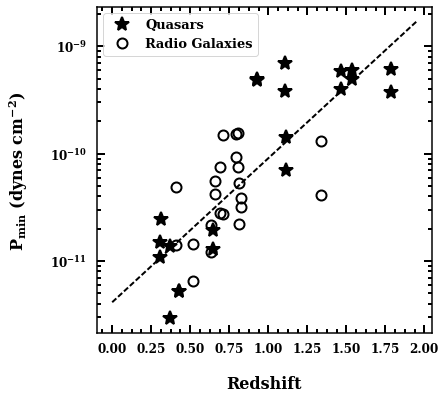}
	\caption{Minimum pressure in the radio lobes vs redshift for quasars and radio galaxies.}
	\label{fig:redshift_pmin}
\end{figure}

\begin{table}
	\caption{Spearman correlation statistics for different properties of the combined quasars and radio galaxy sample.}
    \label{tab:corr_stats}
    \begin{tabular}{cccc}
    	\hline
        Property 1	&	Property 2	&	Correlation	&	Probability\\
        {}	&	{}	&	{coefficient}	&	{p-value}\\
        \hline
        $z$	&	Arm-length Ratio	&	0.2	&	0.4\\
        $z$	&	$\alpha_{HS}$	&	-0.4	&	0.003	\\
        $\alpha_{HS}$	&	$P_{1.4 GHz}$	&	-0.5	&	0.02\\
        $R_c$	&	Arm-length Ratio	&	-0.4	&	0.05\\
        LLS	&	$\alpha_{HS}$	&	-0.3	&	0.2	\\	
        Axial Ratio	&	$\alpha_{HS}$	&	0.3	&	0.01\\
        $R_c$	&	Misalignment Angle	&	0.03	&	0.8\\
        \hline
    \end{tabular}
\end{table}

\section{Low frequency emission in quasars and radio galaxies}
\label{sec:lowfrequency}
TGSS is a radio survey carried out using the GMRT at 150~MHz and covers a declination range of $-55\degr$ to $+90\degr$. It has the highest resolution of $25\arcsec$ in the low frequency regime. The TGSS Alternative Data Release \citep[ADR1,][]{Intema2017} re-processed the TGSS images using a robust automated pipeline SPAM \citep{Intema2014} which employs techniques such as direction dependent calibration of ionospheric phase errors and image-based flagging to deal with the widespread RFI at the low frequency. We use the TGSS ADR1 images to study the morphology of {this} quasar and radio galaxy sample at low frequencies and derive total energy transported by the jets during the lifetime of the AGN. We find that the classical FR~$\mathtt{II}$ type morphology seen in the 1.4 and 5 GHz images is no longer the same at 150~MHz. Several sources (see Figure~\ref{fig:tgss_quasars}) show signatures of wings and extended emission in directions perpendicular {and at an oblique orientation} to the primary lobes observed at GHz frequencies, reminiscent of X-shaped galaxies \citep{Leahy1984,Leahy1992}. Here we simply refer to all such structures as ``distorted''. We note that these wings persist in higher resolution GMRT images at 610~MHz (Vaddi et al. 2018, in preparation), attesting to the reality of these features. These distorted radio morphologies may indicate jet axis re-alignment and/or multiple activity episodes in these AGNs.

\subsection{Jet Kinetic Power}
\label{sec:jet_power}
The synchrotron emitting plasma in the lobes can be used as an indicator of the amount of energy supplied by the jets to the lobes \citep{Godfrey_Shabala2016}. The time-averaged total bulk kinetic jet power, defined as the total energy transported by the jets over the lifetime of the source, is given by (Equation 12 of \cite{Willott1999})
\begin{eqnarray}
Q_{jet} &\approx 3 \times 10^{45} 
[(1+z)^{-(1+\alpha)}~4\pi D_L^2~S_{150}]^{6/7} ~~ erg~s^{-1}
\label{eq:jet_power}         
\end{eqnarray}

where $S_{150}$ is the total flux density at 150~MHz in W~m$^{-2}$~Hz$^{-1}$, $\alpha$ is the spectral index estimated using the flux density at 150~MHz and 1.4 GHz, and $D_L$ is the luminosity distance.
Equation~\ref{eq:jet_power} assumes that {synchrotron radiation losses are negligible. The average quasar jet power is 6.4$\pm$0.07 $\times$ 10$^{45}$ erg~s$^{-1}$.}  The KS test shows that the time-averaged jet kinetic power for quasars and radio galaxies may have similar distribution ($p-value=0.46$).  The jet power ranges from ${(1 - 38)}\times 10^{45}$~erg~s$^{-1}$, with 3C~470 radio galaxy having the highest jet kinetic power (refer Table ~\ref{tab:quasar_basic_prop}, and Table ~\ref{tab:radiogalaxy_prop}). The jet power and redshift are correlated which is expected from a flux limited 3C sample. {For the quasar sample, we notice that jet power is related to the mass of the supermassive black hole such that higher jet powers are associated with more massive black holes.  However, a partial correlation test between jet power and the black hole mass while ignoring the effect of redshift does not show any correlation ($\rho$=-0.3, $p-value$=0.5) between jet power and black hole mass suggesting that the observed trend is likely a manifestation of Malmquist bias.}

{
\section{Possible Biases}
\subsection{Selection Bias}
The \citet{Kharb2008} radio galaxies have radio power > 10$^{28}$ WHz$^{-1}$ at 178 MHz.  Their angular sizes are > 27" and the sources have a redshift range from 0.4 to 1.65.  The quasars were selected to have BH mass determinations, at least one side of the source without a radio jet, and radio powers $>$ 3$\times$10$^{26}$h$^{-1}$ W~Hz$^{-1}$~sr$^{-1}$ which for our chosen {H$_0$} gives 5$\times$10$^{26}$ WHz$^{-1}$~sr$^{-1}$; the sources have redshift between 0.3 and 1.78.  The black hole masses for the quasars were obtained from \cite{Mclure2006}, who studied the FWHM of the broad-line emission and typically found v $>$ 2000~km~s$^{-1}$. {Based on these selection criteria, we can say the following about its effects on our results}.
{ The selected radio galaxies are NLRG that are mostly in the plane of the sky and lack the broad-lines for BH mass estimates through virial methods.} This BH mass selection criteria for quasars, therefore, does not change the results. Since the quasars in our sample are selected such that spectral ageing can be performed, it automatically selects quasars that have larger lobe extents which in turn selects sources that are {intrinsically larger or are at larger angles to} the line of sight. This gives the quasars in our sample a larger average LLS compared to previous studies. This, however, does not change the general conclusion that quasars have statistically smaller projected sizes than the radio galaxies, in line with unification.

\section{Summary and Conclusions}
We have presented the results from our 1.4 and 5~GHz study with the VLA of 11 FR~$\mathtt{II}$ radio-loud quasars. The $1.4-5$~GHz spectral index images reveal spectral steepening along the lobes, moving away from the hotspots. These maps and results will be used in the spectral ageing analysis, the results of which will be presented in Paper II.  We compare these quasars with FR~$\mathtt{II}$ radio galaxies to study the radio-loud unification scheme. We consider 13 narrow-line FR~$\mathtt{II}$ radio galaxy sample that span similar redshift, luminosity, and have matched frequency/resolution data. We see that radio-loud unification largely holds but environmental factors also play an important role. 
We have shown that:
\begin{enumerate}
	\item 
{ The radio core prominence ($R_c$) parameter, which is a statistical indicator of beaming and thereby orientation, suggests that the radio galaxies considered in our study lie at orientation angles between $34\degr$ and $74\degr$, while the quasars lie at orientation angles between $19\degr$ and $38\degr$, broadly consistent with the radio-loud unification scheme.}
    
	\item About 90 per cent of the sources have brighter hotspots on the jetted side and the brighter hotspot is equally likely to be on the shorter or on the longer arm indicating that beaming effects may not be the primary contributor to the observed structural and brightness asymmetry.
	
	\item The hotspot spectral index ($\alpha_{HS}$) distribution is similar for quasars and radio galaxies.  The $\alpha_{HS}$ is strongly anti-correlated with redshift and it is argued to be due to higher synchrotron losses at higher redshifts due to {density-dependent effects while luminosity-dependent or size-dependent effects are not seen}.

	\item The average projected LLS of radio galaxies is 1.6$\pm$0.3 times larger than those of quasars in our sample. {This is comparable to the ratio of the sine of average orientation angles calculated from $R_c$. This is consistent with the orientation based unification scheme and LLS is likely probing orientation.}
   	 
	\item The median axial ratios of quasars are roughly a factor of 1.4 lower than the radio galaxies. This may be explained by projection in length since the differences in lobe widths between quasars and radio galaxies are minimal.  However, larger and significant lobe distortions seen in quasars compared to radio galaxies support the importance of intrinsic/environmental differences.  It is possible that the direction of the quasar jets in our sample is changing with time, {as} in the \qq{dentist drill} model, while radio galaxies have stable long-term collimated jets.  { Alternatively, evolutionary relationship between quasars and radio galaxies, however, cannot be ruled out. }
	
	\item  The misalignment angles for quasars are roughly 1.8 times greater than radio galaxies likely indicating an orientation effect. However, the radio core prominence which is also a statistical indicator of orientation does not show a correlation with the misalignment angle.  Environmental factors may be instrumental in this lack of correlation.  {R$_c$ could be influenced by small-scale environment asymmetries while the misalignment angle could be influenced by both small scale and large scale environment asymmetries.}
    
    \item  {The minimum lobe pressure is strongly correlated with redshift. High redshift sources have lobes with higher minimum internal pressure to maintain pressure equilibrium with the denser ambient gas.}
    
    \item Low frequency emission at 150~MHz with the GMRT reveal the lobes to move away from the classical FR~$\mathtt{II}$ morphologies; many sources reveal extended, winged morphologies that could suggest changes in jet orientation and/or restarted AGN activity in these quasars. 
    
	\item  The time-averaged jet power estimated using the 150~MHz data is similar for quasars and radio galaxies.

\end{enumerate}
{In summary, several of the observed properties of quasars are consistent with the orientation-based radio-loud unification scheme. However, differences and relations between orientation indicators such as axial ratio, misalignment angle, and larger lobe distortions in quasars can be explained by intrinsic/environmental asymmetries.}

\section*{Acknowledgements}
{We thank the referee for constructive and insightful suggestions that have significantly improved the quality of the manuscript.}  The National Radio Astronomy Observatory is a facility of the National Science Foundation operated under cooperative agreement by Associated Universities, Inc.  {The work of Stefi Baum and Chris O'Dea was supported by NSERC (Natural Sciences and Engineering Research Council of Canada). The work of Ruth Daly and Trent Barbusca was supported by Penn State University.} This research has made use of the NASA/IPAC Extragalactic Database (NED), which is operated by the Jet Propulsion Laboratory, California Institute of Technology, under contract with the National Aeronautics and Space Administration. We thank the staff of the GMRT who have made these observations possible. GMRT is run by the National Centre for Radio Astrophysics of the Tata Institute of Fundamental Research.

\newpage

\bibliography{references}
\bibliographystyle{mnras}


\appendix

\section{Spectral Index and continuum Images of the Quasars}
	\begin{figure*}
		\subfloat[3C~14]{\includegraphics[width=0.5\textwidth]{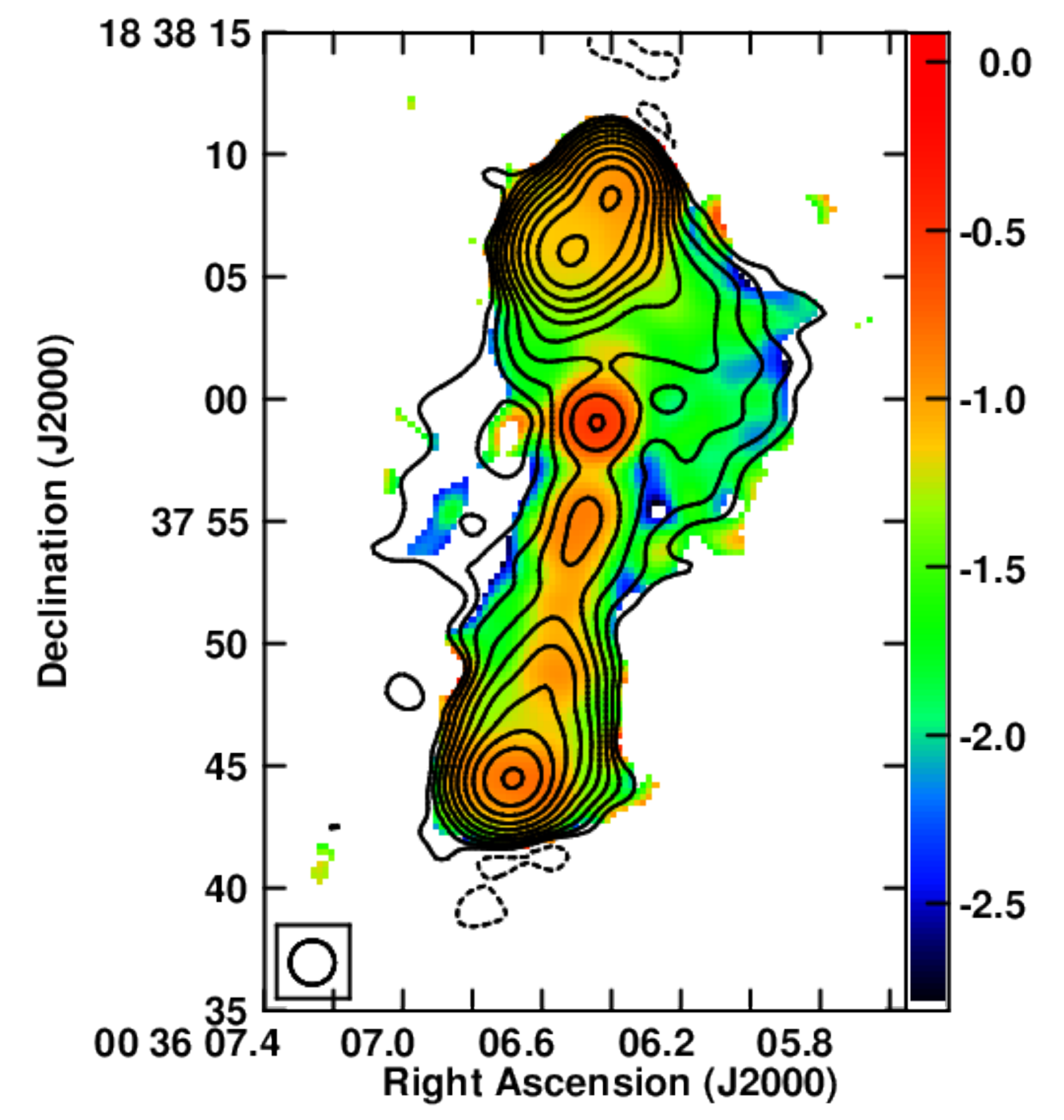}}
		\subfloat[3C~47]{\includegraphics[width=0.5\textwidth]{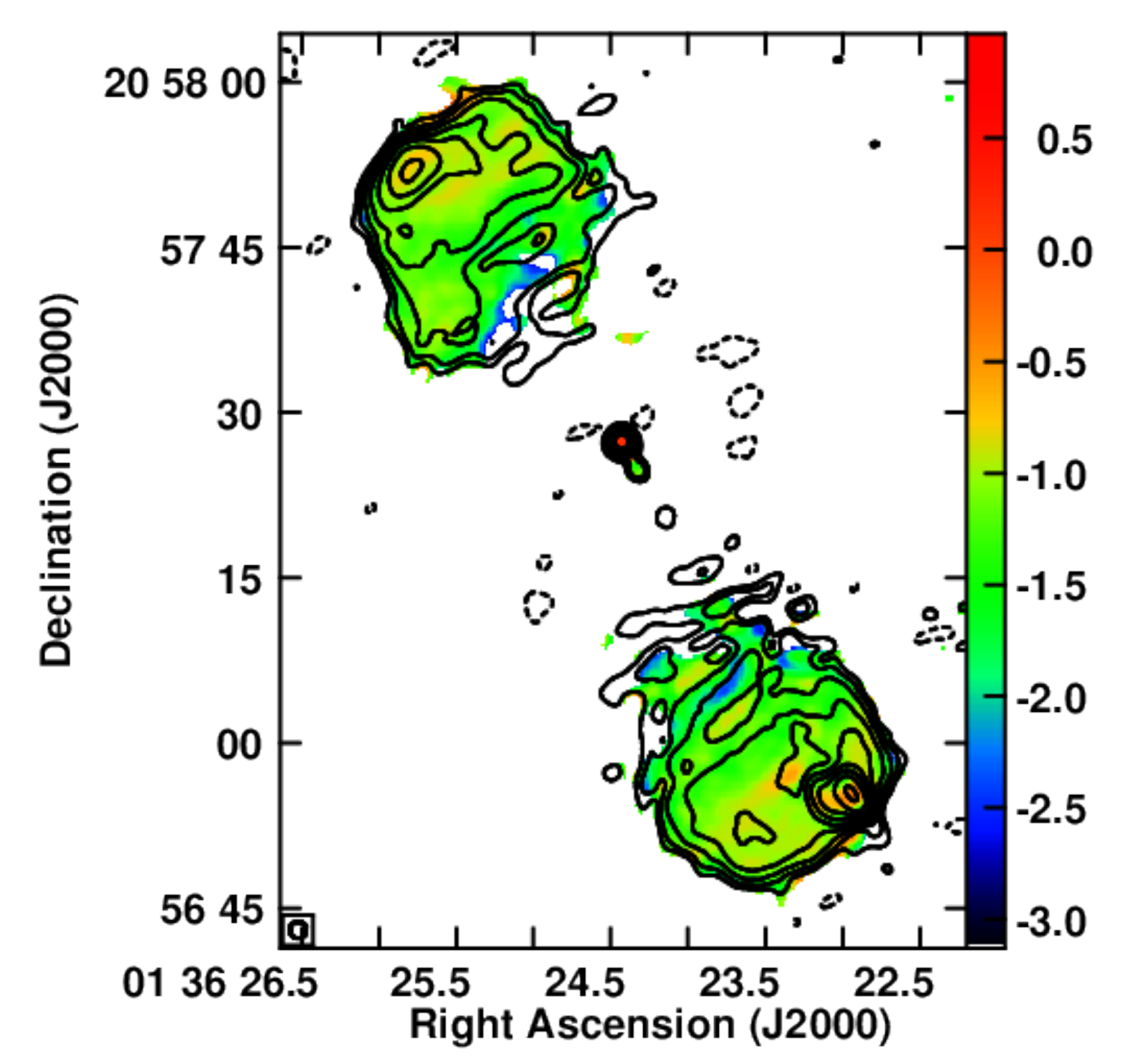}}\\
		\subfloat[3C~109]{\includegraphics[width=0.5\textwidth]{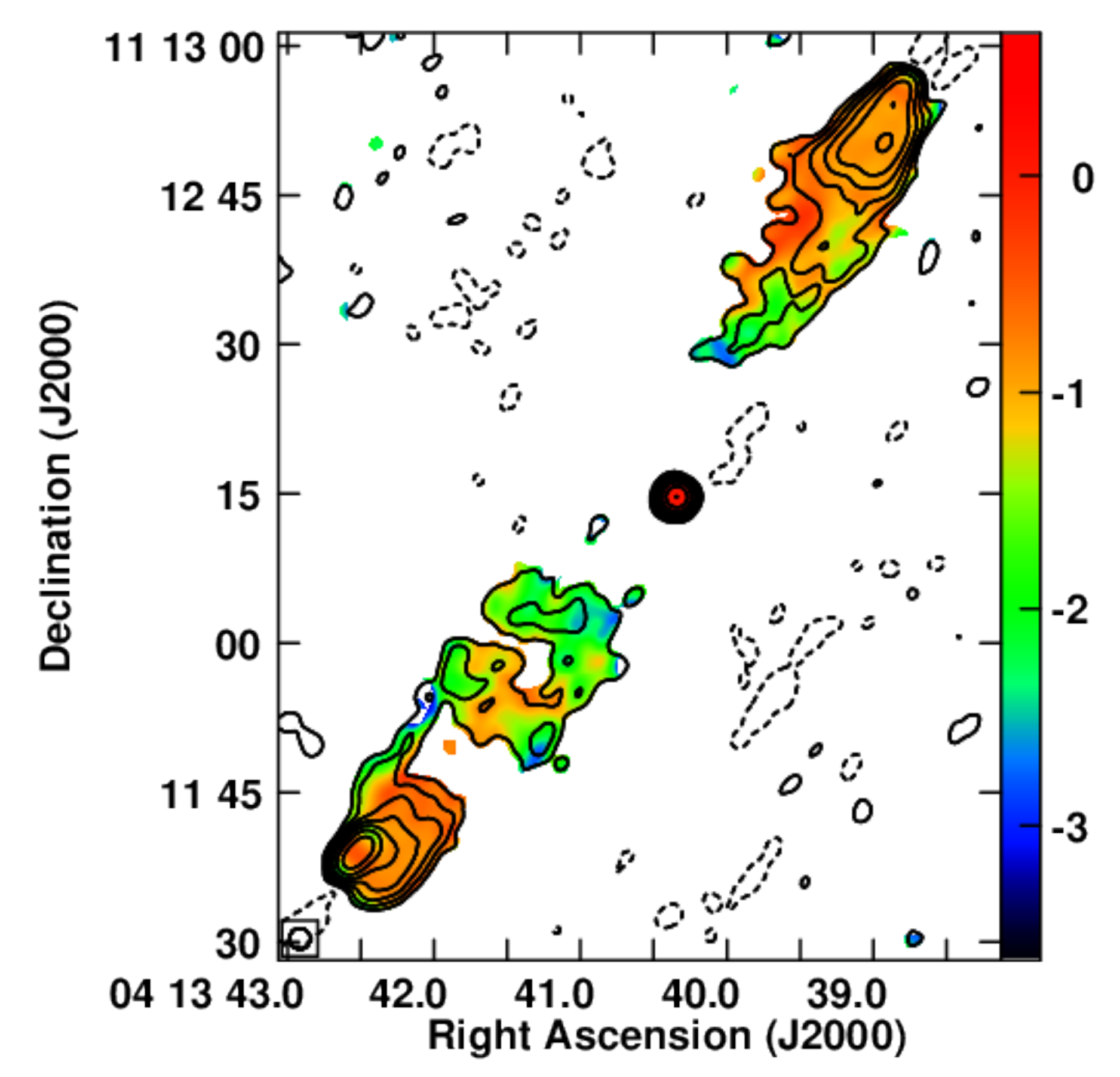}}
		\subfloat[3C~205]{\includegraphics[width=0.5\textwidth]{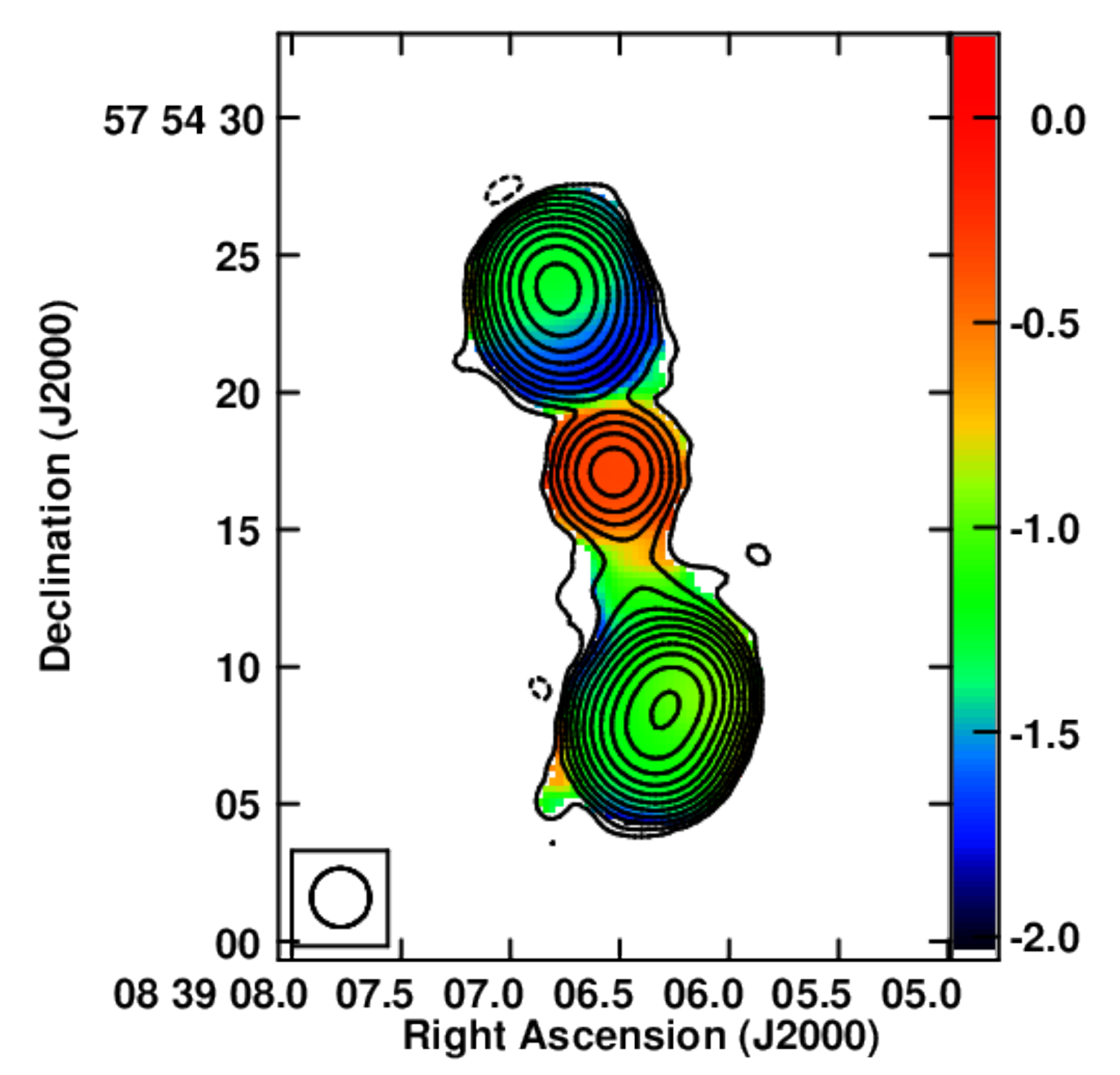}}
		\caption{Spectral index maps of the quasars made from the 1.4 GHz and 5 GHz VLA images.  The contours trace the 1.4 GHz total intensity.  The color bar indicates the variation in the spectral index.  The beam is convolved with the largest FWHM of the 1.4 and 5 GHz total intensity images.}	
	\end{figure*}
		
	\begin{figure*}
		\subfloat[3C~336]{\includegraphics[width=0.5\textwidth]{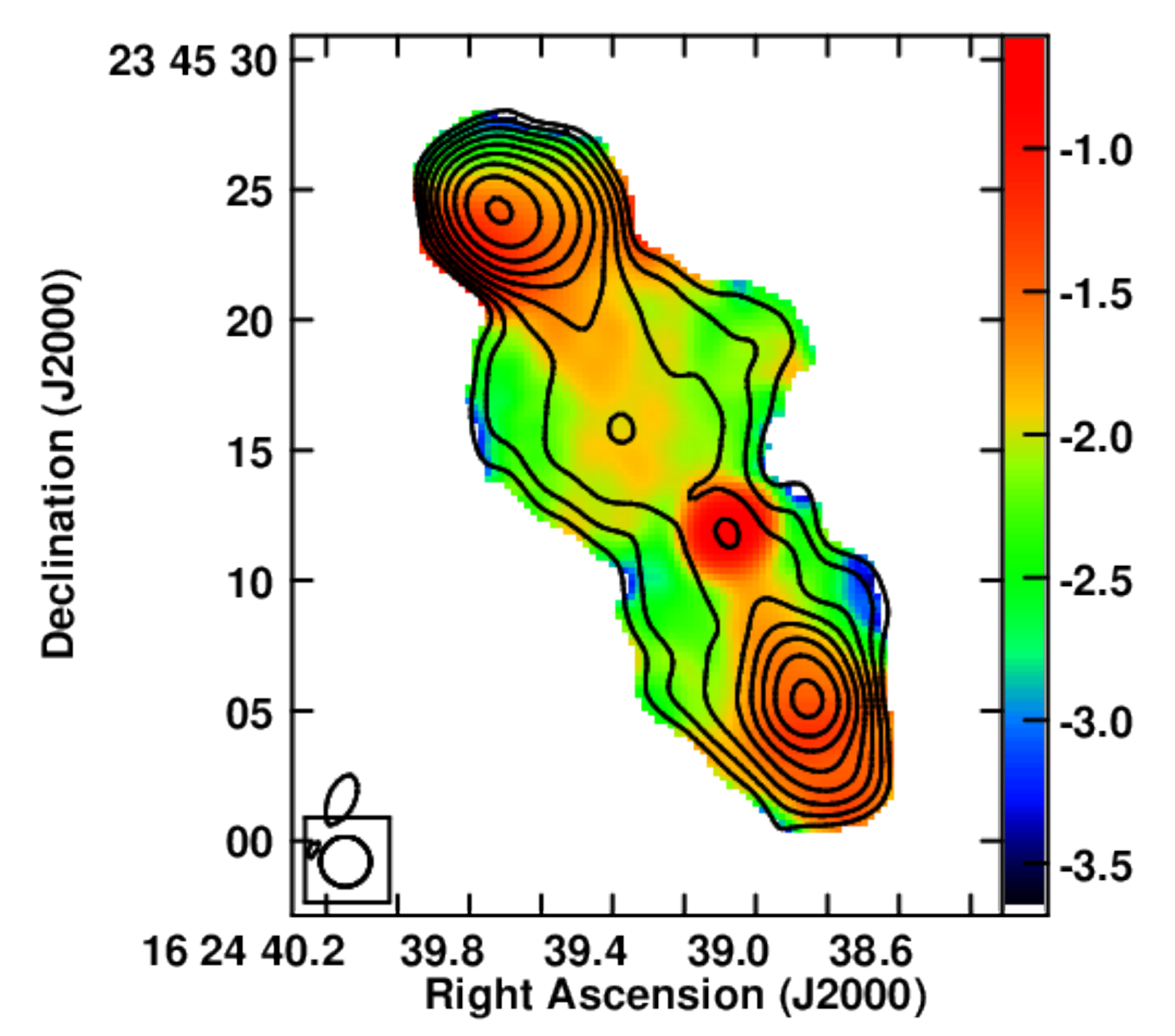}}
		\subfloat[3C~351]{\includegraphics[width=0.5\textwidth]{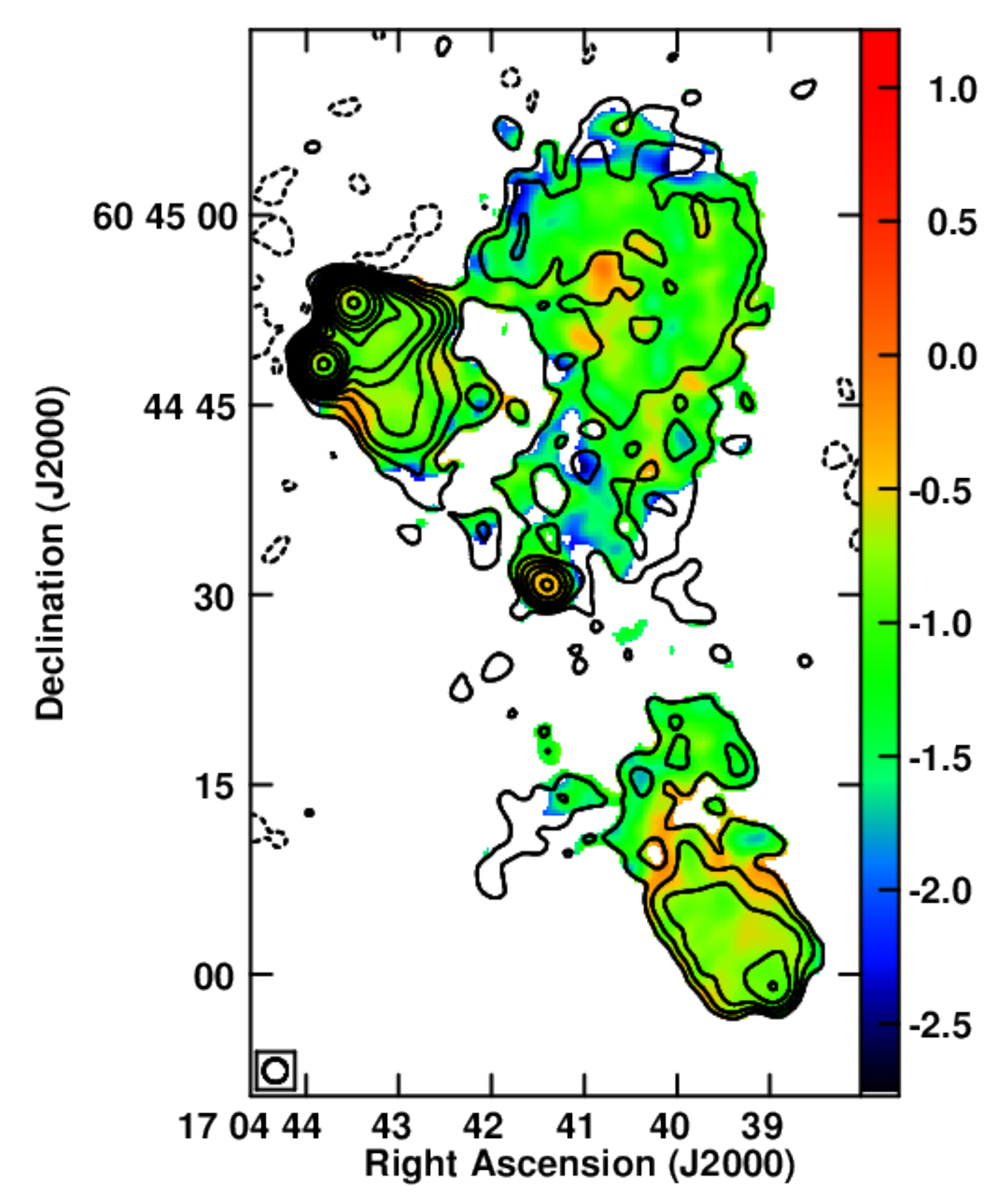}}\\
		\subfloat[3C~204]{\includegraphics[width=0.5\textwidth]{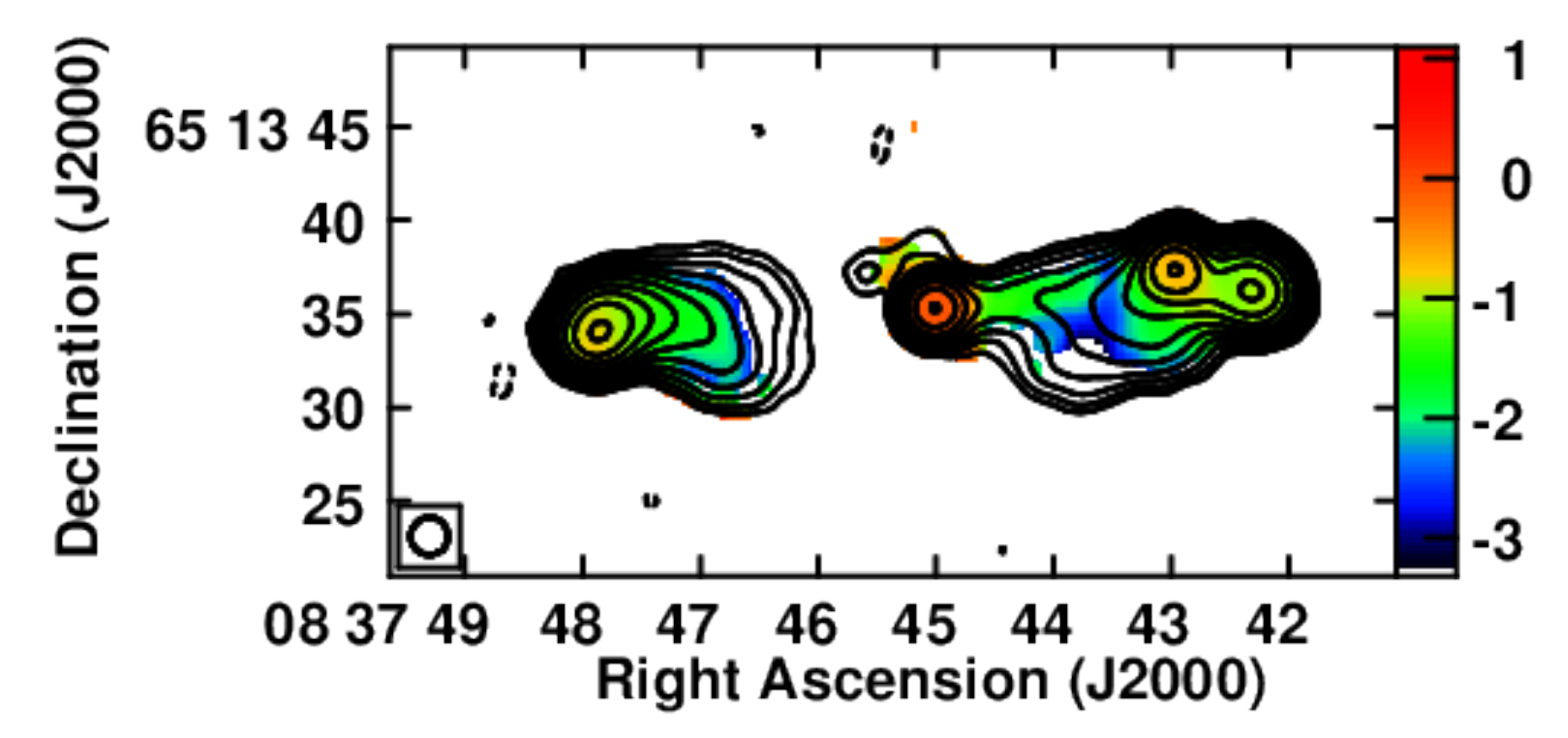}}
		\subfloat[3C~208]{\includegraphics[width=0.5\textwidth]{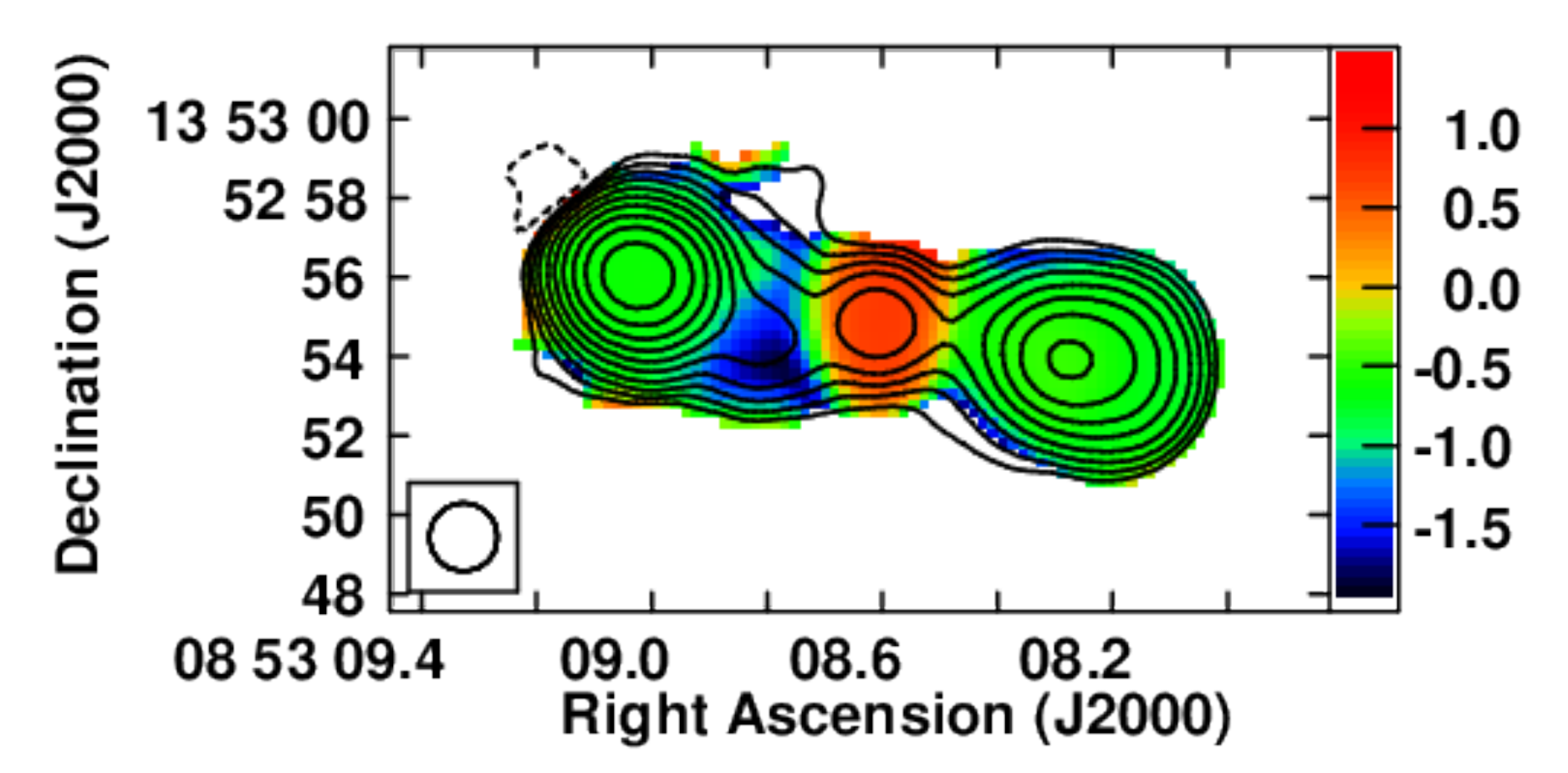}}
		\caption{Spectral index maps of the quasars made from the 1.4 GHz and 5 GHz VLA images.  The contours trace the 1.4 GHz total intensity.  The color bar indicates the variation in the spectral index.  The beam is convolved with the largest FWHM of the 1.4 and 5 GHz total intensity images.}	
	\end{figure*}
	\begin{figure*}
		\subfloat[3C~249.1]{\includegraphics[width=0.5\textwidth]{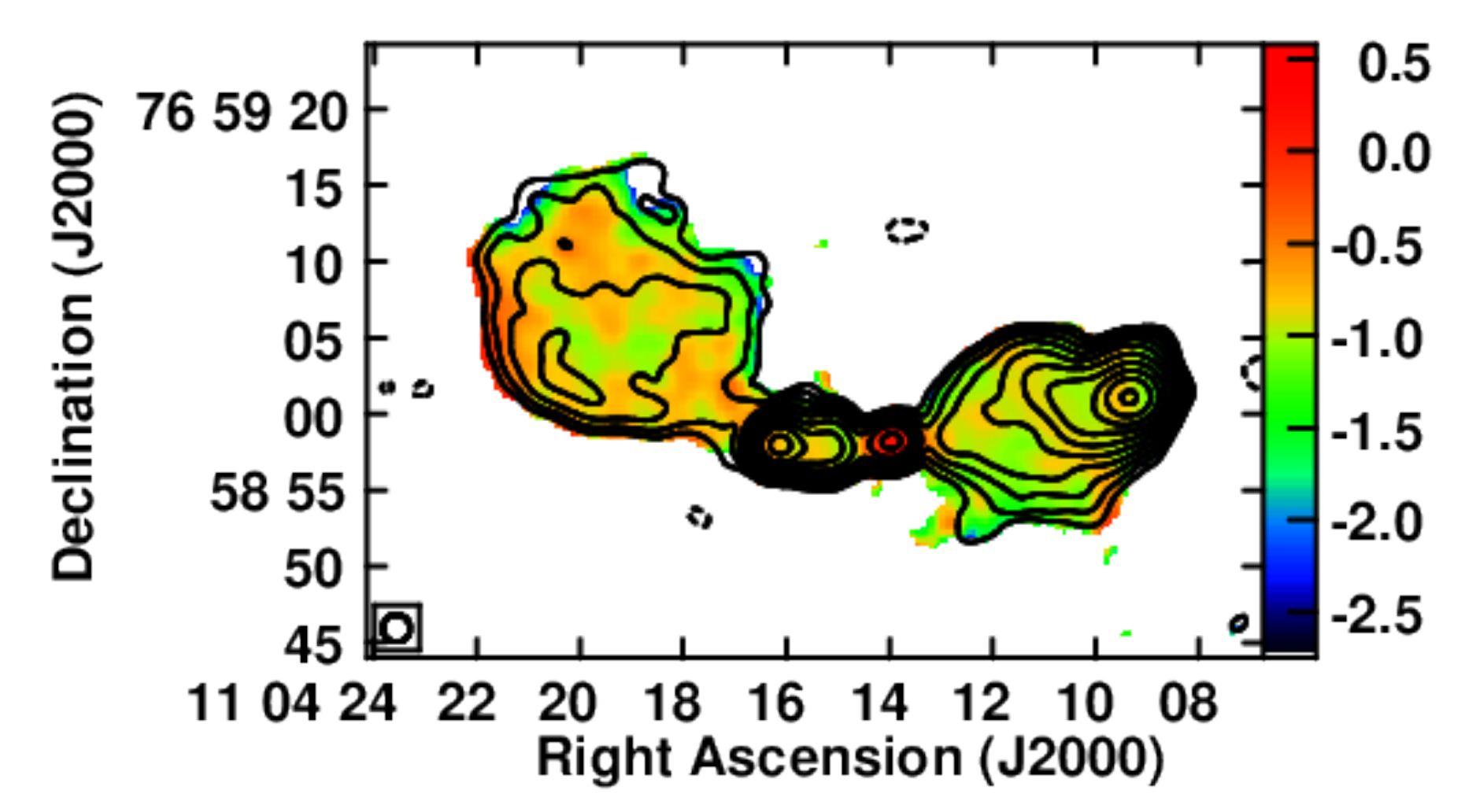}}
		\subfloat[3C~263]{\includegraphics[width=0.5\textwidth]{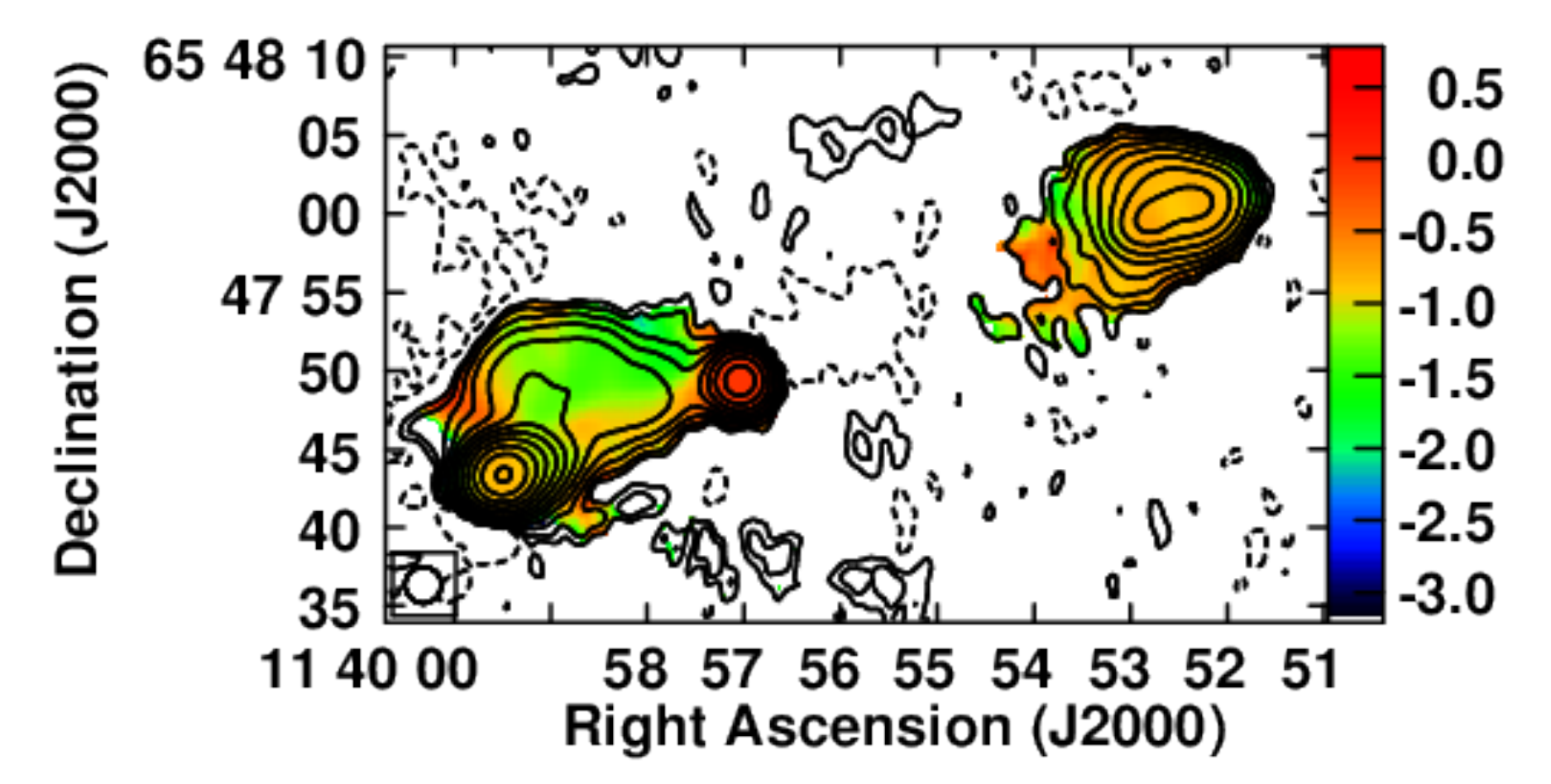}}\\
		\subfloat[3C~432]{\includegraphics[width=0.5\textwidth]{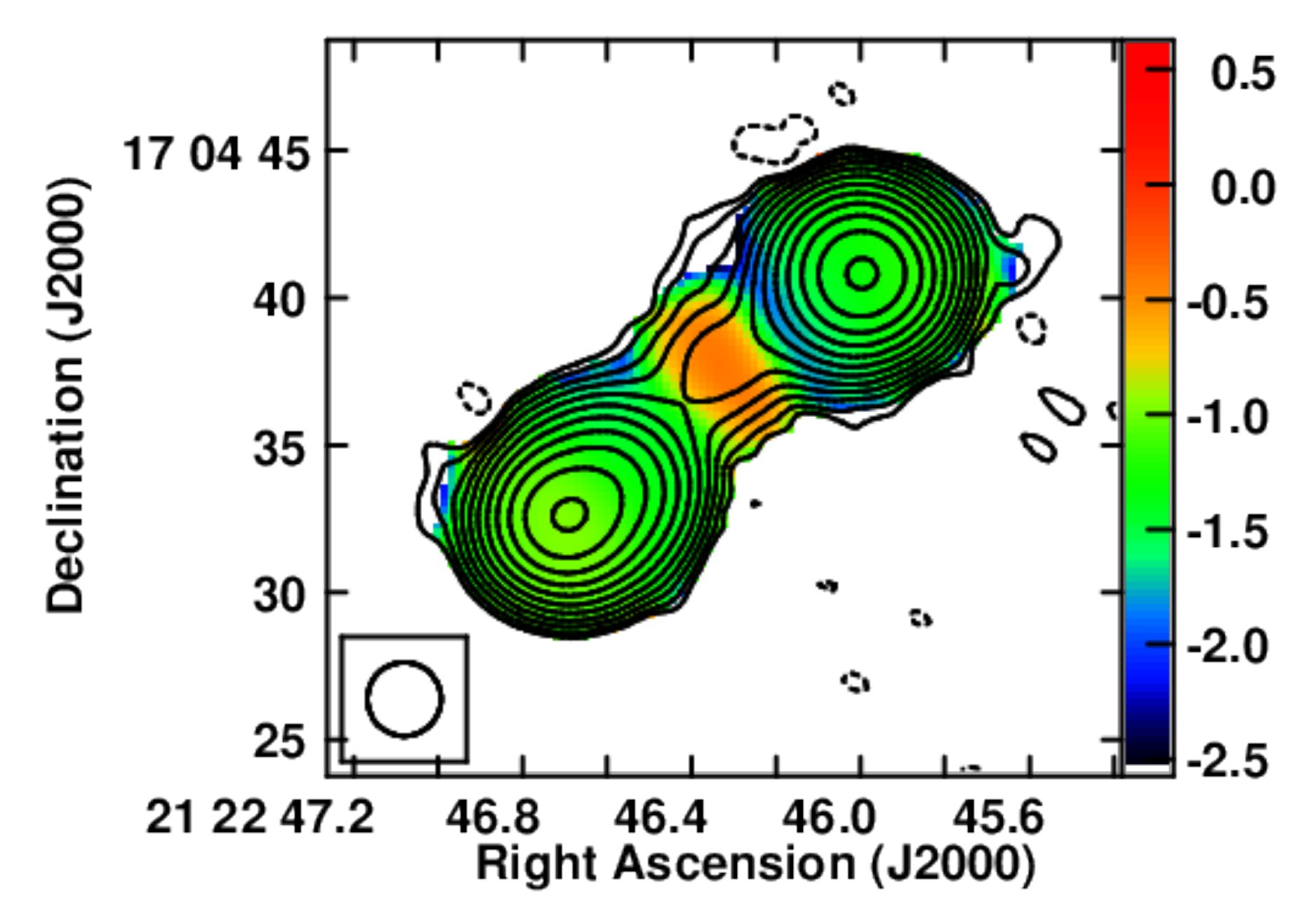}}
		\caption{Spectral index maps of the quasars made from the 1.4 GHz and 5 GHz VLA images.  The contours trace the 1.4 GHz total intensity.  The color bar indicates the variation in the spectral index.  The beam is convolved with the largest FWHM of the 1.4 and 5 GHz total intensity images.}	
	\end{figure*}

	\begin{figure*}
	\centering
	\subfloat[3C 14 :- Total intensity contour map at 1.4 GHz.  The peak surface brightness is 0.275 Jy beam$^{-1}$.  The contours are such that the levels increase in steps of 2 (-0.02 (dashed), 0.02,...,90)\% of the peak brightness.  The CLEAN beam FWHM is 1.50$\times$1.25 arcsec and the P.A. is -1.76 (shown in the inset in the lower left).]{\includegraphics[width=0.4\textwidth]{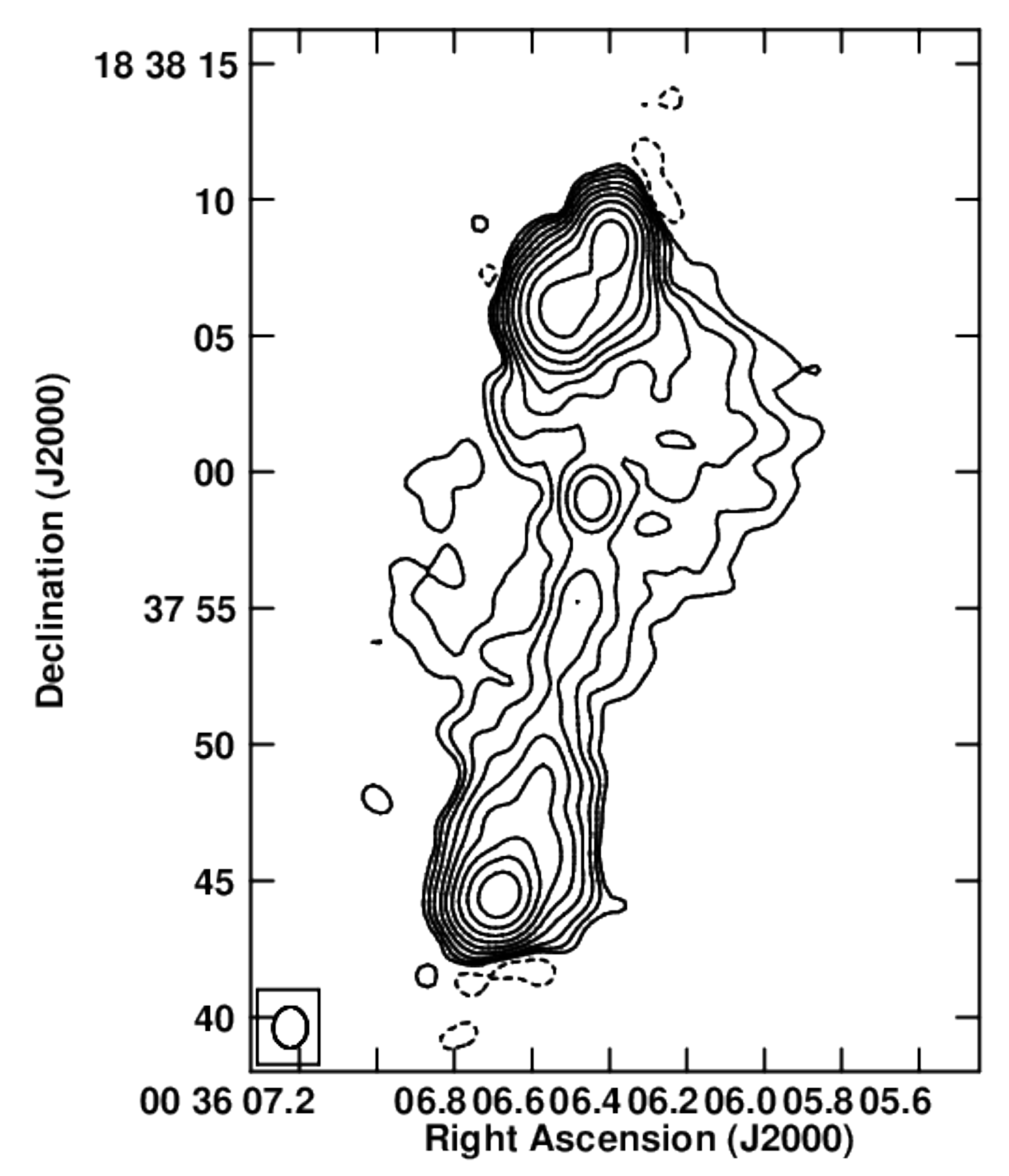}}
	\hspace{10pt}
	\subfloat[3C 14 :- Total intensity contour map at 5 GHz.  The peak surface brightness is 0.112 Jy beam$^{-1}$.  The contours are such that the levels increase in steps of 2 (-0.01 (dashed), 0.01,...,90)\% of the peak brightness.  The CLEAN beam FWHM is 1.81$\times$1.24 arcsec and the P.A. is 71.33 (shown in the inset in the lower left).]{\includegraphics[width=0.4\textwidth]{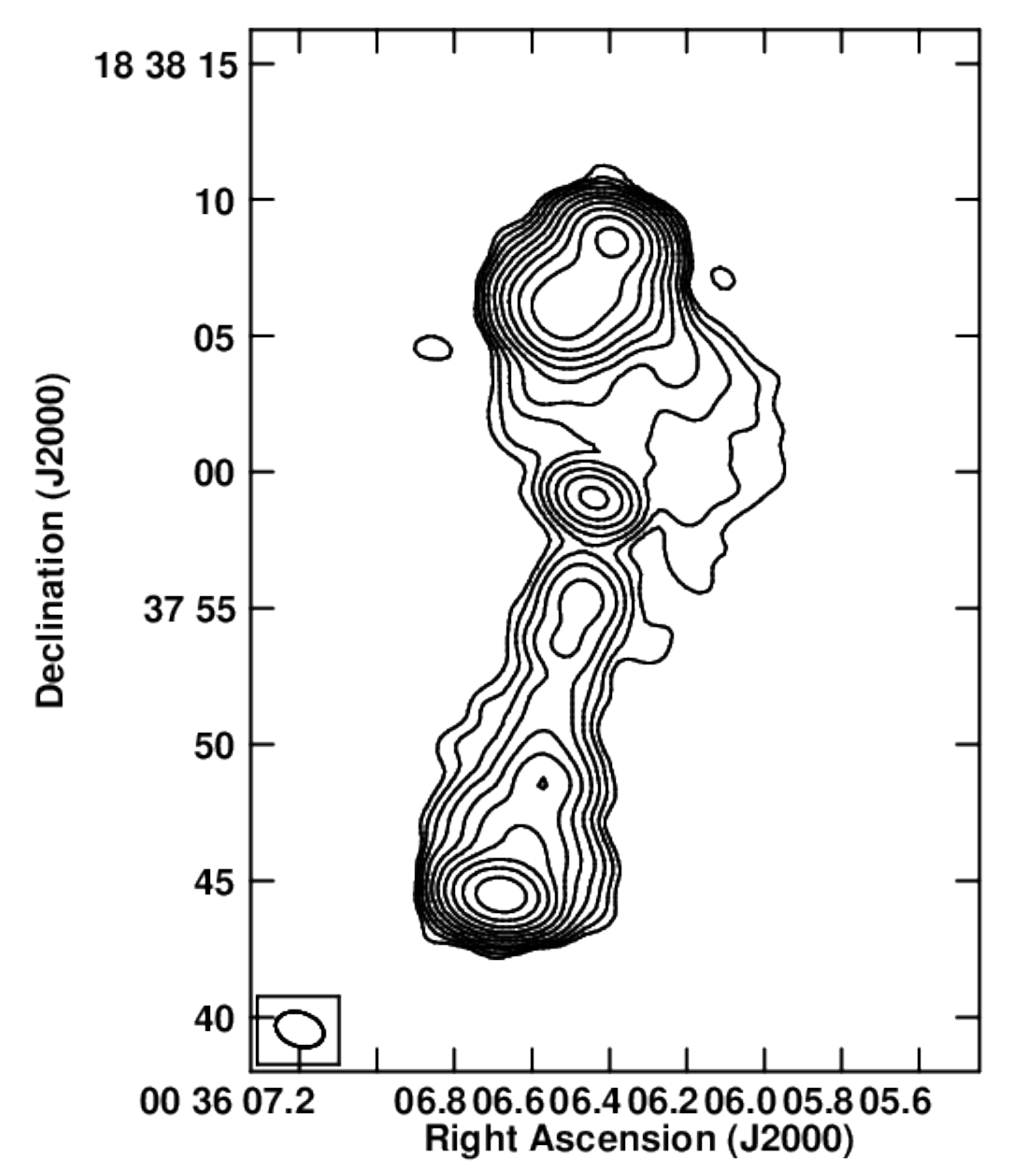}}
	
	\caption{Total intensity maps at 1.4 GHz and 5 GHz.}
	\label{fig:3C14_rgb_tgss}
	\end{figure*}

	\begin{figure*}
	\centering
	\subfloat[3C 47 :- Total intensity contour map at 1.4 GHz.  The peak surface brightness is 0.271 Jy beam$^{-1}$.  The contours are such that the levels increase in steps of 2 (-0.02 (dashed), 0.02,...,90)\% of the peak brightness.  The CLEAN beam FWHM is 1.50$\times$1.10 arcsec and the P.A. is 25.98 (shown in the inset in the lower left).]{\includegraphics[width=0.4\textwidth]{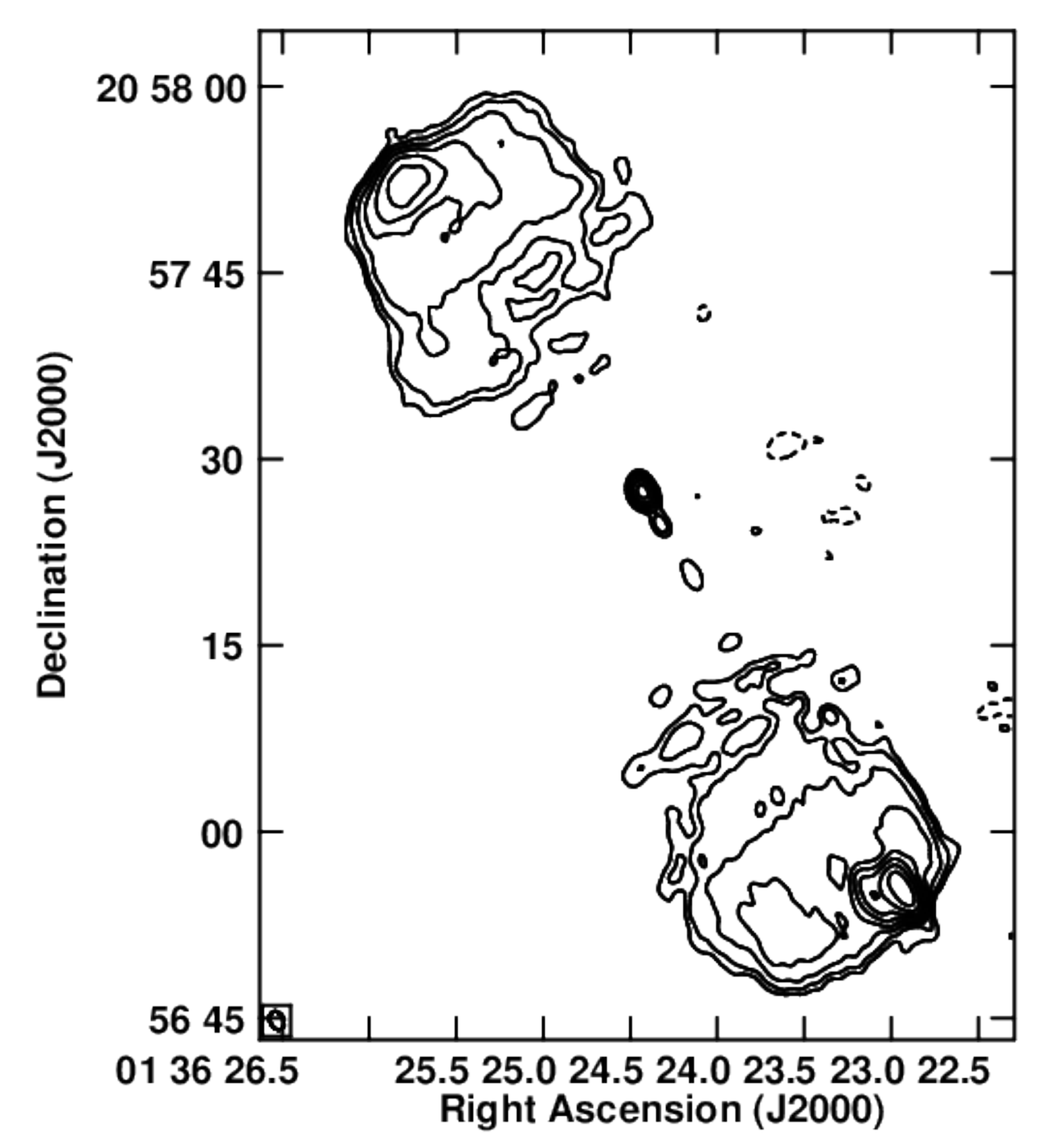}}
	\hspace{10pt}
	\subfloat[3C 47 :- Total intensity contour map at 5 GHz.  The peak surface brightness is 0.161 Jy beam$^{-1}$.  The contours are such that the levels increase in steps of 2 (-0.01 (dashed), 0.01,...,90)\% of the peak brightness.  The CLEAN beam FWHM is 1.50$\times$1.10 arcsec and the P.A. is 23.61 (shown in the inset in the lower left).]{\includegraphics[width=0.4\textwidth]{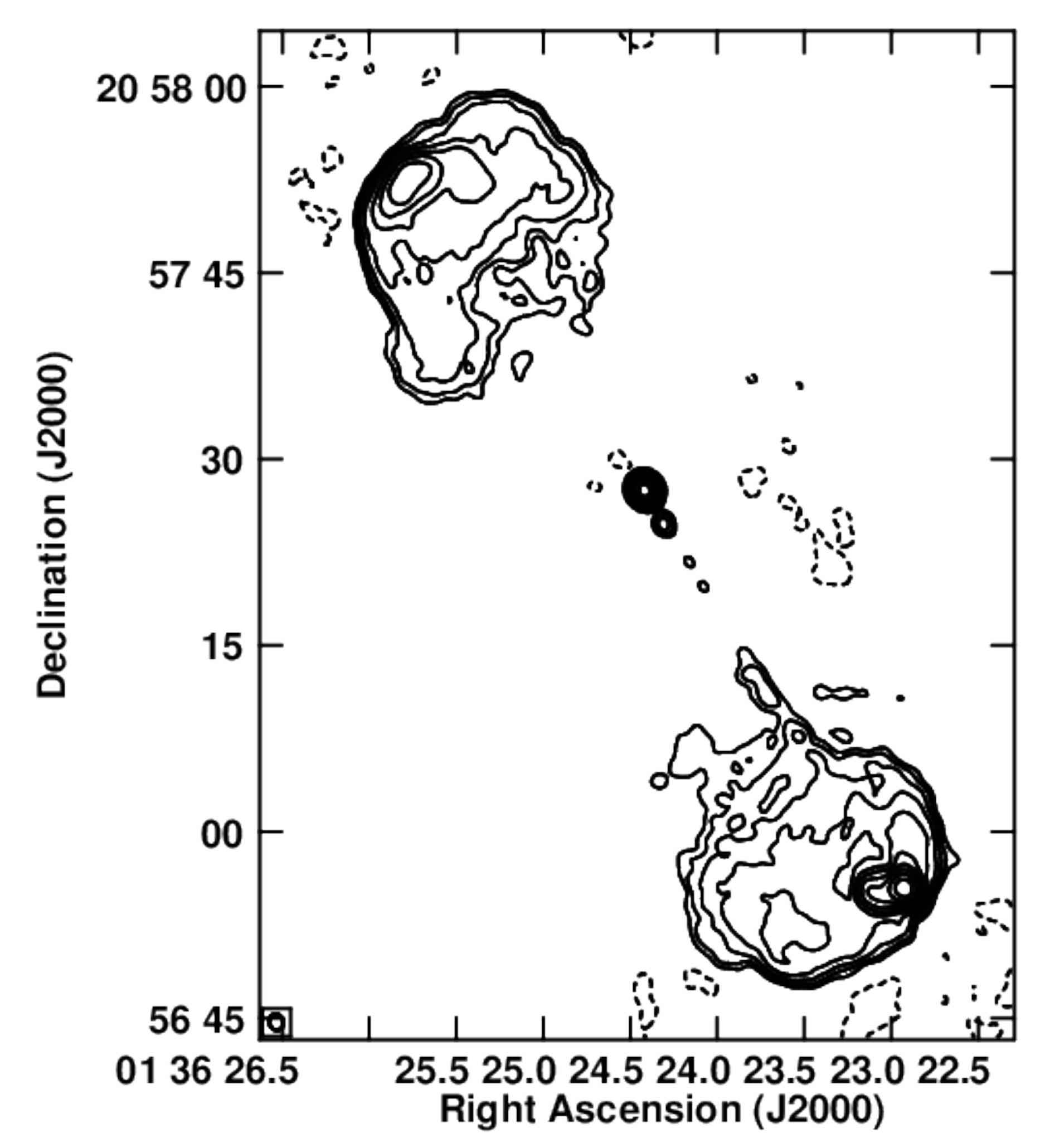}}	
	\caption{Total intensity maps at 1.4 GHz and 5 GHz.}
	\label{fig:3C14_rgb_tgss}
	\end{figure*}		

	\begin{figure*}
	\centering
	\subfloat[3C 109 :- Total intensity contour map at 1.4 GHz.  The peak surface brightness is 0.313 Jy beam$^{-1}$.  The contours are such that the levels increase in steps of 2 (-0.175 (dashed), 0.175,...,90)\% of the peak brightness.  The CLEAN beam FWHM is 1.60$\times$1.21 arcsec and the P.A. is -35.81 (shown in the inset in the lower left).]{\includegraphics[width=0.4\textwidth]{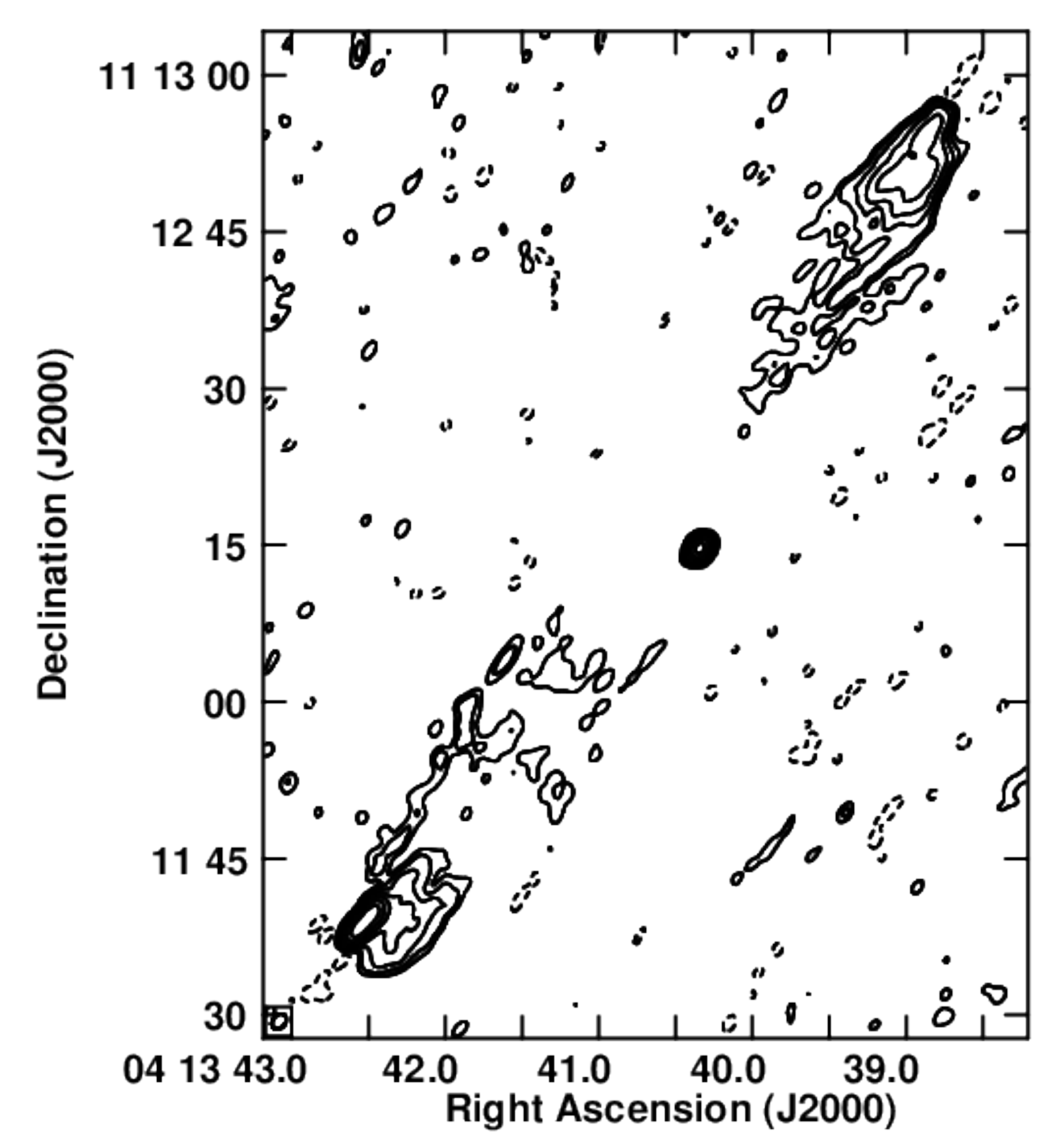}}
	\hspace{10pt}
	\subfloat[3C 109 :- Total intensity contour map at 5 GHz.  The peak surface brightness is 0.266 Jy beam$^{-1}$.  The contours are such that the levels increase in steps of 2 (-0.01 (dashed), 0.01,...,90)\% of the peak brightness.  The CLEAN beam FWHM is 1.50$\times$1.28 arcsec and the P.A. is -54.84 (shown in the inset in the lower left).]{\includegraphics[width=0.4\textwidth]{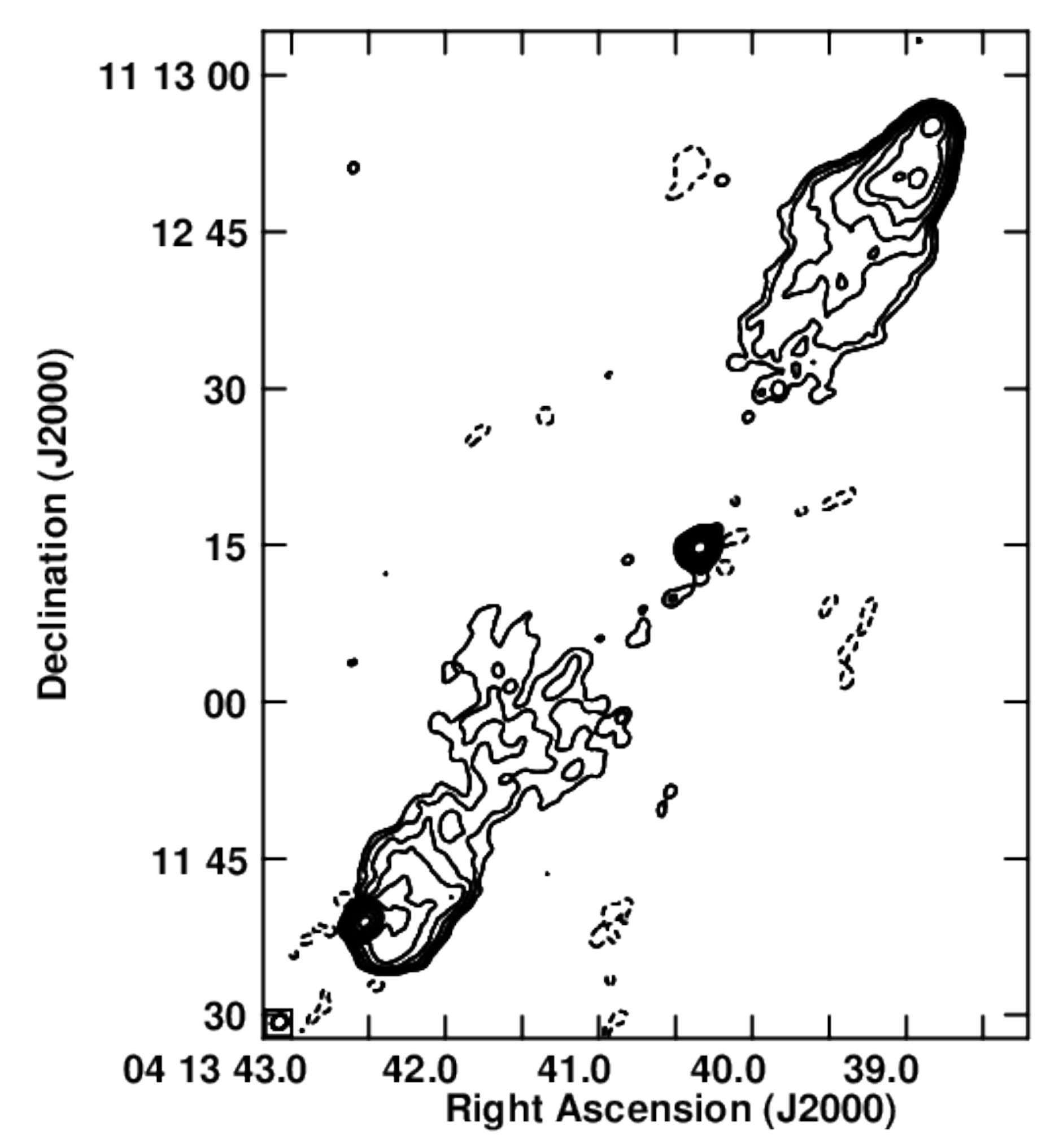}}	
	\caption{Total intensity maps at 1.4 GHz and 5 GHz.}
	\label{fig:3C14_rgb_tgss}
	\end{figure*}	

	\begin{figure*}
		\centering
		\subfloat[3C 204 :- Total intensity contour map at 1.4 GHz.  The peak surface brightness is 0.201 Jy beam$^{-1}$.  The contours are such that the levels increase in steps of 2 (-0.01 (dashed), 0.01,...,90)\% of the peak brightness.  The CLEAN beam FWHM is 1.99$\times$1.16 arcsec and the P.A. is 75.21 (shown in the inset in the lower left).]{\includegraphics[width=0.4\textwidth]{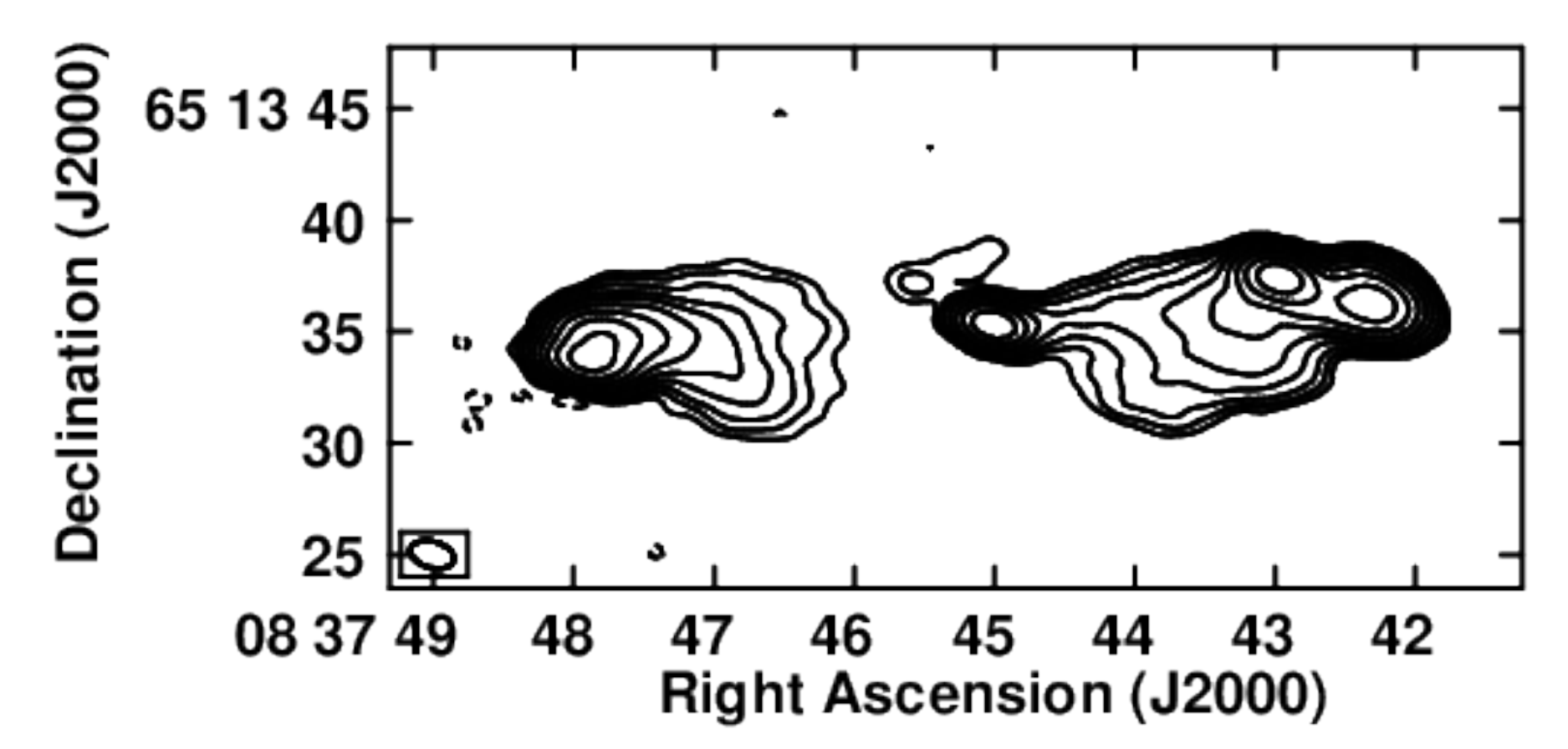}}
		\hspace{10pt}
		\subfloat[3C 204 :- Total intensity contour map at 5 GHz.  The peak surface brightness is 0.0863 Jy beam$^{-1}$.  The contours are such that the levels increase in steps of 2 (-0.02 (dashed), 0.02,...,90)\% of the peak brightness.  The CLEAN beam FWHM is 1.61$\times$1.07 arcsec and the P.A. is 50.76 (shown in the inset in the lower left).]{\includegraphics[width=0.4\textwidth]{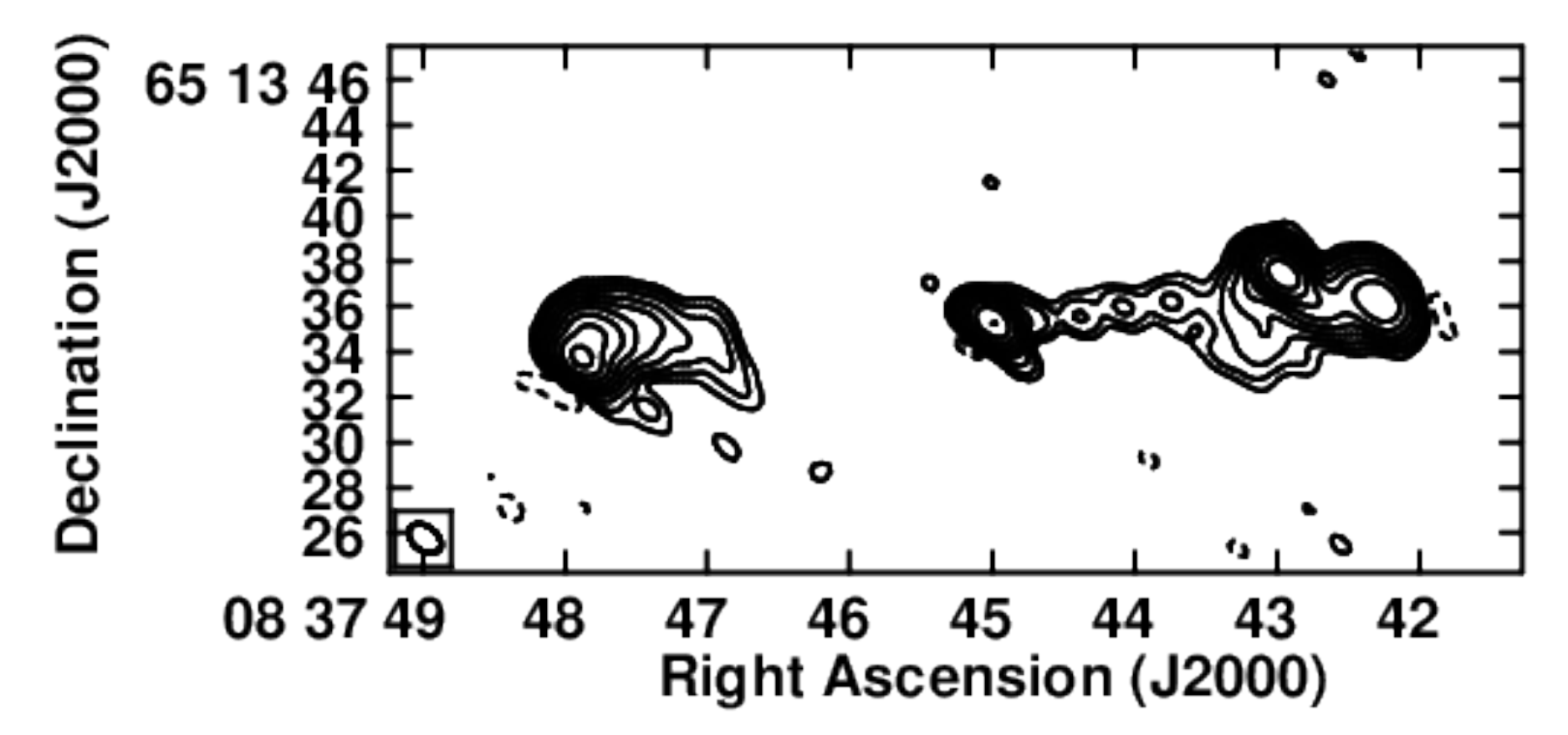}}	
		\caption{Total intensity maps at 1.4 GHz and 5 GHz.}
		\label{fig:3C14_rgb_tgss}
	\end{figure*}	

	\begin{figure*}
	\centering
	\subfloat[3C 205 :- Total intensity contour map at 1.4 GHz.  The peak surface brightness is 0.702 Jy beam$^{-1}$.  The contours are such that the levels increase in steps of 2 (-0.01 (dashed), 0.01,...,90)\% of the peak brightness.  The CLEAN beam FWHM is 2.13$\times$1.33 arcsec and the P.A. is 88.23 (shown in the inset in the lower left).]{\includegraphics[width=0.4\textwidth]{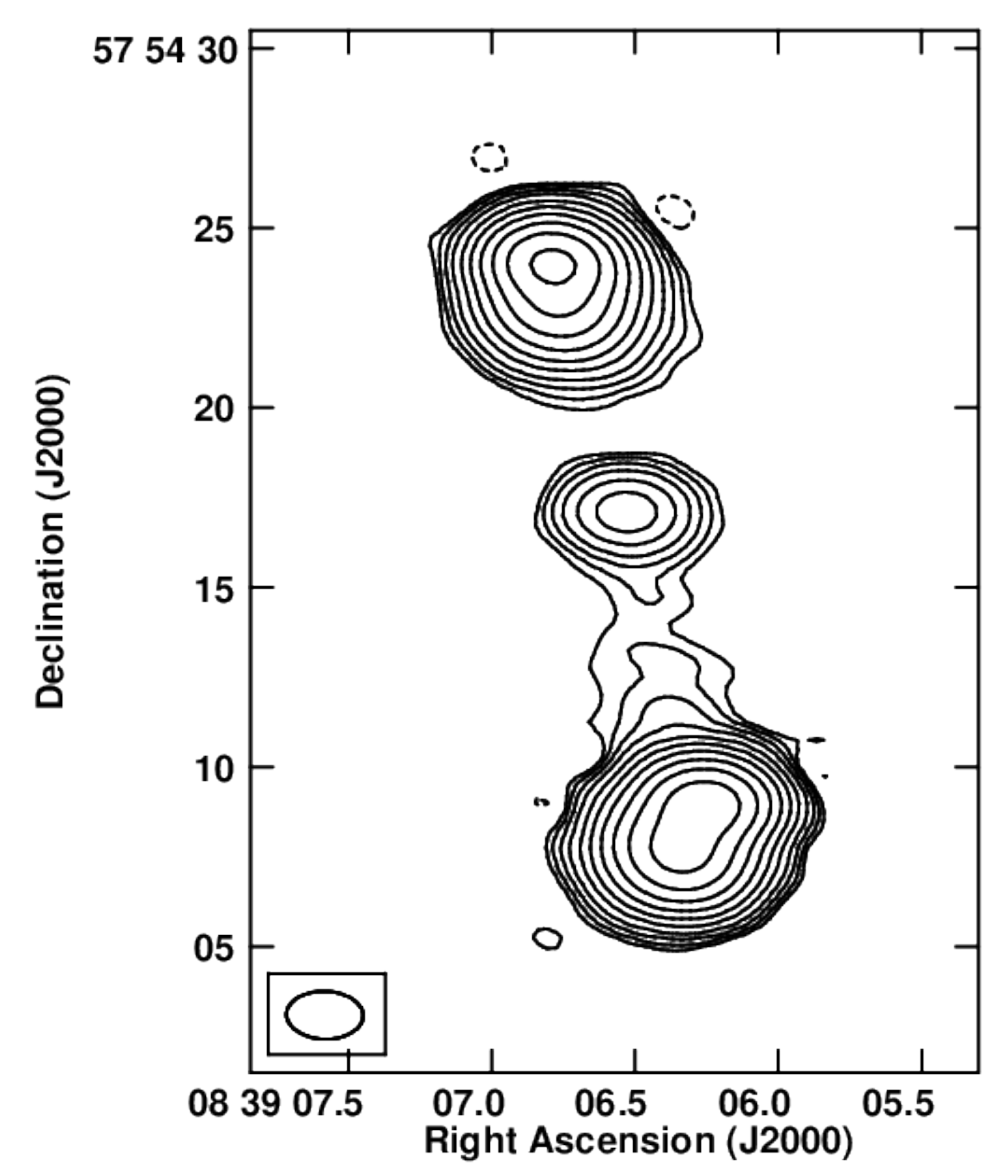}}
	\hspace{10pt}
	\subfloat[3C 205 :- Total intensity contour map at 5 GHz.  The peak surface brightness is 0.196 Jy beam$^{-1}$.  The contours are such that the levels increase in steps of 2 (-0.01 (dashed), 0.01,...,90)\% of the peak brightness.  The CLEAN beam FWHM is 1.87$\times$1.08 arcsec and the P.A. is 86.18 (shown in the inset in the lower left).]{\includegraphics[width=0.4\textwidth]{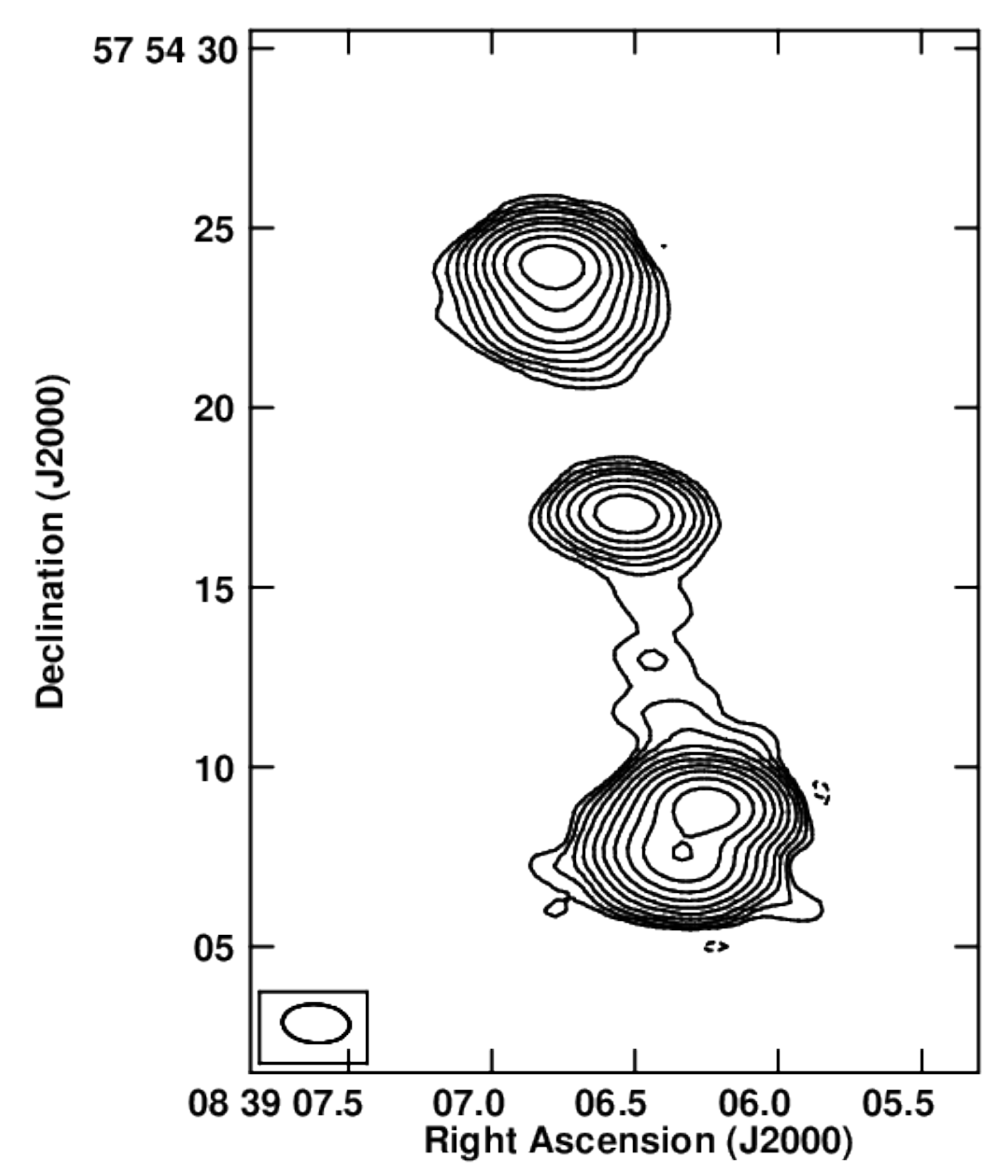}}	
	\caption{Total intensity maps at 1.4 GHz and 5 GHz.}
	\label{fig:3C14_rgb_tgss}
	\end{figure*}	
	\begin{figure*}
	\centering
	\subfloat[3C 208 :- Total intensity contour map at 1.4 GHz.  The peak surface brightness is 1.35 Jy beam$^{-1}$.  The contours are such that the levels increase in steps of 2 (-0.01 (dashed), 0.01,...,90)\% of the peak brightness.  The CLEAN beam FWHM is 1.46$\times$1.35 arcsec and the P.A. is 77.79 (shown in the inset in the lower left).]{\includegraphics[width=0.4\textwidth]{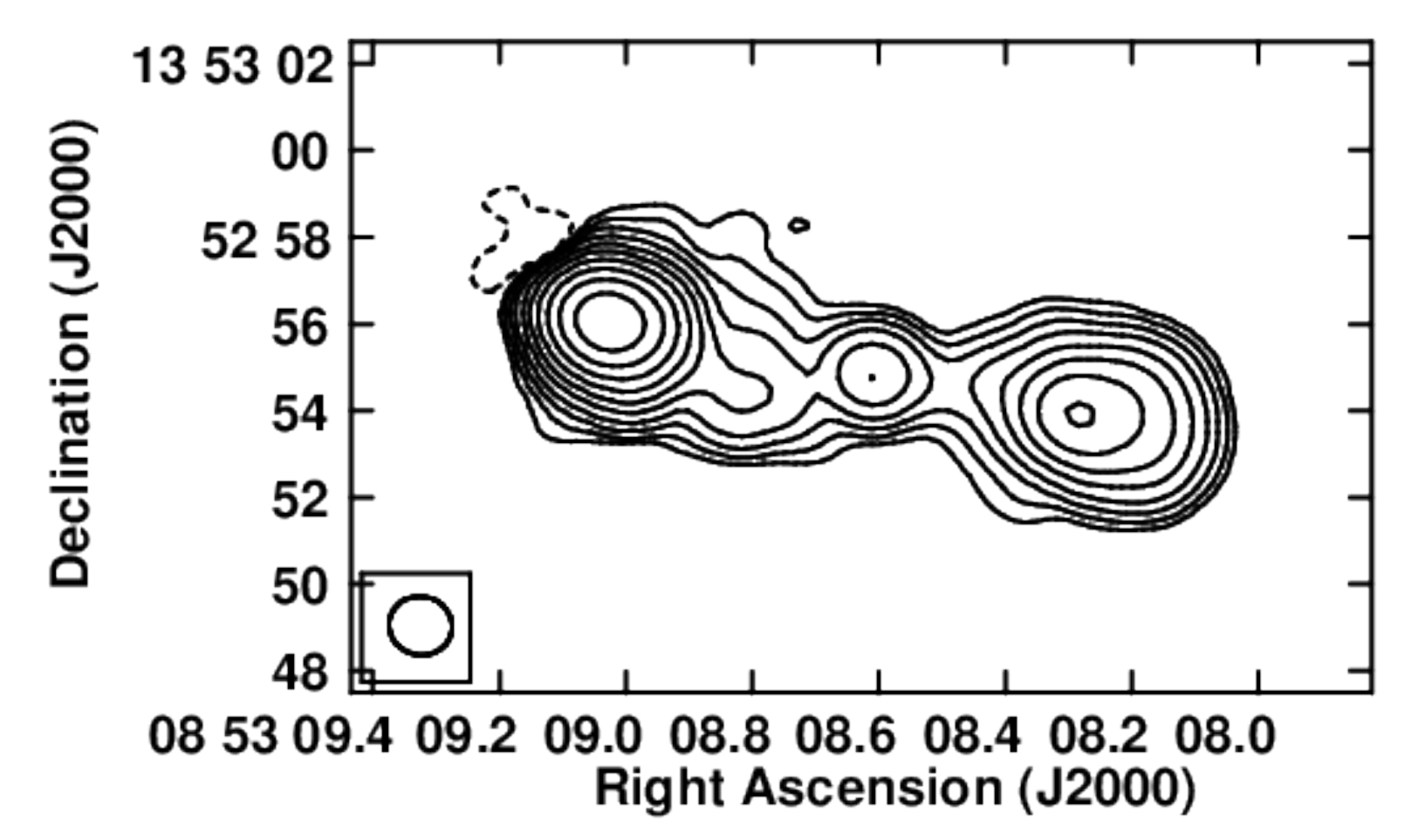}}
	\hspace{10pt}
	\subfloat[3C 208 :- Total intensity contour map at 5 GHz.  The peak surface brightness is 0.196 Jy beam$^{-1}$.  The contours are such that the levels increase in steps of 2 (-0.01 (dashed), 0.01,...,90)\% of the peak brightness.  The CLEAN beam FWHM is 1.87$\times$1.08 arcsec and the P.A. is -66.71 (shown in the inset in the lower left).]{\includegraphics[width=0.4\textwidth]{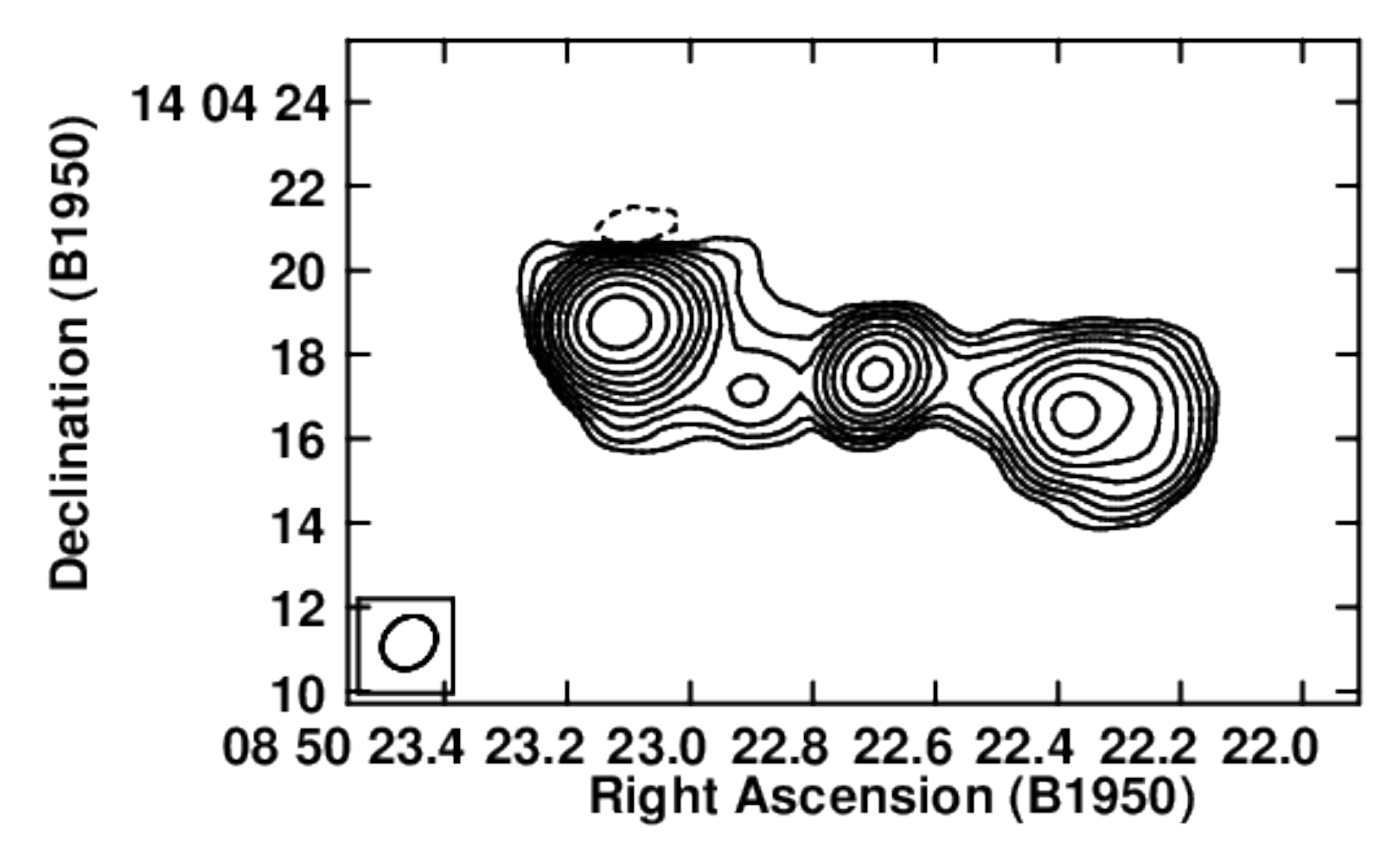}}	
	\caption{Total intensity maps at 1.4 GHz and 5 GHz.}
	\label{fig:3C14_rgb_tgss}
	\end{figure*}

	\begin{figure*}
	\centering
	\subfloat[3C 249.1 :- Total intensity contour map at 1.4 GHz.  The peak surface brightness is 0.329 Jy beam$^{-1}$.  The contours are such that the levels increase in steps of 2 (-0.02 (dashed), 0.02,...,90)\% of the peak brightness.  The CLEAN beam FWHM is 1.80$\times$1.12 arcsec and the P.A. is -29.29 (shown in the inset in the lower left).]{\includegraphics[width=0.4\textwidth]{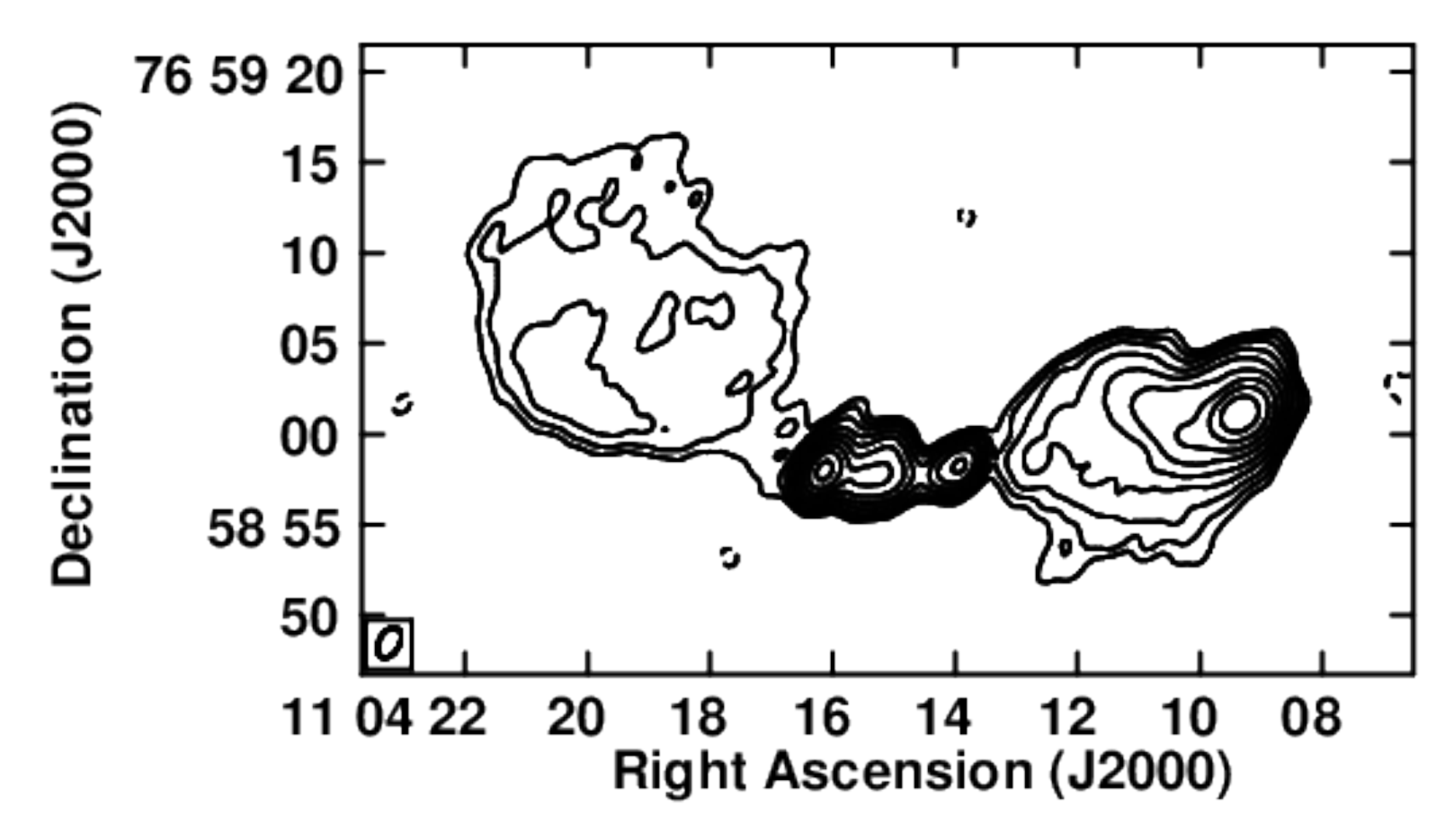}}
	\hspace{10pt}
	\subfloat[3C 249.1 :- Total intensity contour map at 5 GHz.  The peak surface brightness is 0.104 Jy beam$^{-1}$.  The contours are such that the levels increase in steps of 2 (-0.01 (dashed), 0.01,...,90)\% of the peak brightness.  The CLEAN beam FWHM is 1.79$\times$1.01 arcsec and the P.A. is -33.66 (shown in the inset in the lower left).]{\includegraphics[width=0.4\textwidth]{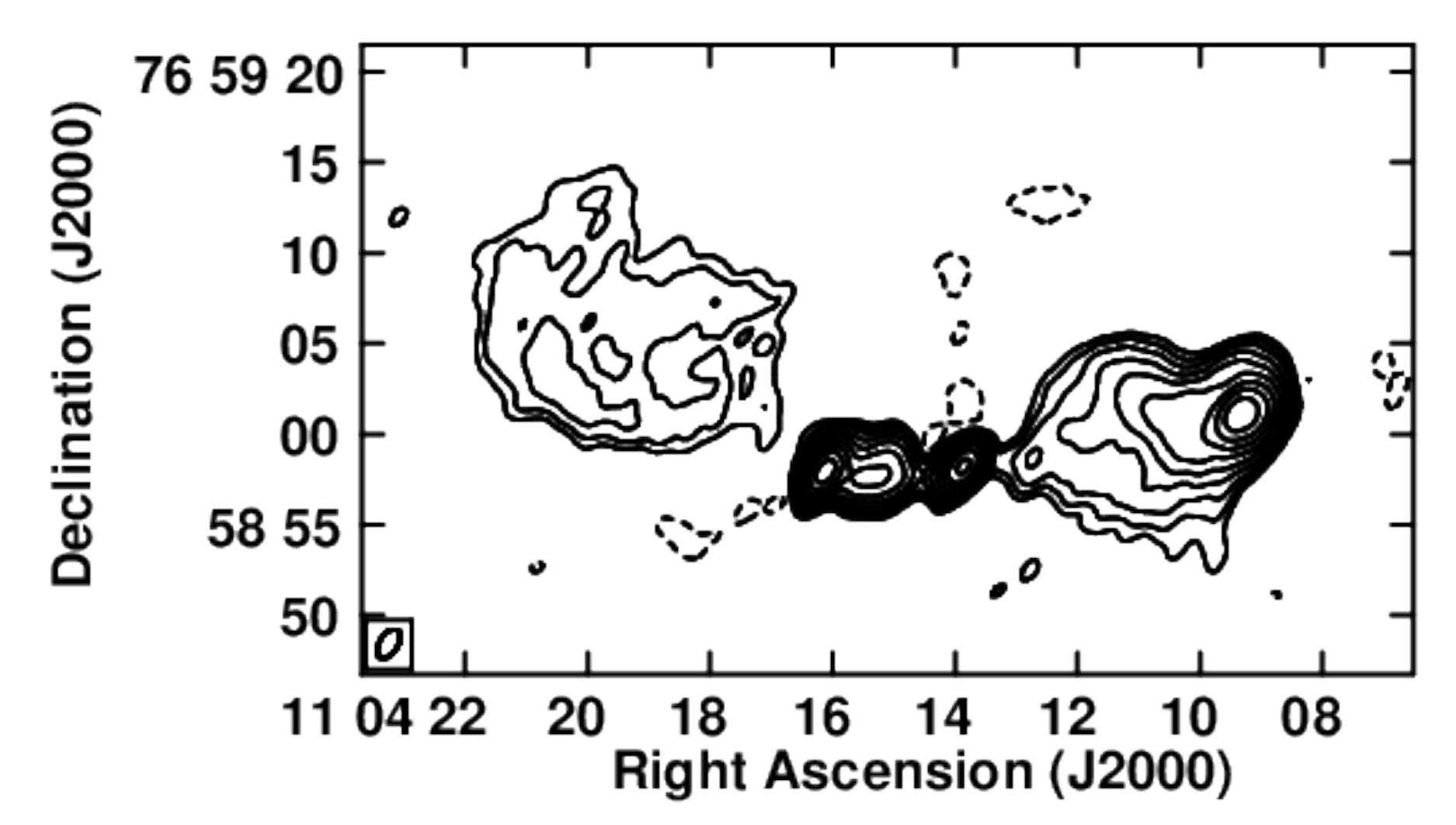}}	
	\caption{Total intensity maps at 1.4 GHz and 5 GHz.}
	\label{fig:3C14_rgb_tgss}
	\end{figure*}

	\begin{figure*}
	\centering
	\subfloat[3C 263 :- Total intensity contour map at 1.4 GHz.  The peak surface brightness is 1.50 Jy beam$^{-1}$.  The contours are such that the levels increase in steps of 2 (-0.01 (dashed), 0.01,...,90)\% of the peak brightness.  The CLEAN beam FWHM is 2.30$\times$1.14 arcsec and the P.A. is -63.55 (shown in the inset in the lower left).]{\includegraphics[width=0.4\textwidth]{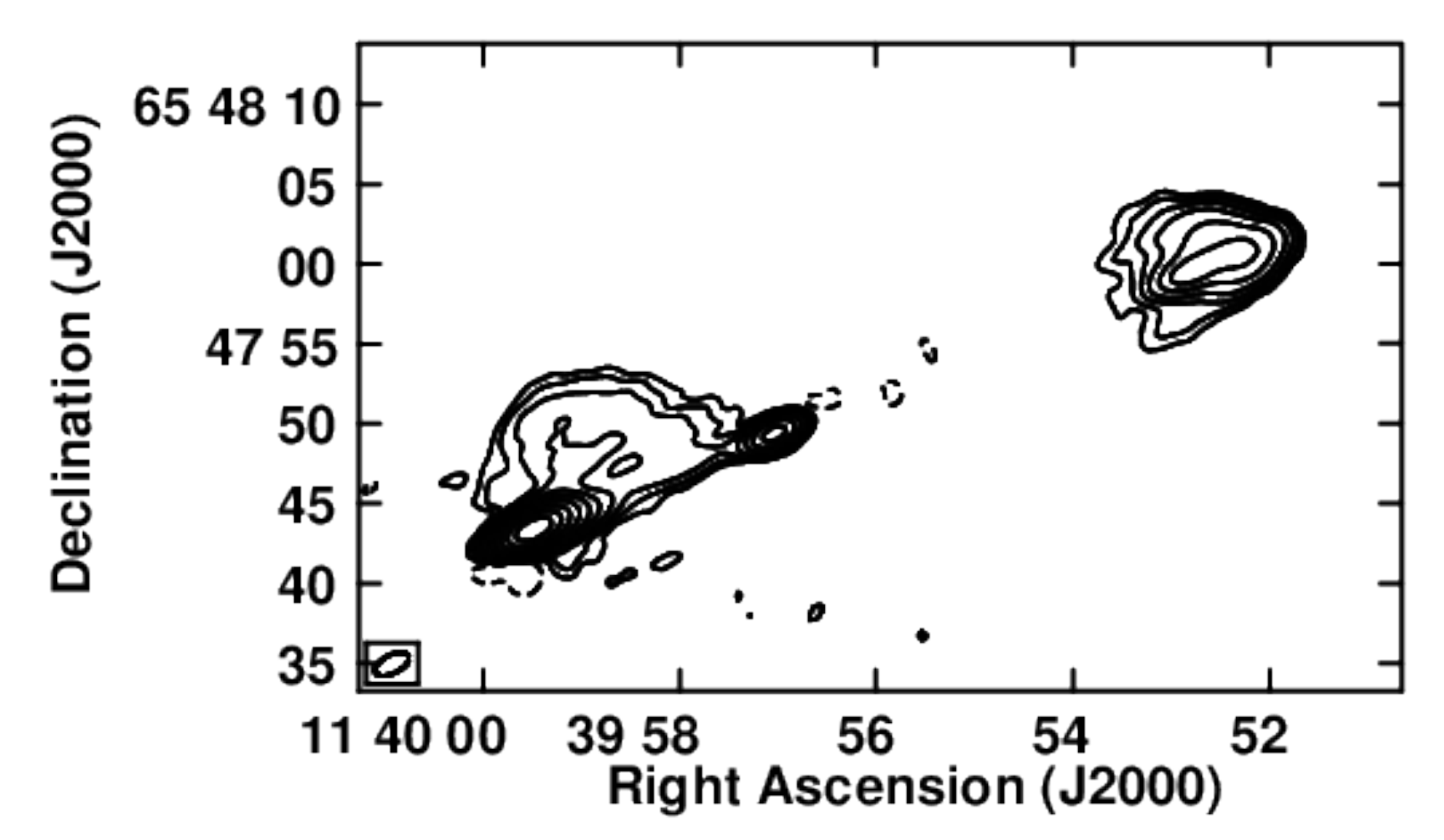}}
	\hspace{10pt}
	\subfloat[3C 263 :- Total intensity contour map at 5 GHz.  The peak surface brightness is 0.517 Jy beam$^{-1}$.  The contours are such that the levels increase in steps of 2 (-0.005 (dashed), 0.005,...,90)\% of the peak brightness.  The CLEAN beam FWHM is 1.83$\times$1.05 arcsec and the P.A. is -62.32 (shown in the inset in the lower left).]{\includegraphics[width=0.4\textwidth]{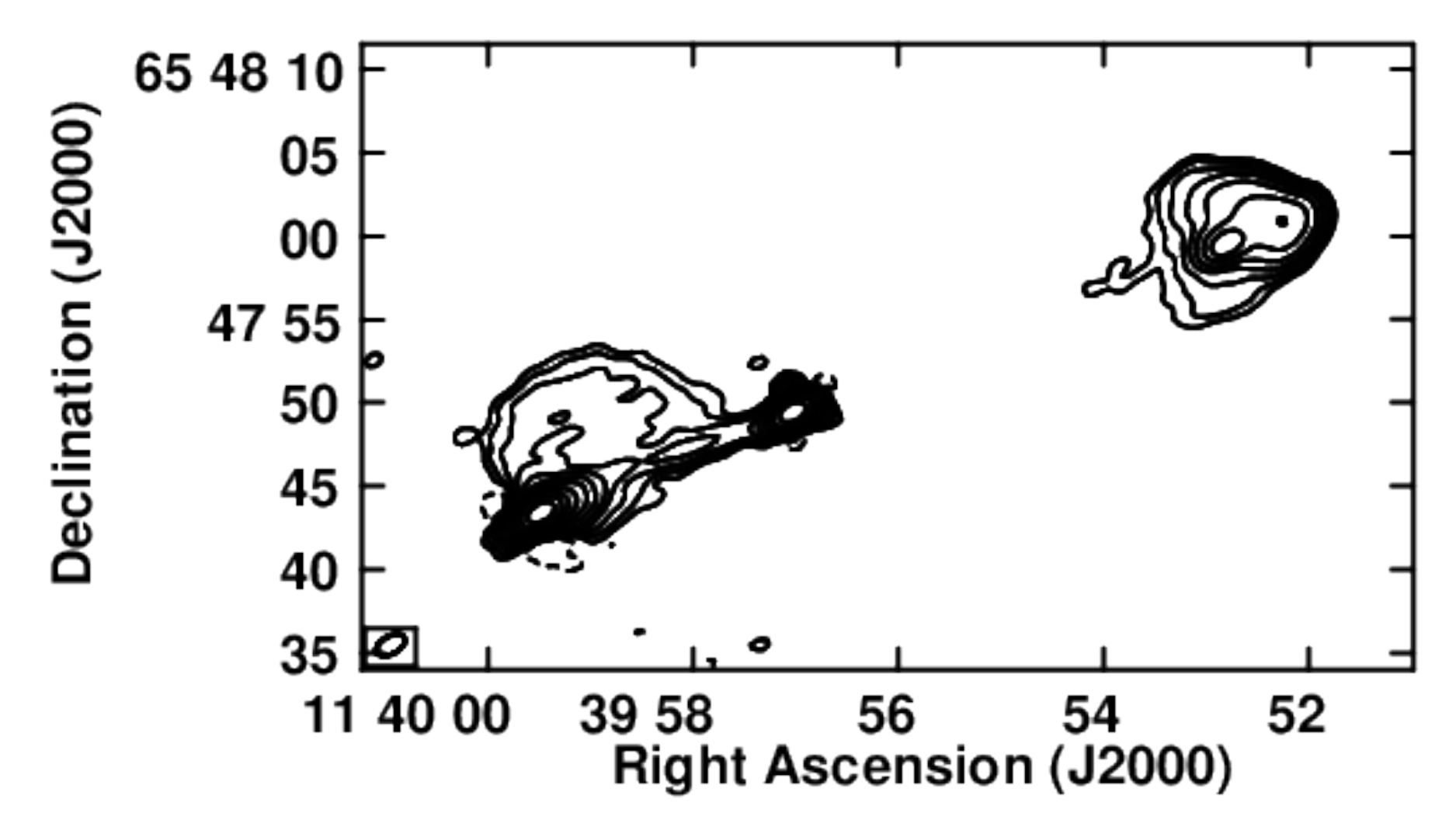}}	
	\caption{Total intensity maps at 1.4 GHz and 5 GHz.}
	\label{fig:3C14_rgb_tgss}
	\end{figure*}

	\begin{figure*}
	\centering
	\subfloat[3C 336 :- Total intensity contour map at 1.4 GHz.  The peak surface brightness is 0.624 Jy beam$^{-1}$.  The contours are such that the levels increase in steps of 2 (-0.02 (dashed), 0.02,...,90)\% of the peak brightness.  The CLEAN beam FWHM is 1.90$\times$1.40 arcsec and the P.A. is -72.57 (shown in the inset in the lower left).]{\includegraphics[width=0.4\textwidth]{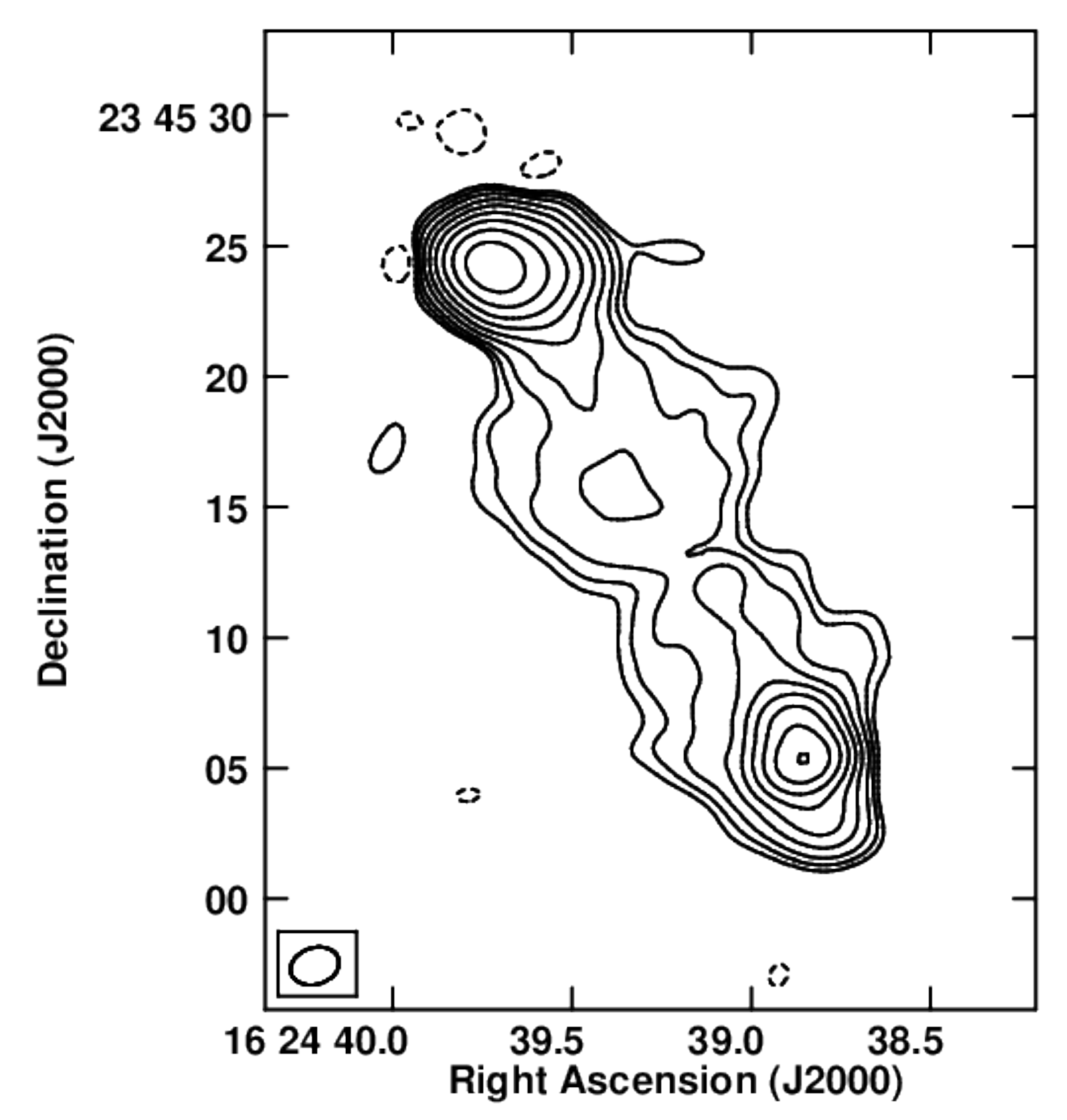}}
	\hspace{10pt}
	\subfloat[3C 336 :- Total intensity contour map at 5 GHz.  The peak surface brightness is 0.0925 Jy beam$^{-1}$.  The contours are such that the levels increase in steps of 2 (-0.01 (dashed), 0.01,...,90)\% of the peak brightness.  The CLEAN beam FWHM is 1.32$\times$1.14 arcsec and the P.A. is -62.93 (shown in the inset in the lower left).]{\includegraphics[width=0.4\textwidth]{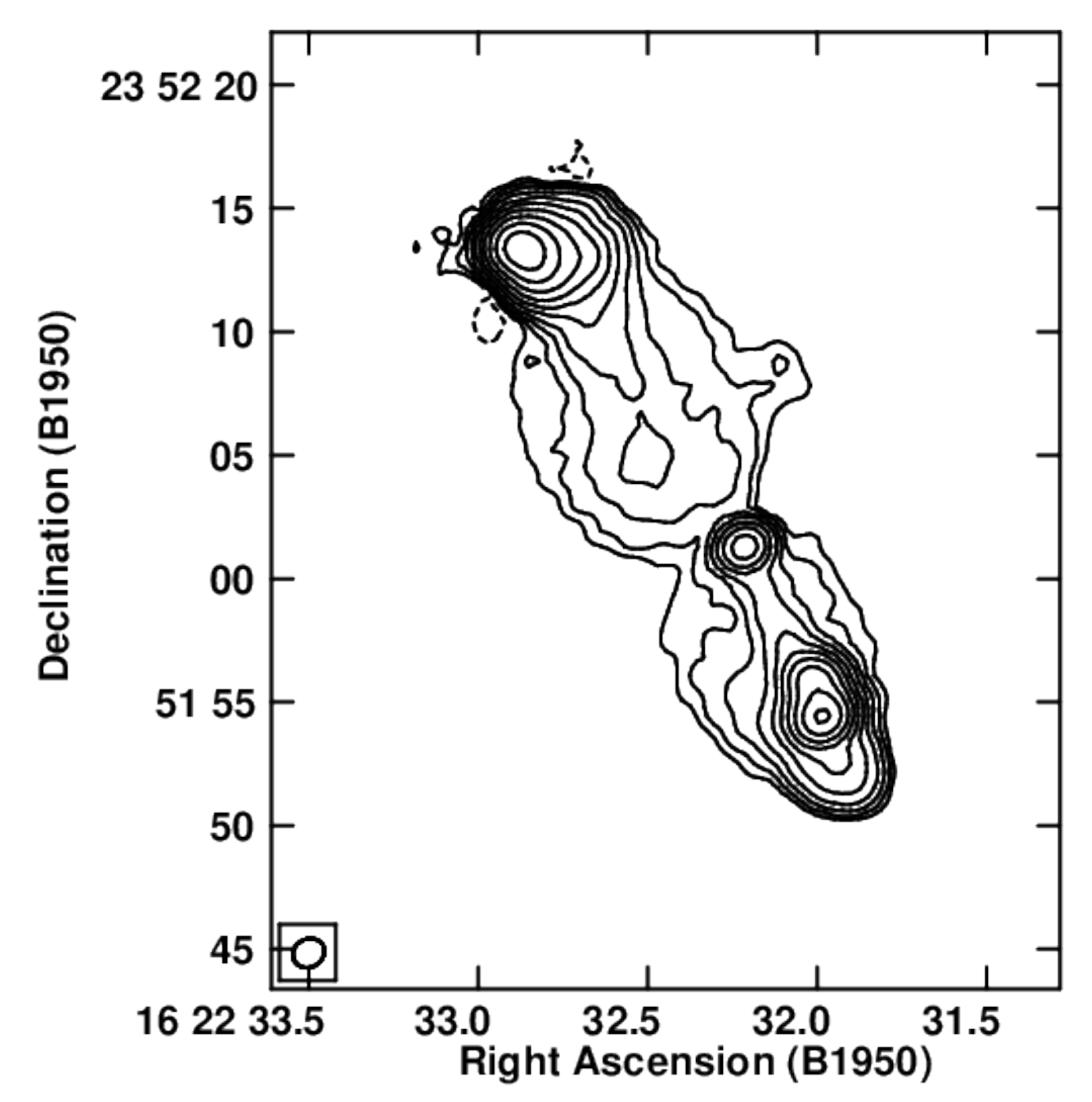}}	
	\caption{Total intensity maps at 1.4 GHz and 5 GHz.}
	\label{fig:3C14_rgb_tgss}
	\end{figure*}

	\begin{figure*}
	\centering
	\subfloat[3C 351 :- Total intensity contour map at 1.4 GHz.  The peak surface brightness is 0.911 Jy beam$^{-1}$.  The contours are such that the levels increase in steps of 2 (-0.005 (dashed), 0.005,...,90)\% of the peak brightness.  The CLEAN beam FWHM is 1.84$\times$1.26 arcsec and the P.A. is 81.23 (shown in the inset in the lower left).]{\includegraphics[width=0.4\textwidth]{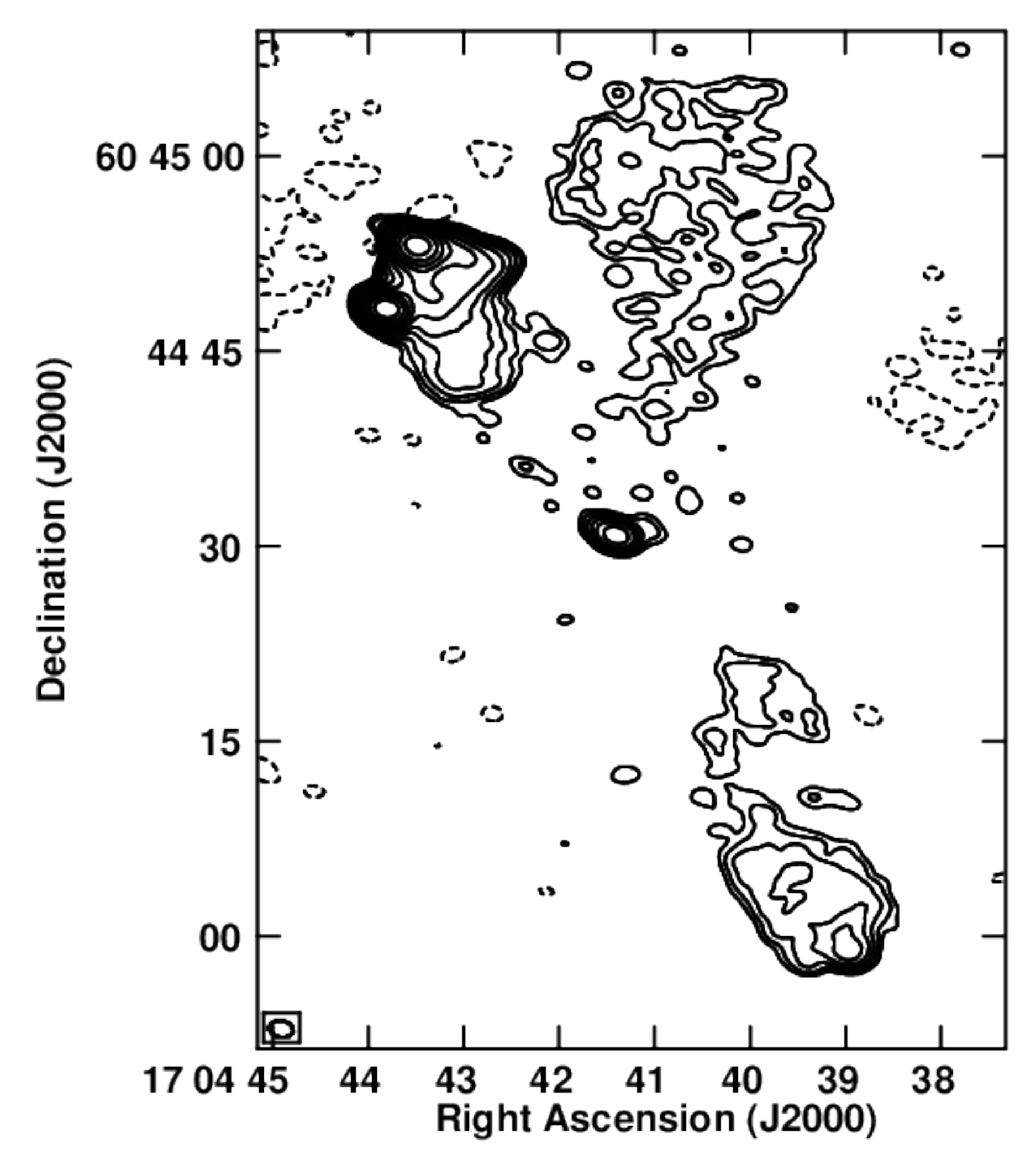}}
	\hspace{10pt}
	\subfloat[3C 351 :- Total intensity contour map at 5 GHz.  The peak surface brightness is 0.322 Jy beam$^{-1}$.  The contours are such that the levels increase in steps of 2 (-0.01 (dashed), 0.01,...,90)\% of the peak brightness.  The CLEAN beam FWHM is 1.40$\times$1.12 arcsec and the P.A. is -37.19 (shown in the inset in the lower left).]{\includegraphics[width=0.4\textwidth]{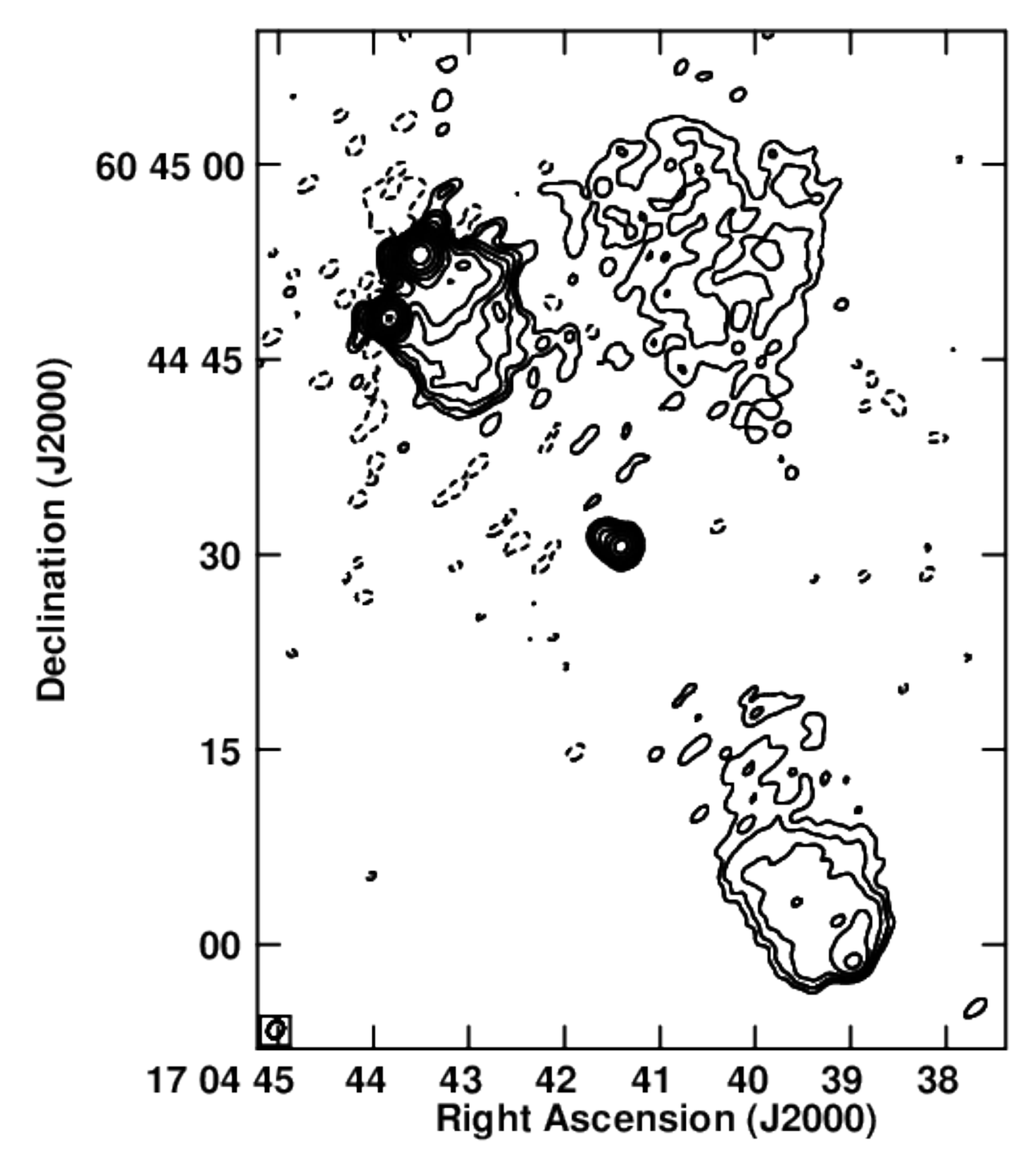}}	
	\caption{Total intensity maps at 1.4 GHz and 5 GHz.}
	\label{fig:3C14_rgb_tgss}
	\end{figure*}
	\begin{figure*}
	\centering
	\subfloat[3C 432 :- Total intensity contour map at 1.4 GHz.  The peak surface brightness is 0.746 Jy beam$^{-1}$.  The contours are such that the levels increase in steps of 2 (-0.005 (dashed), 0.005,...,90)\% of the peak brightness.  The CLEAN beam FWHM is 2.50$\times$1.42 arcsec and the P.A. is 66.75 (shown in the inset in the lower left).]{\includegraphics[width=0.4\textwidth]{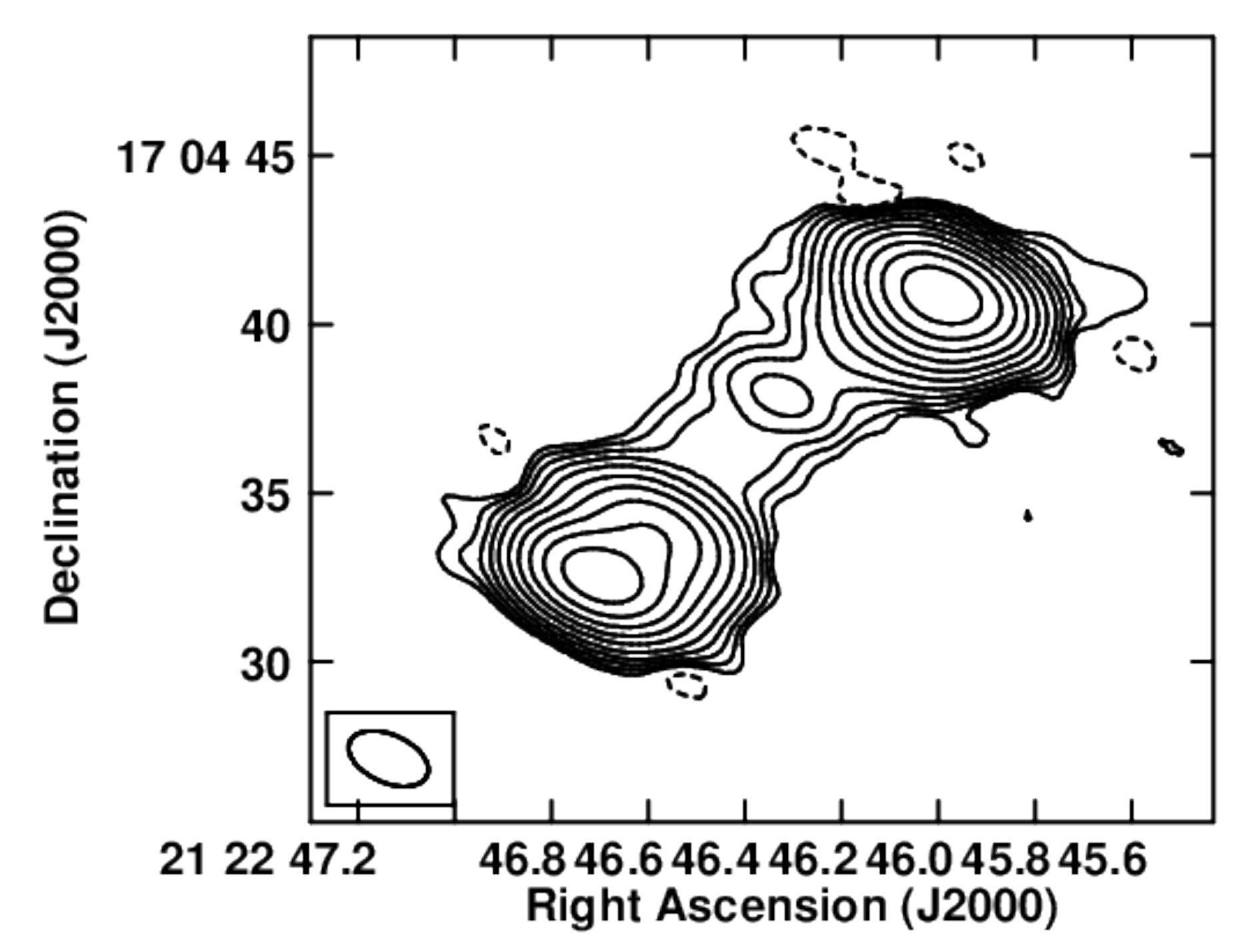}}
	\hspace{10pt}
	\subfloat[3C 432 :- Total intensity contour map at 5 GHz.  The peak surface brightness is 0.121 Jy beam$^{-1}$.  The contours are such that the levels increase in steps of 2 (-0.01 (dashed), 0.01,...,90)\% of the peak brightness.  The CLEAN beam FWHM is 1.44$\times$1.25 arcsec and the P.A. is -85.00 (shown in the inset in the lower left).]{\includegraphics[width=0.4\textwidth]{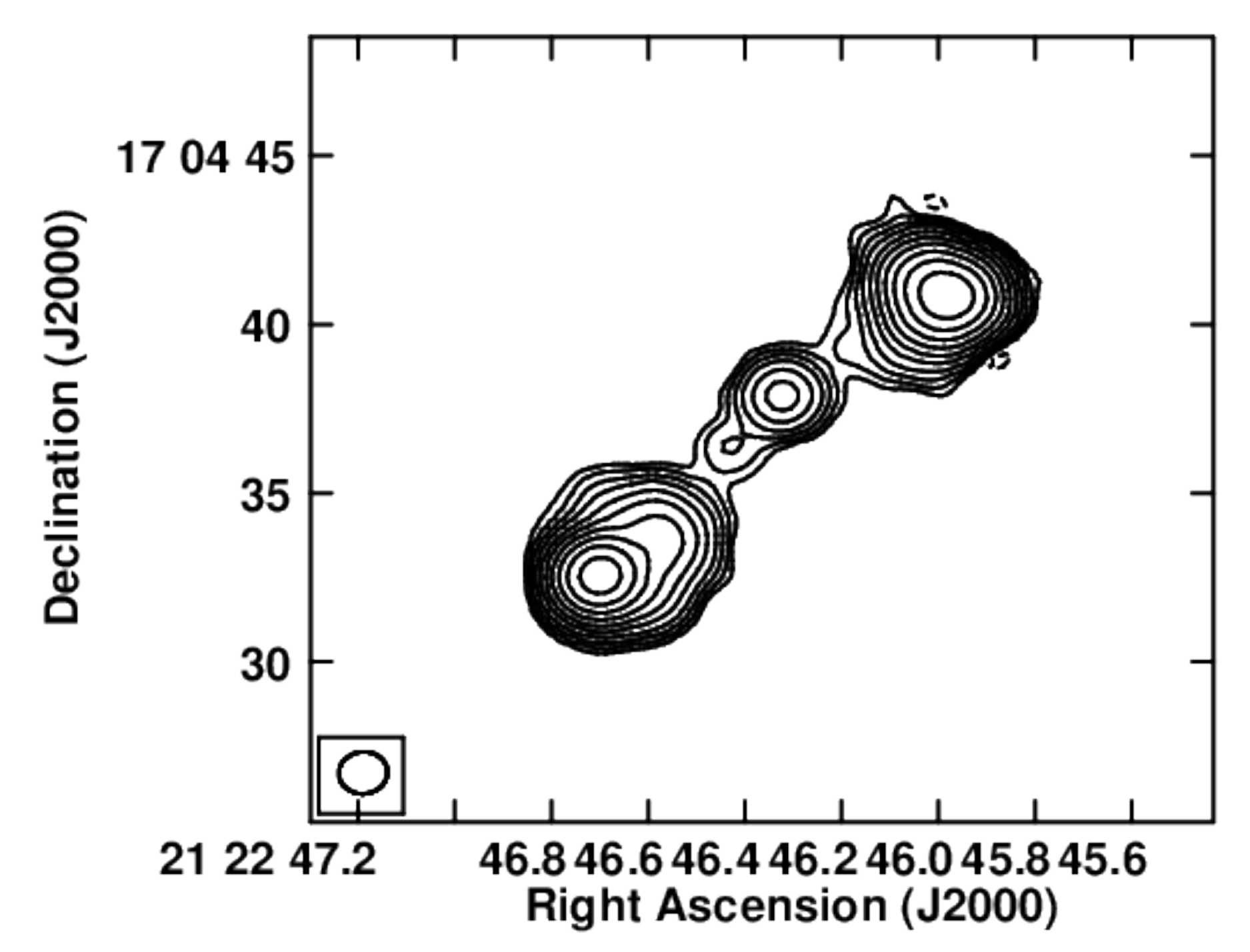}}	
	\caption{Total intensity maps at 1.4 GHz and 5 GHz.}
	\label{fig:3C14_rgb_tgss}
	\end{figure*}

	\begin{figure*}
	\centering
	\subfloat[3C204 -  TGSS contour map.  The lowest TGSS contour is at 0.08 Jy/Beam and the contours increase in steps of 2.    The peak intensity is 7.5 Jy/Beam and the rms is 5.2 mJy/Beam.]{\includegraphics[width=0.4\textwidth]{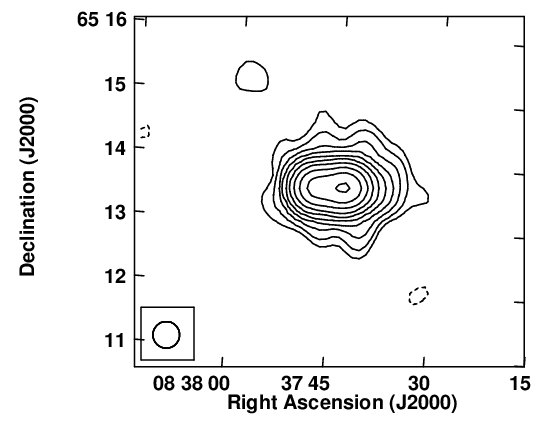}}
	\hspace{10pt}
	\subfloat[3C208 -  TGSS contour map.  The lowest TGSS contour is at 0.08 Jy/Beam and the contours increase in steps of 2.    The peak intensity is 20 Jy/Beam and an rms of 9 mJy/Beam.  TGSS data reveal extended structure that is perpendicular to the VLA 1.4 GHz jet-lobe axis]{\includegraphics[width=0.4\textwidth]{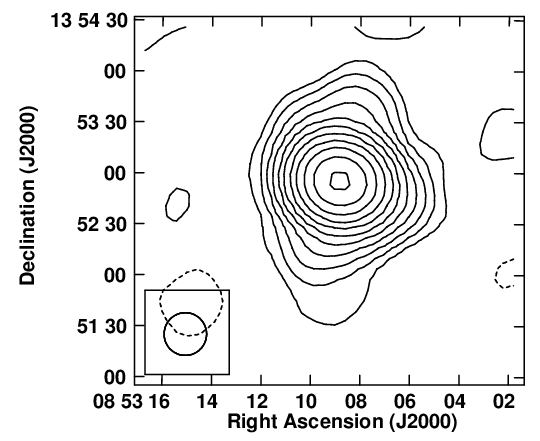}}\\	
	
	\subfloat[3C249.1 -  TGSS contour map.  The lowest TGSS contour is at 0.08 Jy/Beam and the contours increase in steps of 2.    The peak intensity is 11.5 Jy/Beam and an rms of 5 mJy/Beam. This source also shows extended radio emissionat 150 MHz along the axis perpendicular to that of the 1.4 GHz emission.]{\includegraphics[width=0.4\textwidth]{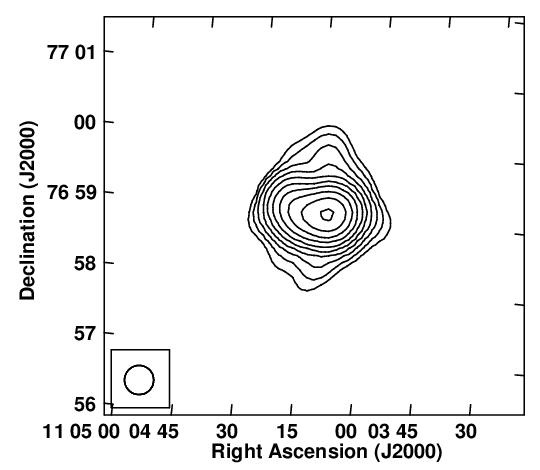}}
	\hspace{10pt}
	\subfloat[3C263 -  TGSS contour map.  The lowest TGSS contour is at 0.04 Jy/Beam and the contours increase in steps of 2.    The peak intensity is 14 Jy/Beam and an rms of 3.1 mJy/Beam.]{\includegraphics[width=0.4\textwidth]{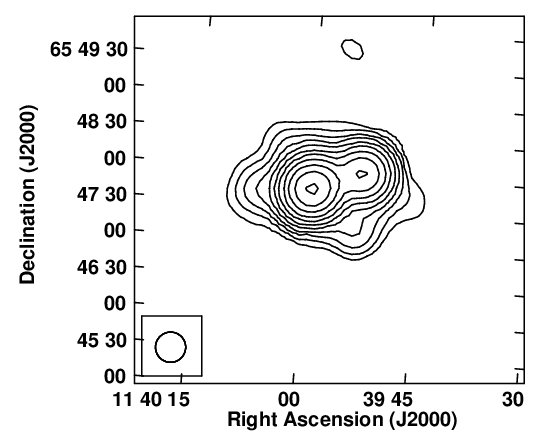}}\\	
	\caption{Total intensity maps at 150 MHz reveal diffuse extended emission much beyond the emission at 1.4 GHz.}
	\label{fig:tgss_quasars}
	\end{figure*}

{
	\begin{figure*}
    	\includegraphics[width=0.9\textwidth]{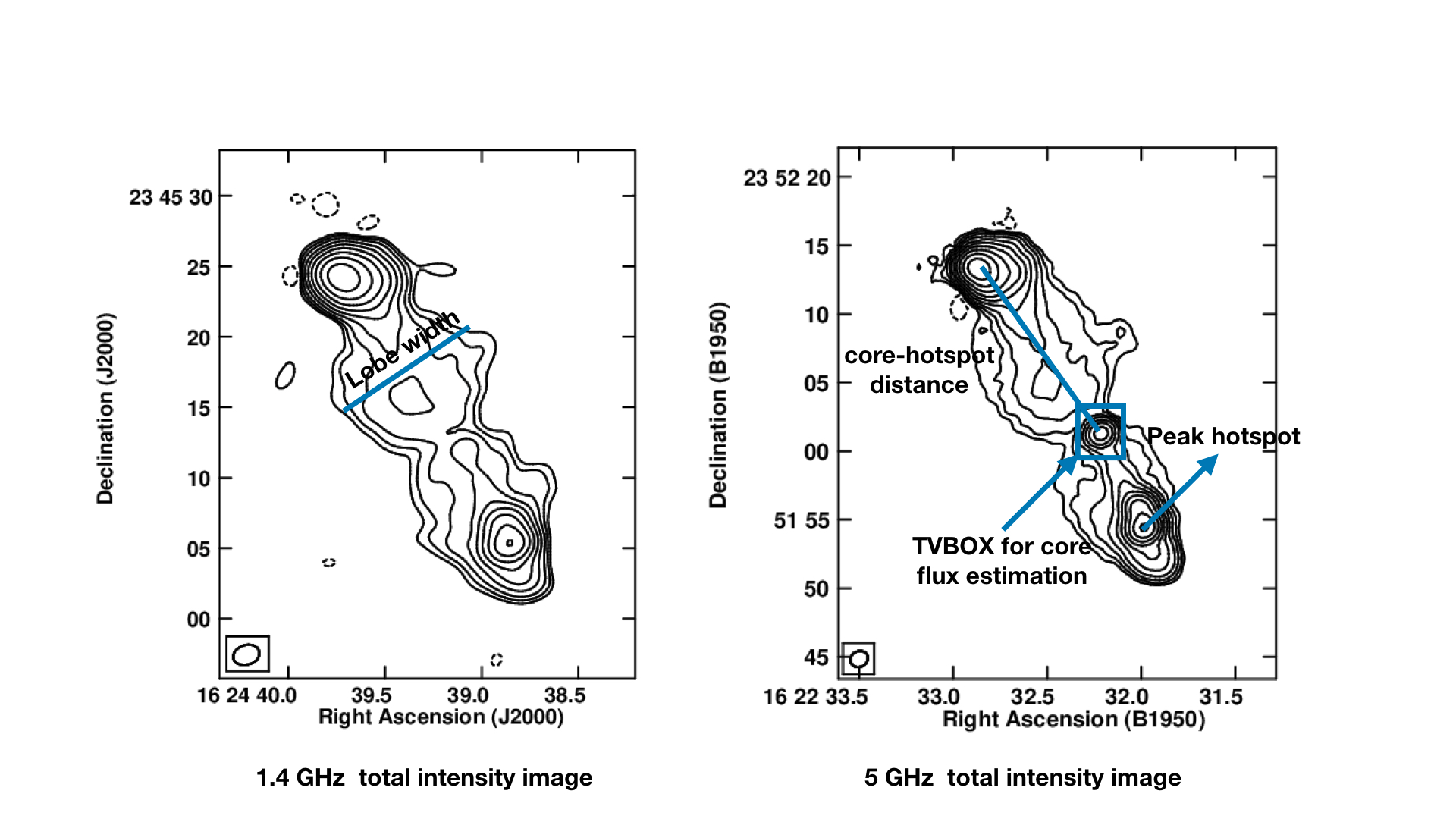}
        \caption{An example figure with labels to different components that are used to estimate different physical parameters discussed in Section ~\ref{sec:param} .}
        \label{fig:example_fig}
    \end{figure*}
}


%
%
%
\bsp	
\label{lastpage}
\end{document}